\documentclass{pasa}%

\usepackage{graphicx}
\def\hi{H{\sc i}}
\def\htwo{H$_{2}$}
\def\Msun{M$_{\odot}$}

\title[Cold gas stripping and quenching in satellites]{The Dawes Review 9: The role of cold gas stripping on the star formation quenching of satellite galaxies}

\author[L. Cortese et al.]{L. Cortese$^{1,2}$, B. Catinella$^{1,2}$, R. Smith$^3$
\affil{$^1$International Centre for Radio Astronomy Research, The University of Western Australia, 35 Stirling Hw, 6009 Crawley, Australia}%
\affil{$^2$ARC Centre of Excellence for All Sky Astrophysics in 3 Dimensions (ASTRO 3D), Australia}
\affil{$^3$Korea Astronomy and Space Science Institute (KASI), 776 Daedeokdae-ro, Yuseong-gu, Daejeon 34055, Republic of Korea}
}
\jid{PASA}
\doi{10.1017/pas.\the\year.xxx}
\jyear{\the\year}

\usepackage{aas_macros}
\usepackage{hyperref} 
\usepackage{url} 
\hypersetup{colorlinks,citecolor=blue,linkcolor=blue,urlcolor=blue}

\hypersetup{draft}

\begin{document}

\begin{frontmatter}
\maketitle

\begin{abstract}
One of the key open questions in extragalactic astronomy is what stops star formation in galaxies. While it is clear that the cold gas reservoir, which fuels the formation of new stars, must be affected first, how this happens and what are the dominant physical mechanisms involved is still a matter of debate. At least for satellite galaxies, it is generally 
accepted that internal processes alone cannot be responsible for fully quenching their star formation, but that environment should play an important, if not dominant, role. 
In nearby clusters, we see examples of cold gas being removed from the star-forming disks of galaxies moving through the intracluster medium, but
whether active stripping is widespread and/or necessary to halt star formation in satellites, or quenching is just a consequence of the inability of these galaxies to replenish their cold gas reservoirs, remains unclear.

In this work, we review the current status of environmental studies of cold gas in star-forming satellites in the local Universe from an observational perspective, focusing on the evidence for a physical link between cold gas stripping and quenching of the star formation. We find that stripping of cold gas is ubiquitous in satellite galaxies in both group and cluster environments. While hydrodynamical mechanisms such as ram pressure are important, the emerging picture across the full range of dark matter halos and stellar masses is a complex one, where different physical mechanisms may act simultaneously and cannot always be easily separated. Most importantly, we show that stripping does not always lead to full quenching, as only a fraction of the cold gas reservoir might be affected at the first pericentre passage. We argue that this is a key point to reconcile apparent tensions between statistical and detailed analyses of satellite galaxies, as well as disagreements between various estimates of quenching timescales.

We conclude by highlighting several outstanding questions where we expect to see substantial progress in the coming decades, thanks to the advent of the Square Kilometre Array and its precursors, as well as the next-generation optical and millimeter facilities.

\end{abstract}

{\it The Dawes Reviews are substantial reviews of topical areas  in  astronomy,  published  by  authors  of  international standing at the invitation of the PASA Editorial Board. The reviews recognise William Dawes (1762-1836), second lieutenant in the Royal Marines and the astronomer on the First Fleet. Dawes was not only an accomplished astronomer, but spoke five languages, had a keen interest in botany, mineralogy, engineering, cartography  and  music,  compiled  the  first Aboriginal-English dictionary, and was an outspoken opponent of slavery.}\\
\vskip 25pt
\end{frontmatter}

{\it This review is dedicated to the Arecibo 305m radio telescope in Puerto Rico, tragically collapsed on December 1st 2020, in recognition to its outstanding contribution to the field of research discussed in this work.}
\section{INTRODUCTION }
\label{sec:intro}

In our current paradigm of galaxy formation and evolution, galaxies are no longer thought of as isolated entities, but as self-regulated systems whose life is governed by the balance between gas replenishment (via inflows from the intergalactic medium), consumption (via star formation) and ejection (outflows) -- the so-called ``gas-regulator'' model (e.g., \citealp{oort70,larson72,rees1977,white91,keres05,bouche10,lilly13}). In star-forming galaxies, gas is converted into stars at a relatively steady rate, maintaining objects on a well-defined relation known as ``star-forming main sequence'' \citep[e.g.,][]{brinchmann04,noeske07,whitaker12}. If this equilibrium is somehow broken, star formation is directly affected and galaxies experience either starbursts or quenching phases. While the starburst phase is generally short-lived ($\sim$10$^{8}$ yr or less, e.g., \citealp{larson78,rieke80,heckman90}), quenching appears to be a critical stage in the life-cycle of galaxies. Very few systems seem able to recover from it and resume a star formation activity typical of the main sequence. Indeed, understanding what makes galaxies passive (e.g., reduce their star formation by a factor $\sim$4 or more, roughly $\gtrsim$2$\sigma$ below the main sequence) is one of the main open questions in galaxy evolution.

A number of physical processes have been invoked to explain how star formation is quenched in galaxies. Even though this question remains unanswered, two key facts have become clear. First, a basic requirement to halt star formation is to affect the galaxy's cold gas reservoir by either removing it, consuming it or keeping it stable against fragmentation. Second, the way that galaxies quench is very different depending on whether they are satellites of a bigger galaxy within their host halo or are centrals.

In the case of central galaxies -- the most massive galaxies within a halo and usually those sitting at (or near) the centre of the dark matter potential well --, ``internal'' processes such as feedback from star formation and/or accreting super-massive black holes are currently the most popular culprits to explain quenching in the local galaxy population (e.g., \citealp{croton06}). Conversely, the fate of satellite galaxies appears to be directly shaped by the external physical processes influencing them while orbiting around the central system in their host dark matter halo (e.g., \citealp{GUNG72,vdbosch08,weinmann09,wetzel12,bluck20}).

{\it How is the gas cycle in satellites affected such that star formation is quenched? }
In this review we focus on the physical mechanisms that halt star formation in satellite galaxies in the local Universe, with particular emphasis on the importance of active stripping of their cold gas reservoirs. By ``cold gas'' we refer to both atomic and molecular hydrogen (\hi\ and \htwo, with typical temperatures $T<10^{4}$ K and $T<50$ K, respectively) in the interstellar medium (ISM) of galaxies, which fuel star formation. We aim to review the progress made in this field in the past few decades, and summarize the status of our knowledge before the next-generation \hi\ surveys with the Square Kilometre Array \citep[SKA;][]{SKA} and its pathfinders open a new era for studies of gas in galaxies. While doing so, we endeavour to clarify common misconceptions, as well as perceived tensions between theoretical and observational approaches attempting to identify \underline{the} physical process driving galaxies outside the star-forming main sequence.

To set the background, we begin by reviewing the quenching mechanisms usually invoked to  break the gas-star-formation cycle in satellite galaxies (\S~2). Next, we summarise and contrast the various definitions of {\it gas poorness} or {\it deficiency} adopted in the literature, as this is a key property to quantify environmental effects on the cold gas reservoirs of galaxies that has led to substantial confusion in the field (\S~3). At this point, we are well equipped to critically examine the observational evidence for cold gas stripping, its effect on star formation and the important, related question of quenching timescales for satellite galaxies in nearby clusters (\S~4) and groups (\S~5), as well as the potential role of pre-processing and large-scale structure in gas stripping and quenching (\S~6). This will allow us to build a comprehensive picture of how environment affects the cold gas and star formation of satellites (\S~7). While this is mostly an observational review, we link our discussion to predictions of theoretical models throughout, and towards the end provide a high-level view of what can be learned on this topic from state-of-the-art semi-analytic and hydrodynamical simulations of galaxy formation and evolution (\S~8). We conclude by looking ahead and highlighting future prospects and challenges for this field (\S~9).

\noindent

\section{Quenching mechanisms in satellite galaxies}
When it comes to studying the effect of nurture on galaxy evolution, it has been common practice to conceptually categorize 
complex physical processes into separate boxes, and to force either observations or models to fit into the single box able to better reproduce the observed properties of galaxies. This has led not only to debates focused on 
perceived dichotomies (e.g., ram pressure vs. starvation, hydrodynamical vs. tidal interactions, stripping vs. evaporation, etc.), but also to the use of the same word (e.g., starvation, see below) to refer to different physical processes, creating some confusion among the community.

As today's passive satellites have typically spent at least a few billion years in their host halos (e.g., \citealp{mcgee09,han18}), during such a long time it is inevitable that most of the primary quenching mechanisms invoked in the literature must have played a role in shaping their properties. 
Thus, the question of {\it which} physical process affects satellite galaxies can be re-framed in terms of {\it how}, {\it when} and {\it for how long (i.e., on which timescale)} each process contributes to halt the gas star-formation cycle.

It is indeed now clear that in most cases multiple processes are at play simultaneously, and that 
it may not always be possible to isolate a dominant one, at least observationally, especially when we consider the population of passive satellites in the local Universe rather than individual objects. Instead, a more pragmatic approach is to first identify which step(s) of the star-formation cycle has(have) been broken, causing star formation to halt, and then assess the role and impact of the various physical processes that might have led to it. As shown schematically in Fig.~1, there are several mechanisms that could decrease or even halt star formation, by stopping the supply of gas that replenishes the ISM or by removing cold gas from the star-forming disk (either via external processes or internal ones, such as heating/ejection by nuclear activity and stellar feedback). Cold gas could also simply be consumed by star formation, or stabilised against fragmentation and thus unable to form new stars. We briefly review these scenarios below.

\begin{figure*}[t]
\includegraphics[width=17.7cm]{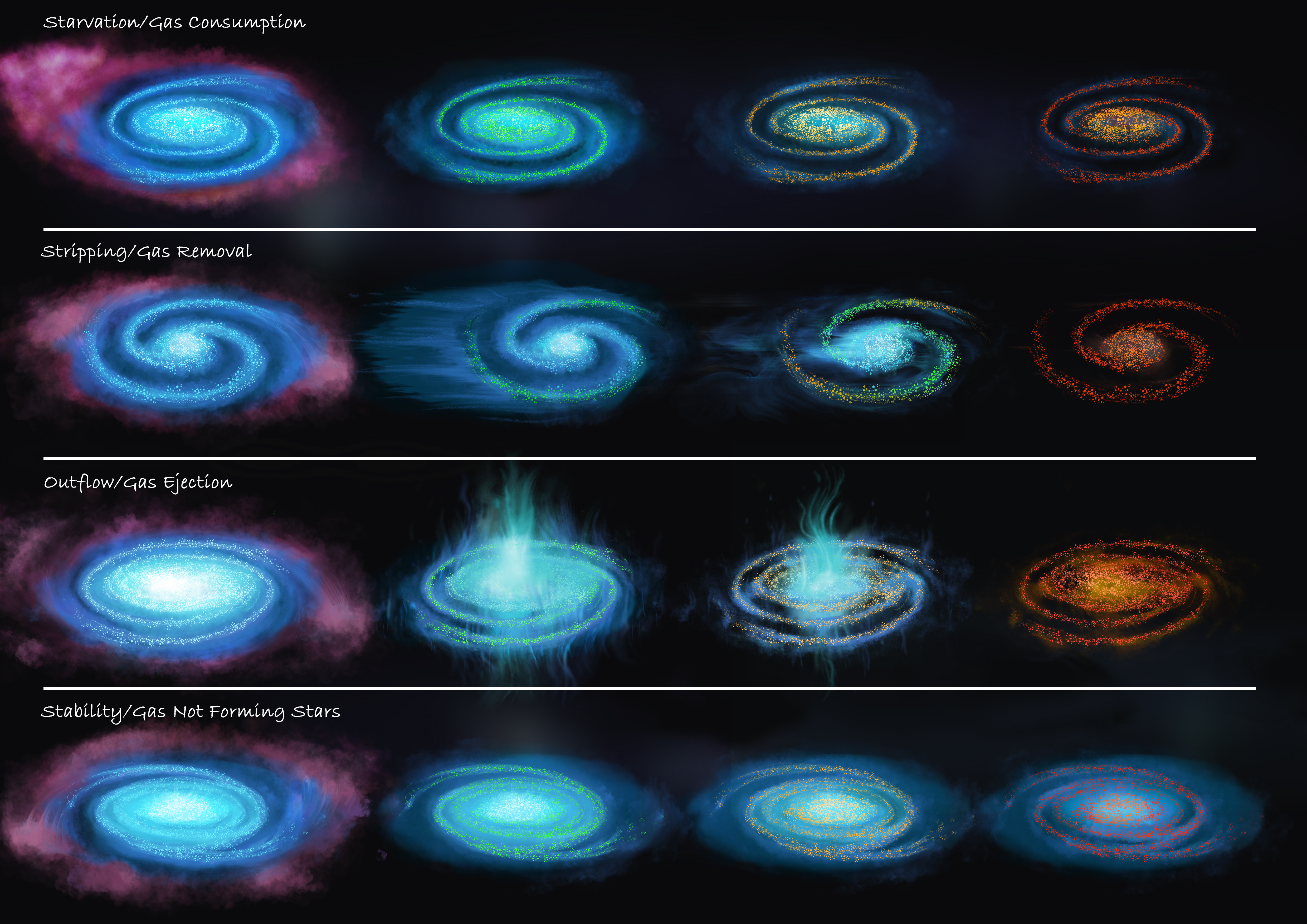}
\caption{Illustration showing the various quenching pathways discussed in \S~2, with particular emphasis on what happens to the cold gas component of the ISM (diffuse blue). Each `quenching sequence' starts with the galaxy losing its ability to accrete gas from the surrounding CGM/IGM (pink). The color of the stars (from blue/young to red/old) indicates the stage of quenching.}
\label{Fig_cartoon}
\end{figure*}

\subsection{Starvation -- halting gas accretion}
At least for galaxies with stellar masses greater than $\sim$10$^{9}$ M$_{\odot}$, the typical timescale for cold gas depletion (i.e., the ratio of gas mass to star formation rate, $\tau_{dep}\equiv M_{gas}/SFR$) is a few billion years for both molecular and atomic phases, assuming a constant SFR \citep{kennicut83,boselli01,saintonge17}. This is generally used to argue that, in order to keep forming stars at a constant rate, the cold reservoir of galaxies in star-forming disks needs to be continuously replenished. Support to this picture also comes from the fact that, at least for molecular hydrogen, depletion times appear to have been even shorter at higher redshift \citep{schinnerer16,scoville16}. 

While how gas gets into galaxies is still unclear, cosmological simulations suggest that, in the last few billion years, cold accretion via filaments and/or smooth accretion from the intergalactic medium (IGM) have gradually become inefficient \citep{dekel09,alcazar17,grand19}, as cold streams no longer reach the galactic disk. Instead, either accretion of gas-rich companions (i.e., minor or major mergers) or wind recycling (e.g., direct cooling of the hot circumgalactic medium via galactic fountain; \citealp{fraternali08,marinacci10}) appear to be the most likely accretion channels to keep local star-forming galaxies active. Regardless of which physical mechanism(s) is(are) responsible for quenching star formation in the first place, further accretion of cold gas needs to be prevented in order to avoid that quenched galaxies restart their star formation. Indeed, while rare, rejuvenation cases are both observed in the local Universe \citep{cortese09,thilker10} and predicted by numerical simulations \citep{birnboim07,nelson18}. Thus, in general, a galaxy that no longer has access to a source of gas will eventually run out of fuel and stop forming stars (Fig.~\ref{Fig_cartoon}, top row).

In the literature, the lack of replenishment of cold gas in the ISM is generally referred to as {\it starvation}. This general idea was originally proposed by \cite{larson80} to explain the existence of passive disk galaxies outside clusters, where stripping of the ISM could not be effective. In this case, accretion was thought to take place via either minor mergers or infall of clouds of cold gas (i.e., primarily atomic hydrogen), and galaxies with no access to such a supply would simply run out of gas. With the advent of semi-analytic cosmological models, this picture has significantly evolved into the removal of the hot gas component (only) in the halo of the galaxy (i.e., the primary reservoir of gas directly available to the galaxy to support star formation;  \citealp{cole00}). This also corresponds to the introduction of the term {\it strangulation} \citep{balogh00}, and now both {\it starvation} and {\it strangulation} are interchangeably used to refer to the cessation of accretion from both IGM and circumgalactic medium (CGM).

Regardless of how exactly gas infall is halted, it is obvious that this is the necessary condition to turn a star-forming galaxy into a quiescent system and keep it passive for the rest of its life. {\it 
Thus, we assume that gas accretion is prevented and focus on the ISM}. Here, the challenge is to determine 
whether gas already in the disk is simply consumed by star formation, or other physical processes \underline{directly} affect its ability to feed star formation.

\subsection{Stripping -- removing cold gas from the disk}
A more dramatic way to deprive a galaxy of its fuel for star formation is to directly remove the cold ISM already in its disk (Fig.~\ref{Fig_cartoon}, second row). For satellite galaxies, the interaction with the environment of their host halo provides several channels to remove star-forming gas. Broadly, these can be divided into two general classes. 

(1) Hydrodynamical mechanisms. These involve
the direct interaction between the ISM and the intra-group, or intra-cluster medium (ICM), and include {\it ram pressure stripping} (i.e., the removal of the ISM due to the pressure exerted by the ICM while a satellite is moving through its host halo; \citealp{GUNG72}), {\it viscous stripping} (i.e., the removal of the outer layer of the ISM due to the viscosity momentum transfer with the ICM; \citealp{NUL82}) and {\it thermal evaporation} (increase of temperature, and subsequent evaporation of the cold ISM at the interface 
with the hot ICM; \citealp{COWS77}). While the vast majority of past work has focused on ram pressure, we have observational evidence (as well as theoretical predictions) that both viscous stripping and thermal evaporation can play a role (e.g., \citealp{bureau02,roediger05,review,randall08}). Unfortunately, it is extremely challenging to 
separate these three physical processes observationally, and it is expected that all three -- if present -- affect simultaneously the cold ISM of galaxies. It is also important to remember that, technically, starvation (i.e., the cessation of gas infall) is a hydrodynamical mechanism. It is, in practice, a mild ram pressure that is able to affect the gas in the halo but not that in the disk. This is primarily why starvation and ram pressure have been considered in the literature as two separate mechanisms, despite the physical process being most likely the same.

(2) Gravitational mechanisms.
The ISM can also be removed from the disk by the gravitational pull affecting satellites while orbiting in groups and clusters \citep{review}. The key element here 
is the variation (temporal and/or spatial) of the external gravitational potentials that baryons in the disk are subjected to \citep{binney87}.
This could be caused by fly-bys of companion satellites and/or interaction with the central galaxy (generally divided between low- and high-speed interactions depending on how the relative speed of the two galaxies compares to their rotational velocity), or by the interaction with the whole gravitational potential of groups and clusters (often referred to as the galaxy-cluster interaction; e.g., \citealp{bird90}). 
The combined effect of multiple high-speed galaxy-galaxy interactions over long timescales (i.e., longer than a typical crossing time, a mechanism generally referred to as {\it harassment}) has also been invoked to explain the properties of satellite galaxies in clusters \citep{harrassment,Smith2010}.

As discussed later, discriminating between hydrodynamical and gravitational processes on the ISM is not always straightforward, as both can perturb the cold ISM in the disk of galaxies in a similar way. Arguments based on geometry of disturbed features, on the comparison between gas and stellar distribution/kinematics, timescale of the interaction and distribution of star-forming regions can definitely help in the process but, as we will see, there is circumstantial evidence that both processes may sometime act simultaneously in galaxies and that {\it direct stripping} of the ISM from the disk is a fundamental step in the pathways of satellite galaxies towards quiescence. 

\subsection{Cold gas removal/heating by internal feedback mechanisms}
Another way to remove cold ISM from galaxies is via internal feedback related to 
the star-formation process (e.g., supernova feedback, stellar winds) or to the presence of an active galactic nucleus (AGN) at the centre of galaxies. In both cases, cold gas could be ejected from the disk and/or heated up, becoming unavailable for star formation (Fig.~\ref{Fig_cartoon}, third row).

From a theoretical point of view, internal feedback is arguably the most popular mechanism invoked to explain the existence of passive central galaxies and to reproduce key statistical properties such as the SFR-stellar mass relation and the stellar mass function of galaxies (e.g., \citealp{croton06,delucia06,hopkins08}). However, the fact that satellite galaxies appear to have lower cold gas content (e.g., \citealp{brown17}), less star formation, and a higher passive fraction at fixed stellar mass with respect to centrals (e.g., \citealp{vdbosch08,davies19}) has led to the conclusion that feedback may not always be the primary mechanism driving satellites into quiescence. Interestingly, it is certainly possible that feedback could {\it still play a role} by increasing the efficiency of other gas removal processes (see e.g. \S~4.1 and \S~8), as also suggested by cosmological models \citep{zoldan17,delucia19,stevens19}. While neglected in the past due to a lack of observations, this is a scenario that may turn out to be more common than originally thought, thanks to the improvement of ground- and space-based facilities making it easier to identify gas outflows (e.g., via narrow-band imaging or integral field spectroscopy).

\subsection{Stability of the cold gas against fragmentation and star formation}
The processes discussed so far are related to the very first step of the star-formation cycle in galaxies (i.e., the availability of cold gas), based on the notion that, in order to stop forming stars, a galaxy must somehow run out of its atomic and molecular gas reservoirs. A different pathway to quenching is to affect the next steps of the cycle, i.e. either the \hi-to-\htwo\ conversion or the collapse of \htwo\ into stars. Although the formation of molecular gas might not be strictly required to form stars (it could just be a by-product of the gravitational collapse of atomic gas on its way to star formation; \citealt{glover12}), this phase traces remarkably well the physical conditions under which new stars are formed \citep{krumholz11}. In this sense, star formation could be halted by preventing the atomic gas reservoir from condensing into molecules and/or collapsing to form stars (Fig.~\ref{Fig_cartoon}, bottom row).
More specifically, atomic hydrogen could be kept stable on a rotating disk below the typical column density threshold for \htwo\ formation, thus halting star formation (e.g., \citealp{blitz06,bigiel08,leroy08}). Additional factors regulating the rate of condensation of atoms into molecules could be the metallicity of the ISM and the intensity of the interstellar radiation field (e.g., \citealp{krumholz09,gnedin11}). Typical examples of cases in which atomic hydrogen does not appear to be efficiently converted into molecular phase include the outer parts of star-forming disks \citep{bigiel10}, low surface brightness galaxies \citep{wyder09}, early-type galaxies with significant \hi\ reservoirs (e.g., \citealp{knapp85,serra12}) and the even more extreme case of the so-called {\it HI-excess} systems (galaxies with
unexpectedly large \hi\ reservoirs based on their SFR and morphological structure; \citealp{gereb16,gereb18}).

Alternatively, even if molecular clouds can be formed, these could remain stable against collapse due to 
adverse local conditions (e.g., high gas turbulence and/or high stellar velocity dispersion), significantly reducing the efficiency of star formation. 
Examples of this scenario are the inner parts of the Milky Way \citep{longmore13} and of nearby, bulge-dominated systems \citep{martig09}, where turbulence is driven by gravitationally induced pressure from the bulge. In other words, the stability of the gas disk is enhanced by the presence of a dispersion-supported bulge component. However, as we will see, observations seem to exclude this path as an important one for satellite galaxies in groups and clusters. This is also because the reduction in star formation efficiency is generally not large enough to fully quench galaxies (e.g., \citealp{davis14}).\\

Of course, as mentioned at the beginning of this section, in order to be efficient in fully quenching star formation, gas stripping, gas ejection and stability must be accompanied by a halt in gas replenishment onto the star-forming disk.

In what follows, we discuss the observational evidence in support or against these quenching scenarios and show that, after nearly half a century of observational work, we are in a position to build  a coherent picture of the physical processes affecting the gas reservoir and star formation quenching in satellite galaxies at $z\sim$0. 
As a necessary step to quantify environmental effects on cold gas content, we begin by reviewing the various definitions of gas deficiency and their implications.

\section{Quantifying gas poorness -  The concept of gas deficiency} 
Quantifying the degree of cold gas poorness, or {\it deficiency}, of galaxies located in group and cluster environments, compared to similar systems in the field, has been critically important to develop our understanding of quenching mechanisms. Interestingly, several techniques have been adopted to quantify gas deficiency, with often different implicit assumptions on the physics affecting the cold gas reservoirs of galaxies -- sometimes leading to conflicting conclusions. Indeed, as discussed below, galaxies can simultaneously be gas-poor and gas-rich, depending on which definition is used. 

Historically, most of the work in this area has focused on atomic hydrogen deficiency, mainly due to the lack of large statistical surveys of molecular hydrogen in galaxies. However, the arguments presented below are applicable also to molecular hydrogen. 

{\subsection{Definitions based on optical diameter}}
Early studies of cluster galaxies started quantifying gas deficiency by comparing total \hi\ mass-to-luminosity ratios (distance-independent quantities) of cluster and field systems at fixed luminosity and/or morphology \citep{robinson65,davies73,huchtmeier76}. However, it soon became clear that a more robust approach was to control galaxies by their morphological type and optical diameter \citep{chamaraux80}. Indeed, the optical size vs. \hi\ mass relation turned out to be the tightest scaling relation (among those investigated at that time) linking \hi\ mass to optical galaxy properties. While \cite{chamaraux80} were among the first to introduce the concept of \hi\ deficiency, the first statistical definition of the concept of \hi\ normalcy and \hi\ deficiency (and arguably the foundation of this entire field) was presented by \citet[][hereafter HG84]{haynes}, in a study of a sample of 324 isolated galaxies extracted from the \cite{karachentseva73} Catalog of Isolated galaxies and the Uppsala General Catalog \citep{ugc} and observed with the Arecibo 305-m radio telescope. They showed that, at fixed morphological type, the optical property most strongly correlated with global \hi\ mass is the galaxy isophotal optical diameter.
This relation for isolated galaxies defines `\hi\ normalcy', and can be used to predict the total \hi\ mass within $\sim$0.2-0.3 dex. Thus, \hi\ deficiency is defined as:
\begin{equation}
DEF_{\rm HI}= \log(M_{\rm HI}^{pred}[T,D_{opt}]) - \log(M_{\rm HI}^{obs}) \\
\end{equation}
where $M_{\rm HI}^{pred}[T,D_{opt}]$ is the \hi\ mass expected for an isolated galaxy with the same morphological type (T) and optical diameter ($D_{opt}$) as the observed galaxy, i.e.:
\begin{equation}
\log(h^{2}M_{\rm HI}^{pred}[T,D_{opt}])= a_{T} + b_{T}\log(hD_{opt})^2
\end{equation}
where $h$=H$_{0}$/100, H$_0$ is the Hubble constant, and $a_{T}$, $b_{T}$ are the best-fitting coefficients tabulated as a function of morphological type. 

In the past few decades, several independent works have presented revised versions of this calibration, primarily motivated by improvements in data quality, number statistics, morphological spread and/or definition of isolated galaxy (e.g., \citealp{solanes96,bosellicomaI,toribio11,denes14,jones18}). Remarkably, the most recent re-calibration of the optical diameter-based \hi\ deficiency presented by \cite{jones18} -- based on the Analysis of the interstellar Medium in Isolated GAlaxies 
\citep[AMIGA;][]{amiga} project -- has shown that the original coefficients presented by HG84 are still very much consistent with those based on modern datasets. An additional variant of the HG84 definition is to ignore the morphological dependence and assume that all galaxies have the same average \hi\ surface density \citep{chung09}. While qualitatively consistent with the original calibration, this implementation is less robust as the typical \hi\ surface density of galaxies is not constant \citep{bigiel12}, but admittedly it has the advantage of not relying on morphological classification and all potential biases associated with this.

The fact that the isophotal diameter turned out to be more strongly correlated with total \hi\ mass than other optical properties (such as luminosity) should not come as a surprise. Firstly, for a self-gravitating disk it is easy to show that the gas mass of the disk is more tightly connected to the specific angular momentum than the luminosity of a system (or even its stellar mass, see also \S~3.3 and 3.4).  Indeed, at fixed total mass, the scatter in cold gas mass is primarily driven by the spread of spin parameter\footnote{The spin parameter, $\lambda$, is a dimensionless quantity used to parameterize the deviation from centrifugal equilibrium (or in other words the amount of rotation) of a dark matter halo.} \citep{mo98,boissier00,huang12,maddox15,obreschkow16} and, observationally, sizes are more directly linked to spin than optical luminosities. Secondly, for disk galaxies, scaling relations involving isophotal radii are always tighter than those based on scale-lengths and/or effective radii (e.g., \citealp{saintonge11c,hrsgalex,sanchezalmeida20}; see also \S~4.3), as they better trace the outer parts, and thus the full extent, of the disk component of galaxies\footnote{See also \citealp{stevens19b} for an extensive discussion on the origin of the small observed scatter of the \hi\ mass vs. \hi\ size relation.}.

A key point to note is that, {\it{by defining \hi\ deficiency at fixed morphology and optical diameter, we implicitly make strong assumptions on the type of physical processes that this indicator is sensitive to.}} Namely, we indirectly imply that the mechanism(s) removing the \hi: (a) does not change the morphology of the galaxy, and (b) does not affect the extent of the stellar disk. Or, at the very least, the implication is that any morphological and/or size transformations take place on timescales significantly longer than those needed to make the galaxy \hi\ deficient. Thus, while ideal to trace most hydrodynamical mechanisms (including outflows and starvation), this definition may not always be suited to identify stripping events where gas and stars are affected simultaneously (e.g. tidal stripping), or to follow the long term evolution if the stellar distribution or morphology do change with time.\\

\subsection{Definitions based on color and structural properties}

Despite its popularity over the past few decades, the HG84 definition of \hi\ deficiency has turned out not to be the ideal approach for an application to large datasets and comparison with cosmological simulations. With the advent of the Sloan Digital Sky Survey (SDSS, \citealp{york2000}) in the early 2000s, isophotal radii have gradually become `less popular' than effective radii, and reliable estimates have been rarely available for large datasets overlapping with \hi\ observations. At the same time, accurate determination of morphological types (e.g., the ability to discriminate between Sa, Sb and Sc) for large statistical samples became more challenging already at distances beyond those of the Coma cluster ($\sim$100 Mpc, $z\sim$0.023), potentially boosting the uncertainty in the estimate of deficiency. Moreover, morphological types based on visual classification are also not a standard and/or consistent output of numerical simulations, which parameterize galaxy morphology using more quantitative indicators. These issues prompted the community to develop new \hi\ deficiency estimators (and related definitions of \hi\ {\it normalcy}) which, in many instances, have also been used as basis to predict \hi\ masses for large samples of galaxies without 21 cm observations available.


One of such modern versions of \hi\ deficiency is the so-called \hi\ {\it gas fraction plane} introduced by \cite{catinella10}. By taking advantage of data 
from the GALEX Arecibo SDSS Survey (GASS), they show that the best predictor of \hi-to-stellar mass ratio is a linear combination of the observed (i.e., not corrected for internal dust attenuation) near-ultraviolet minus r-band colour (NUV-r) and stellar mass surface density ($\mu_{*} \equiv~$M$_{*}$/(2$\pi$ R$_{50,z}^{2})$, where M$_{*}$ is the total stellar mass and R$_{50,z}$ is the radius containing 50\% of the galaxy light in the $z$-band), with a typical scatter of $\sim$0.3 dex (see also \citealp{catinella13}):
\begin{equation}
\log(M_{\rm HI}/M_{*})= a_1(NUV-r) + a_2\mu_{*} +a_3
\end{equation}
where $a_i$ are the coefficients of the best fit. The potential advantage of this definition 
is that it does not require a morphological classification and is based on quantities that are now readily available for hundreds of thousands of nearby galaxies. 

Interestingly, so far this technique has primarily been used to identify \hi-excess systems
(see \S~2.4)
and only \cite{cortese11} have tested it as a potential indicator for \hi\ deficiency. They show that, while the deviations from the gas fraction plane and the HG84 \hi\ deficiency are correlated, the scatter can be significant ($\sim$0.4 dex), with important systematic offsets. While this is partially due to the sample used to calibrate the scaling relations (e.g., isolated galaxies vs. field objects vs. centrals), there is a fundamental physical reason why these two techniques do not always agree. The position of a galaxy in the gas fraction plane is primarily driven by its NUV-r colour (i.e., a proxy for its unobscured SFR, traced by the NUV emission, per unit stellar mass, traced by emission in $r$ band), with a secondary dependence on optical size (via $\mu_*$). Indeed, it has been shown by several works that \hi\ is much more directly correlated with the star formation traced by the ultraviolet than with the bulk of (mostly obscured) star formation in galaxies (e.g.,\citealp{bigiel10b,catinella18}). Thus, contrary to the HG84 definition, {\it{the underlying assumption in this case is that \hi\ is affected but star formation is not} (at least within the timescale traced by the NUV-r colour, i.e., $\sim$10$^{8}$ yr)}. This implies that this parametrisation is sensitive only to processes that either: (a) remove cold gas from galaxies on a significantly shorter timescale than that needed to affect the star formation, or (b) act only on the part of the atomic gas reservoir that is not directly fueling star formation (e.g., the outer disks of galaxies).

\subsection{Definitions based on broad-band luminosity or stellar mass}
Another approach to quantify gas deficiency is via correlations between gas mass and broad-band optical/near-infrared luminosities (or even stellar masses; \citealt{chamaraux80}).
This technique has recently regained some popularity 
thanks to \cite{denes14}, who estimated scaling relations between \hi\ mass and luminosities in B, $r$, K, H and J bands for galaxies detected by the \hi\ Parkes All Sky Survey \citep[HIPASS;][]{hipass}.
Within the HIPASS sample, these relations have a typical scatter of $\sim$0.3 dex, comparable to that of other \hi\ deficiency proxies, and are used by the authors to identify \hi-excess and \hi-deficient galaxies.

While this approach has the advantage of requiring only a single band luminosity, recent works have highlighted its significant limitations (see also \citealp{bottinelli74} and HG84 for early discussions on why this approach may not be ideal). We now know that the optical luminosity/stellar mass vs. \hi\ mass relation is primarily driven by the fact that there are increasing numbers of star-forming (gas-rich) galaxies compared to passive (gas-poor) systems as we shift to considering lower stellar masses \citep{brown15}. Indeed, at fixed color or SFR, the correlation becomes significantly weaker, with galaxies following shallower, parallel relations on the \hi\ gas fraction-stellar mass plane, and star-forming systems being systematically more gas-rich.
Thus, this relation is not a direct physically-motivated correlation compared to the other correlations with star formation rates or galaxy sizes, and it has a scatter significantly larger (0.5 dex or higher; \citealp{catinella18}).


The unusually small scatter of the \cite{denes14} calibration is simply due to a selection effect -- because of its limited sensitivity, HIPASS detects only the most gas-rich galaxies in the local Universe (see also \citealp{lutz17}).
As we discuss later in this work, this is a common issue for samples detected by \hi-blind surveys, which are primarily biased towards gas-rich galaxies within their volumes (e.g., \citealp{huang12}).
The fact that the \hi\ mass vs. luminosity/stellar mass scaling relation for a blind \hi\ survey is not representative of the global galaxy population, but offsets towards gas-rich systems, implies that galaxies with normal \hi\ content (according e.g. to the HG84 definition) may be labelled as \hi-deficient. Moreover, from a physical point of view, the use of a deficiency parameter based only on mass/luminosity significantly complicates the interpretation of any trends. As all galaxy properties correlate with stellar mass, it is unclear whether gas deficiency with respect to mass is tracing environmental effects or more simply the bimodality of star-forming properties of galaxies, which has been shown to be well established in all environments (e.g., \citealp{baldry06}).

\begin{figure*}
\includegraphics[width=17.2cm]{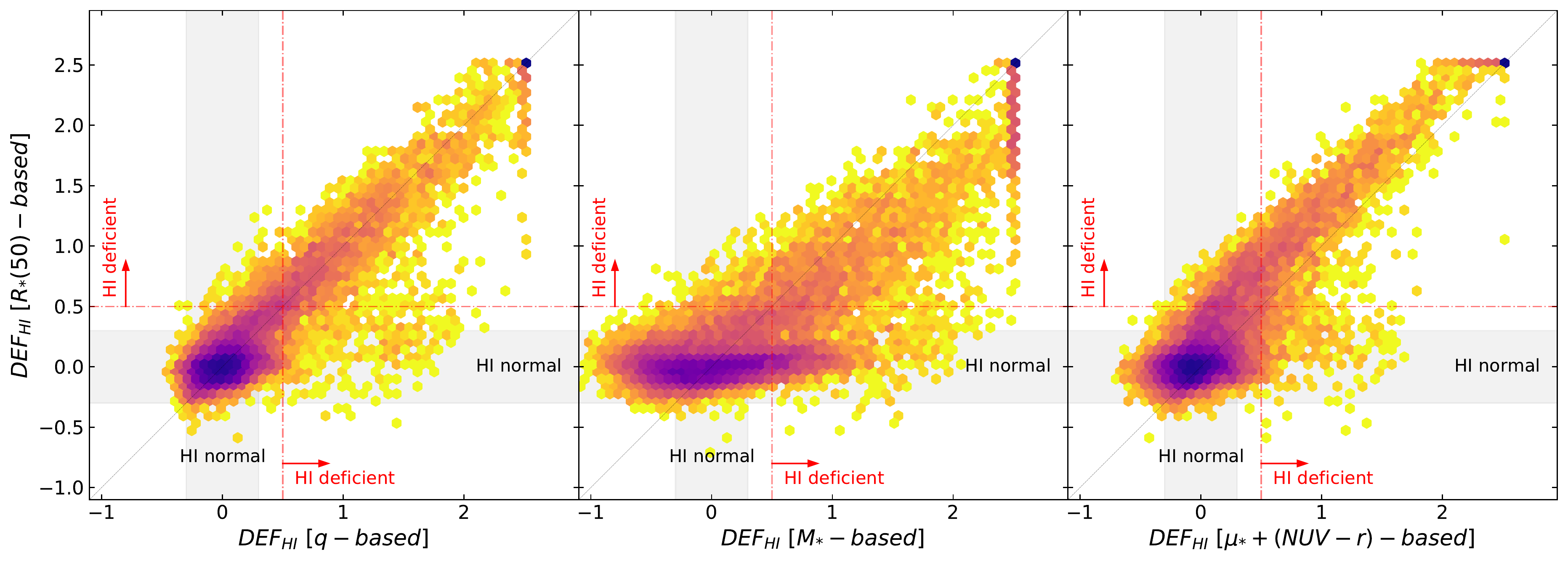}
\caption{Comparison between different estimates of \hi\ deficiency ($DEF_{HI} \equiv \log(M_{HI,pred})-\log(M_{HI})$) calibrated on the {\sc Shark} semi-analytical model of galaxy formation. From left to right, the panels show the specific angular momentum/stability parameter (q)-based, the stellar mass (M$_{*}$)-based and gas fraction plane-based ($(NUV-r)+\mu_{*}$) definitions against the one calibrated on the \hi\ mass vs. optical disk size relation. The solid line shows the 1-to-1 relation, the gray band shows the region of `\hi-normalcy' defined by the typical scatter in the scaling relations (i.e., $-$0.3$<DEF_{HI}<$0.3) and the dotted-dashed lines show the threshold of $DEF_{HI}$=0.5 that we use in the rest of this paper to isolate bona-fide \hi-deficient galaxies. See Sec.~\ref{hidef_com} for details on how each parameter has been estimated.} \label{sharkdef}
\end{figure*}

\subsection{Definitions based on specific angular momentum}
In recent years, independent works have put forward a new definition 
of gas deficiency based on the angular momentum properties of galaxies. This idea has been proposed by a few authors throughout the last few decades (e.g., \citealp{zasov89,safonova11}), and most recently expanded by \cite{jieli20}. Their approach is based on the assumption that, for a flat exponential disk with circular rotation, the maximum allowed, stable value of gas fraction strongly depends on the mass and kinematics of the disk. One way to parametrise this is through the dimensionless {\it global disk stability parameter} $q$ \citep{obreschkow16}:
\begin{equation}
q=\frac{j\sigma}{GM}    
\end{equation}
where $j$ is the baryonic (gas plus stars) specific angular momentum (i.e., angular momentum per unit mass) of the disk, $\sigma$ is the gas velocity dispersion, $M$ the total baryonic mass and $G$ the gravitational constant. The relation with neutral atomic (\hi\ plus helium) gas fraction, $f_{atm}=M_{atm}/M_{baryons}$, can then be generally approximated by a truncated power law of the form
\begin{equation}
f_{atm} = min\{1, 2.5q^{1.12}\}    
\end{equation}
While this is arguably a more physically-motivated definition of \hi\ deficiency (but see also \citealp{romeo18} for some potential limitations of this approach), the basic assumptions are the same behind the original HG84 definition. Indeed, \cite{jieli20} show that (at least in the case of galaxies in the Virgo cluster) the two definitions provide remarkably consistent results.

From a practical point of view, the main short-coming of this technique is that, in addition to being calibrated on pure-disk galaxies only, it requires information on both the two-dimensional distribution and velocity field for at least one baryonic component of galaxies, as well as the total baryonic mass, which is much more difficult to obtain than optical diameter measurements. Regardless, this method implicitly assumes that the process affecting the gas does not change significantly the kinematic properties of the baryons in the galaxy (i.e., the specific angular momentum is not altered), which is not so different from the basic assumption of the optical size-based deficiency. In other words, as already hinted in \S~3.1, this definition can be viewed as a physical explanation for the success of the optical diameter-based approach.

\subsection{A direct comparison of different gas deficiency definitions}
\label{hidef_com}
One of the key reasons why the above limitations have rarely been discussed in the literature is that we are still lacking galaxy samples that simultaneously cover wide ranges of environments (to include both gas-rich and gas-deficient systems),  
are representative of the local galaxy population (to calibrate reliable scaling relations) and for which sizes, luminosities and angular momentum can be homogeneously estimated (see however HG84, \citealp{cortese11}, \citealp{teimoorinia17}, \citealp{jieli20} for discussions on different gas deficiency definitions and gas predictors in general).

Thus, in order to clarify the points 
discussed here, we take advantage of the semi-analytic cosmological model of galaxy formation and evolution {\sc Shark} \citep{lagos18b}. The point of this exercise is not to reproduce the observed scaling relations\footnote{Indeed, while {\sc Shark} shows improvements over previous semi-analytical models in reproducing galaxy scaling relations, there are still clear limitations when it comes to matching the gas properties of galaxies across environments (e.g., \citealp{hu20}).} but to show that, even if we calibrate gas deficiency definitions using exactly the same sample and assumptions, major biases still remain.

We compare the optical size, total stellar mass ($M_{*}$), specific angular momentum (via the $q$ parameter) and (NUV-r)+$\mu_{*}$-based approaches for galaxies with stellar masses between 10$^{9}$ and 10$^{11}$ M$_{\odot}$. The first three definitions are calibrated on central galaxies within 1.5$\sigma$ from the {\sc Shark} main sequence of star-forming galaxies, whereas the (NUV-r)+$\mu_{*}$ plane is estimated for all central galaxies with an atomic gas fraction greater than 1\%, following \cite{catinella10}. For the optical size, we use the radius including 50\% of the stellar mass of the disk ($R_{*}(50)$), whereas we estimate $q$ from the total baryonic specific angular momentum and baryonic mass. For each galaxy in the model, we then calculate deficiency as the logarithmic difference between the \hi\ content predicted from each scaling relation and the one given  ($DEF_{HI}\equiv \log(M_{HI,pred})-\log(M_{HI})$). 
Galaxies with $DEF_{HI}>$2.5 are set to 2.5 to encapsulate the inability of observations to detect cold gas in extremely gas-poor objects.

The results are shown in Fig.~\ref{sharkdef}. As expected, the size and $q$-based estimates generally agree and are not too offset from the 1-to-1 line (left panel), whereas larger systematic offsets are seen for the stellar mass (middle) and (NUV-r)+$\mu_{*}$ (right) approaches. In particular, a significant fraction of \hi-normal galaxies (i.e., $-$0.3$<DEF_{HI}<$0.3, indicated by the gray band) according to the size-based calibration would be classified as \hi\ deficient using the stellar mass-based calibration (i.e., $DEF_{HI}>$0.5 to the right of the vertical dotted-dashed line in the middle panel), or even \hi-rich or \hi-excess galaxies (e.g., $DEF_{HI}<-$0.5). Conversely, many \hi-deficient systems according to the size-based approach would appear \hi-normal for the gas-fraction plane definition (see also \citealp{cortese11}). As mentioned above, this is just due to the different samples used to calibrate the two recipes (all central galaxies with gas fraction above a threshold versus star-forming centrals only). Thus, to first order, \hi-deficient galaxies would move along the plane.  If we calibrate the plane on just star-forming galaxies, the dependence on star formation disappears almost entirely (simply by selection: i.e., all galaxies have similar SFRs) and the plane becomes equivalent to the gas mass-optical size relation. 

In summary, some care must be taken when it comes to the \hi\ deficiency parameter, as its definition and implications for environmental studies may differ significantly from one work to another. As we reviewed in this section, different scaling relations between cold gas content and another galaxy property (or a combination of two or more) can be used to define \hi\ {\it normalcy}, and offsets from it. The interpretation of such an offset as an indicator of \hi\ deficiency depends critically on three elements: (1) the choice of scaling relation, which implies underlying assumptions on the physical process that removes the gas; (2) the reference sample used to calibrate the scaling relation itself, typically isolated or field galaxies with \hi\ measurements. Clearly, reference samples biased towards gas-rich systems (such as detections from \hi-blind surveys) may lead us to misclassify \hi-normal systems as \hi-deficient ones; (3) the choice of a threshold separating \hi-poor from \hi-normal systems (typically from 0.3 to 0.5 dex, which is $\sim$1-1.5 times the scatter of these relations, see also Fig.\ref{sharkdef}).

\begin{figure*}
\includegraphics[width=17.7cm]{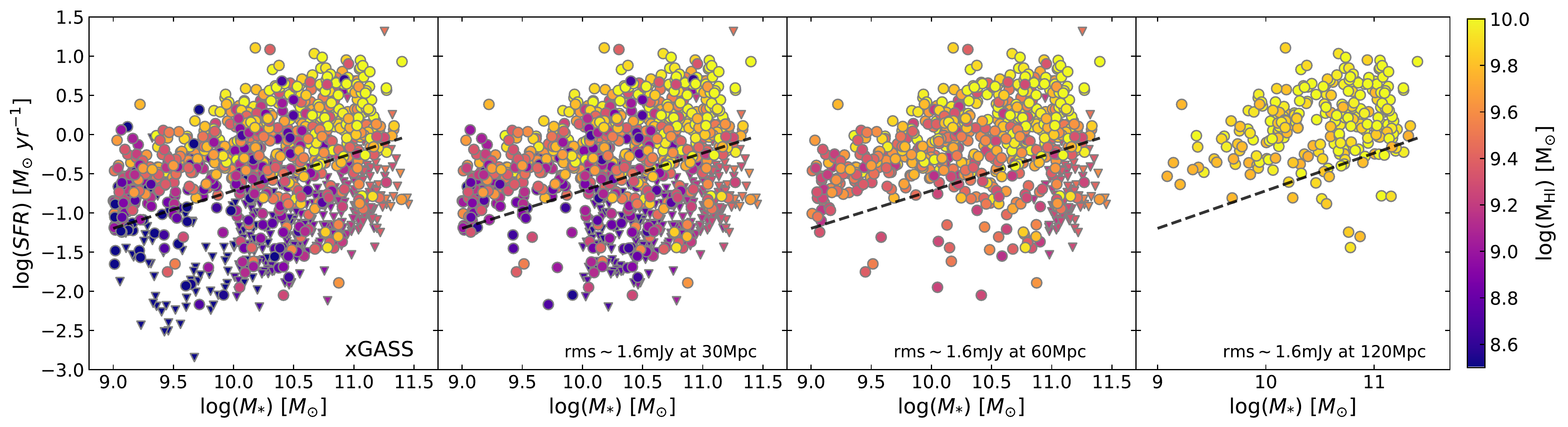}
\caption{The ability of \hi\ surveys to detect galaxies across the stellar mass vs. SFR plane. The left-most panel shows the distribution of galaxies in the xGASS survey, which we use as input. Circles and triangles indicate \hi\ detections and non-detections, respectively. Galaxies are color-coded according to their \hi\ mass (circles), or provided upper limit (triangles). The dashed-line shows 2$\sigma$ below the star-forming main sequence. The remaining panels show which galaxies would be detected (at 5$\sigma$ level assuming a velocity width of 200 km/s) by a survey with 1.6 mJy rms noise (close to the expected sensitivity of the WALLABY survey) for distances varying between 30 and 120 Mpc. We conservatively assume xGASS non detections at their upper-limit. It is clear that, above $\sim$40-50 Mpc, most of the passive population starts to disappear and at distances higher than $\sim$100 Mpc only galaxies in the main sequence are detected.} \label{wallabypred}
\end{figure*}

Definitions of \hi\ deficiency based on the size of the stellar disk or on its global specific angular momentum are most sensitive to stripping processes, which may affect both gas content and star formation, hence are of particular relevance for this review (however, optical sizes are more readily available for large samples of galaxies).
Choosing instead NUV-r colour (a proxy for unobscured specific SFR, see \S~3.2) as a scaling parameter implicitly assumes that gas is removed but star formation is not affected. We discourage the use of definitions based on stellar mass or luminosity alone, which have very large scatter (strongly dependent on SFR) and are often inconsistent with the ``classical'', size-based \hi\ deficiency, {\it even when based on the same calibration sample} (Fig.~\ref{sharkdef}, middle panel).

Much of the literature on the effects of environmental processes on the \hi\ content of galaxies makes use of the optical size-based \hi\ deficiency parameter, and this is what we adopt in the rest of this review (using a conservative threshold of 0.5 dex to safely identify \hi-deficient galaxies, unless otherwise noted). However, these types of binary classifications, although initially convenient, never capture the complexity of galaxies nor the continuum of properties that we observe, thus we hope that the advent of new large-area blind surveys will see the community abandon this historical terminology and refer instead to offsets with respect to scaling relations, making sure that the calibration samples are representative of the galaxy population.

\section{Cold gas removal in nearby clusters -- M$_{halo}>$10$^{14}$ M$_{\odot}$}
\subsection{The case for \hi\ stripping}
We start by focusing on clusters of galaxies, as these are the environments where most of the progress has been made in the last few decades. In reality, as it will become immediately clear, our knowledge of cold gas in these environments is heavily based on studies of one single cluster, Virgo, in particular when it comes to resolved analysis. At a distance of just $\sim$16-18 Mpc \citep{mei07,planckvirgod} from Earth, and sitting in the middle of the northern Spring sky, Virgo has been the primary target of most gas surveys focusing on understanding the role of environment on the gas cycle of galaxies. 
Given that Virgo may not be the average nearby cluster, at least in terms of accretion history and number of substructures \citep{sorce19}, this is something to keep in mind. 

The reason for our limited knowledge of gas content in cluster galaxies is that \hi-blind surveys, which have provided the largest samples of global \hi\ measurements, only detect the most gas-rich galaxies in their volumes (i.e., the star-forming population), hampering our ability to study systems in the process of being affected by the environment. To demonstrate this point, we
take advantage of xGASS (the extended GASS survey; \citealt{catinella18}), a stellar mass-selected, gas fraction-limited survey of $\sim$1200 galaxies that is the current benchmark for a representative sample in terms of \hi\ properties of nearby galaxies. In Fig.~\ref{wallabypred}, we show which parts of the stellar mass-SFR plane are typically detected by \hi-blind surveys, in comparison with xGASS (left panel). As a reference for \hi-blind surveys we consider the upcoming Widefield ASKAP L-band Legacy All-sky Blind surveY \citep[WALLABY;][]{wallaby}, which, for unresolved sources, will have a factor $\sim$2 better sensitivity than the current state of the art, the Arecibo Legacy Fast ALFA \citep[ALFALFA;][]{alfaalfa05} survey.
We assume a typical root mean square (rms) noise for WALLABY of $\sigma=1.6$~mJy and, for each distance shown in Fig.~\ref{wallabypred}, we plot galaxies that will be detected above a 5$\sigma$ threshold assuming a typical velocity width of 200 $\rm km~s^{-1}$. 
Given that not even xGASS detects all passive galaxies, here we take a conservative approach by plotting the upper limits of \hi\ non detections (triangles). It is clear that, while at the distance of Virgo both the star-forming and quiescent populations are mapped, beyond $\sim$60-100 Mpc 
a WALLABY-like survey will not be able to detect a large number of galaxies below the star-forming main sequence.

\begin{figure}
\includegraphics[width=8.5cm]{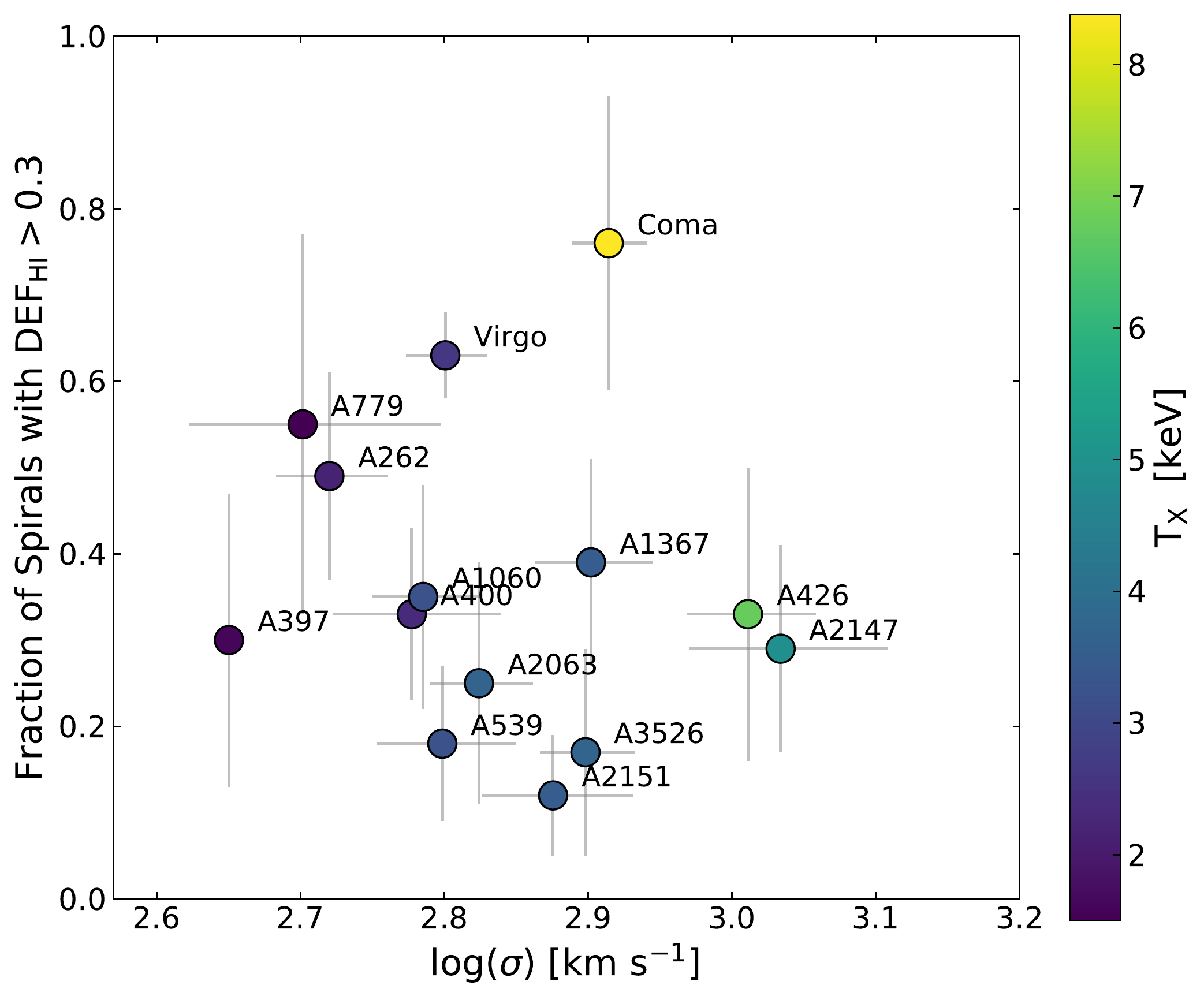}
\caption{The fraction of \hi-deficient spiral galaxies as a function of cluster velocity dispersion (a proxy for cluster mass) for the sample of nearby clusters of galaxies studied in \cite{solanes01}. Points are color-coded by X-ray temperature, $\rm T_X$. It is clear that the fraction of \hi-deficient spirals in clusters does not strongly depend on $\rm T_X$ or mass of the cluster.} \label{solanes}
\end{figure}

From the early works by \cite{davies73} and \cite{huchtmeier76}, every investigation of clusters of galaxies has found that the late-type population has significantly less \hi\ content than galaxies in isolation at fixed morphology, size and/or optical luminosity. Most of the early statistical works in this area were carried out with single-dish antennae like the Arecibo and Nan\c{c}ay radio telescopes (e.g., \citealp{sullivan78,chamaraux80,giovanelli81,giova85,haynes86}), which simply provided global \hi\ masses, but no information on the spatial distribution of the gas within the galaxy. Despite this limitation, 21cm deep surveys of several clusters in the local Universe (e.g., Virgo, Coma, Abell 262, Abell 1367, Cancer, Abell 2147/2151) have shown that \hi\ deficiency is a wide-spread phenomenon in clusters and that, while deficiency does not strongly correlate with any galaxy or cluster property (see Fig.~\ref{solanes}), there is a common trend for the degree of deficiency to increase with decreasing distance from the cluster centre.

Nevertheless, the \hi\ deficiency vs. cluster-centric distance relation has a large scatter \citep{solanes01,review}, with \hi-deficient galaxies observed up to a few Mpc projected distance from the centre, and both the shape and scatter of the correlation appear to depend on the dynamical state of the cluster (e.g., with Coma having a better defined correlation than Virgo). This result, combined with the initial evidence that gas-poor late-type galaxies follow more radial orbits than gas-rich systems \citep{dressler86,giraud86,solanes01}, provided the foundation to the idea that, in clusters, atomic hydrogen is directly removed from the disk during the infall, with hydrodynamical effects such as ram pressure being the primary suspect. However, it is worth noting that the same trends could also be explained via galaxy-galaxy interactions (e.g., \citealp{valluri90,valluri91}). In other words, direct stripping of the cold ISM seemed needed to explain the significant lack of gas observed in cluster spirals, but the primary driver remained elusive to single-dish observations.

The advent of radio interferometric observations dramatically improved the situation. Since the first observations of Virgo cluster galaxies by \cite{warmels88} and \cite{cayatte90}, it became clear that not only the total amount of gas, but also its radial distribution changes when galaxies plunge into clusters. \hi-deficient cluster galaxies have surface density profiles significantly less extended than their optical sizes, as most of the hydrogen is missing from the outer parts of the disk, whereas the inner regions show little (e.g., within a factor of $\sim$2) variations in gas surface density. The transition between inner (gas-normal) and outer (gas-deficient) parts of the \hi\ disk is much sharper than observed in field spirals and can drop from a healthy $\geq$ 5 M$_{\odot}$ pc$^{-2}$ to less than 1 M$_{\odot}$ pc$^{-2}$ within just a couple of kiloparsec. This unique feature - generally referred to as truncation of the gas disk - represents one of the clearest (if not the clearest) observational pieces of evidence supporting the idea that gas is directly stripped from the disk. Indeed, it is difficult to imagine a scenario in which a simple cessation of gas infall and/or outflows could produce such a remarkable feature \citep{n4569}, without also affecting the stellar disk, which instead generally shows no sign of strong environmental perturbations in cluster galaxies.

\begin{figure}
\includegraphics[width=8.5cm]{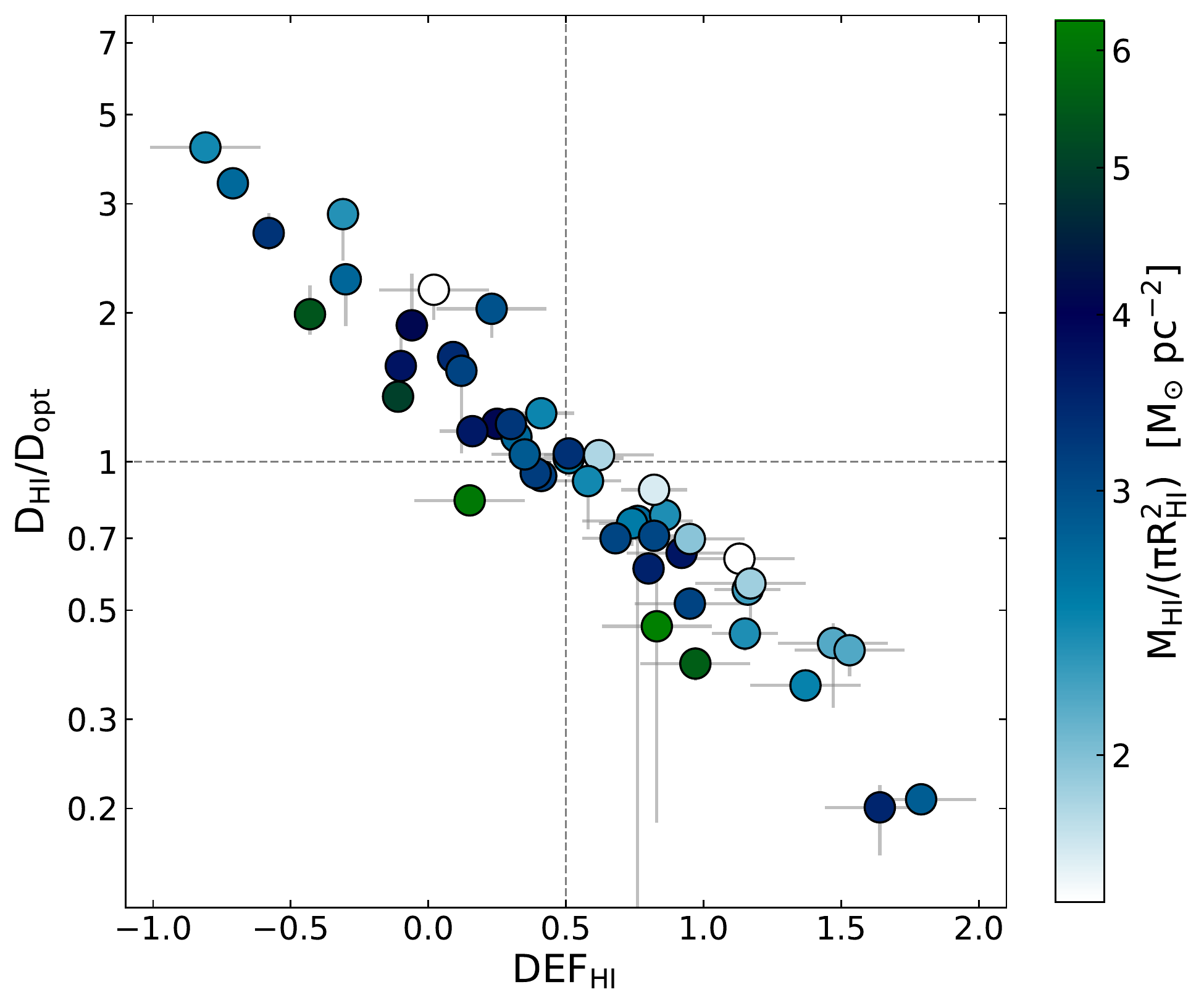}
\caption{The \hi-to-optical isophotal diameter as a function of \hi\ deficiency for galaxies in the Virgo cluster included in the VIVA survey \citep{chung09}. Points are color-coded by average \hi\ surface density. Vertical and horizontal dotted lines are shown to guide the eye and highlight that $DEF_{HI}\sim$0.5 roughly corresponds to gas removal up to the optical isophotal radius.} \label{viva}
\end{figure}

In Fig.~\ref{viva}, we take advantage of data from the VLA Imaging of Virgo in Atomic gas \citep[VIVA;][]{chung09} survey to show the strong correlation between \hi\ deficiency and the extent of the \hi\ disk normalised by the optical diameter. \hi\ diameters are iso-density diameters measured at 1 M$_{\odot}$ pc$^{-2}$ level, while optical ones are measured at the 25th magnitude per square arcsecond from the Third Reference Catalogue of Bright Galaxies \citep[RC3;][]{rc3}. This illustrates how high deficiencies are associated with smaller \hi\ disks, confirming that gas removal happens preferentially outside-in. Points are colour-coded by the average \hi\ surface density within the \hi\ size. It is clear that there is not a strong difference in surface densities between \hi-deficient and \hi-normal galaxies (see also \citealp{wang16}), as surface density appears to regulate the scatter of the relation for {\it all} deficiency values.

The detailed information provided by the resolved \hi\ maps of Virgo galaxies made it possible to start modeling the interaction with the cluster environment, thus getting closer to isolating the physical mechanism(s) responsible for the observed \hi\ deficiency. Fundamental in this area have been the works carried out by \cite{vollmer01,vollmer03,vollmer04} on individual Virgo galaxies, showing that stripping via a hydrodynamical process such as ram pressure is necessary to reproduce the three-dimensional distribution of atomic hydrogen in these systems. A similar conclusion was reached by interferometric studies of other nearby clusters such as Coma and Abell 1367 \citep{hector1,hector2,scott10}. More recently, the discovery of one-headed \hi\ tails, in some cases pointing away from the X-ray centre of clusters, has provided even more direct proof for the importance of stripping in removing cold hydrogen from the star-forming disk \citep[e.g.,][]{osterloo05,chung07,ramatsoku19,deb20}. Indeed, for the typical ICM density of $\sim$10$^{-3}$ cm$^{-3}$ inferred from X-ray observations, infalling velocities of $\sim$1000 km~$s^{-1}$ are enough for ram pressure to strip at least part of the atomic hydrogen content from the star-forming disk of galaxies \citep{GUNG72,review,koppen18}.

While all the observational evidence collected so far indicates that active stripping due to the interaction between the galactic ISM and the ICM is likely the main reason for the lack of \hi\ in disk galaxies, this does not mean that ram pressure is sufficient to explain the transformation of satellite galaxies. We now know plenty of galaxies for which at least one of the alternative environmental mechanisms described in the Introduction needs to be invoked to explain their properties. These include gravitational interactions needed to reproduce the kinematic properties of the stripped gas (e.g., NGC4254, \citealp{duc08}; NGC4438, \citealp{vollmer05,kenney08}; CGCG97-073/79, \citealp{GAVB01}), and feedback from star formation (e.g., NGC4424, \citealp{bos4424}). Moreover, there are also individual cases where gravitational interactions may be the dominant mechanism responsible for \hi\ stripping (e.g., NGC 4532/DDO 137 in Virgo, \citealp{koopmann08}; NGC1427A in Fornax, \citealp{leewaddell18}; the Abell cluster 1367, \citealp{scott10}). This should not come as a surprise, as the pericentre of the orbit of a satellite in a cluster is precisely where ram pressure, cluster tides and number of high-speed encounters are expected to reach their peak intensity (e.g., \citealp{Smith2010}). Once again, this highlights the complexity of galaxy evolution in clusters and, while focusing on a single physical mechanism might still prove useful for individual objects (where a given process might indeed explain the main current features of a galaxy, see e.g. \S~8), it would be reductive to expect that it alone could account for the full galaxy transformation over long times.

\subsection{What about molecular hydrogen?}
According to our current view of star formation in galaxies, it is now accepted that molecular hydrogen (\htwo) is more directly connected to star formation than \hi\ on local (i.e. kiloparsec or smaller) scales \citep{bigiel08,leroy08},  with \hi\ simply providing the reservoir out of which molecular clouds are formed. As such, the link between \hi\ stripping and quenching of star formation is not  straightforward. Is molecular gas directly stripped as well? Or is this denser phase of the ISM not strongly perturbed by the environment and simply consumed by star formation?

Answering these questions has been challenging because of the lack of molecular gas measurements for large statistical samples of galaxies, spanning different environments. The first complication is that, being a symmetric molecule with low mass, \htwo\ does not emit at the low temperatures typical of molecular clouds ($T<$50 K), and thus requires an indirect indicator. The most widely used tracer of \htwo\ is carbon monoxide (CO), the most abundant molecule after \htwo, whose rotational lines are easily excited in cold molecular clouds (see e.g., \citealp{bolatto13} for an extensive review on this topic).
The second complication is that, unlike for \hi, there are no large-area, blind surveys of CO emission in the local Universe, and targeted samples are small. At the time of writing, quantification of global \htwo\ scaling relations is limited to a few hundred galaxies (e.g., \citealp{saintonge11,saintonge17,bothwell14,cicone17}), a number that drops even more when it comes to resolved properties \citep{heracles}. Luckily, the Atacama Large Millimeter/submillimeter Array \citep[ALMA;][]{alma} is already having a tremendous impact on this field, especially for galaxies at higher redshift.

The lack of representative samples suitable to define \htwo\ normalcy for galaxies has meant that any potential quantification of \htwo\ deficiency is made more challenging, due to selection biases and large scatter in scaling relations \citep{boselligdust}. Nevertheless, all initial investigations suggested that galaxies in rich clusters have on average the same molecular hydrogen content as systems in isolation (e.g., \citealp{kenney86,kenney88,stark86,casoli91,casoli98,bos97}). From a theoretical point of view, this is fully consistent with an outside-in stripping scenario, where only the more diffuse gas phase (i.e., \hi) in the outer parts of the star-forming disk can be significantly affected by environment. Indeed, it can be easily shown analytically that giant molecular clouds are too dense to be appreciably stripped by the ram-pressure wind (e.g., \citealp{yamagami11}).

\begin{figure}
\includegraphics[width=8.5cm]{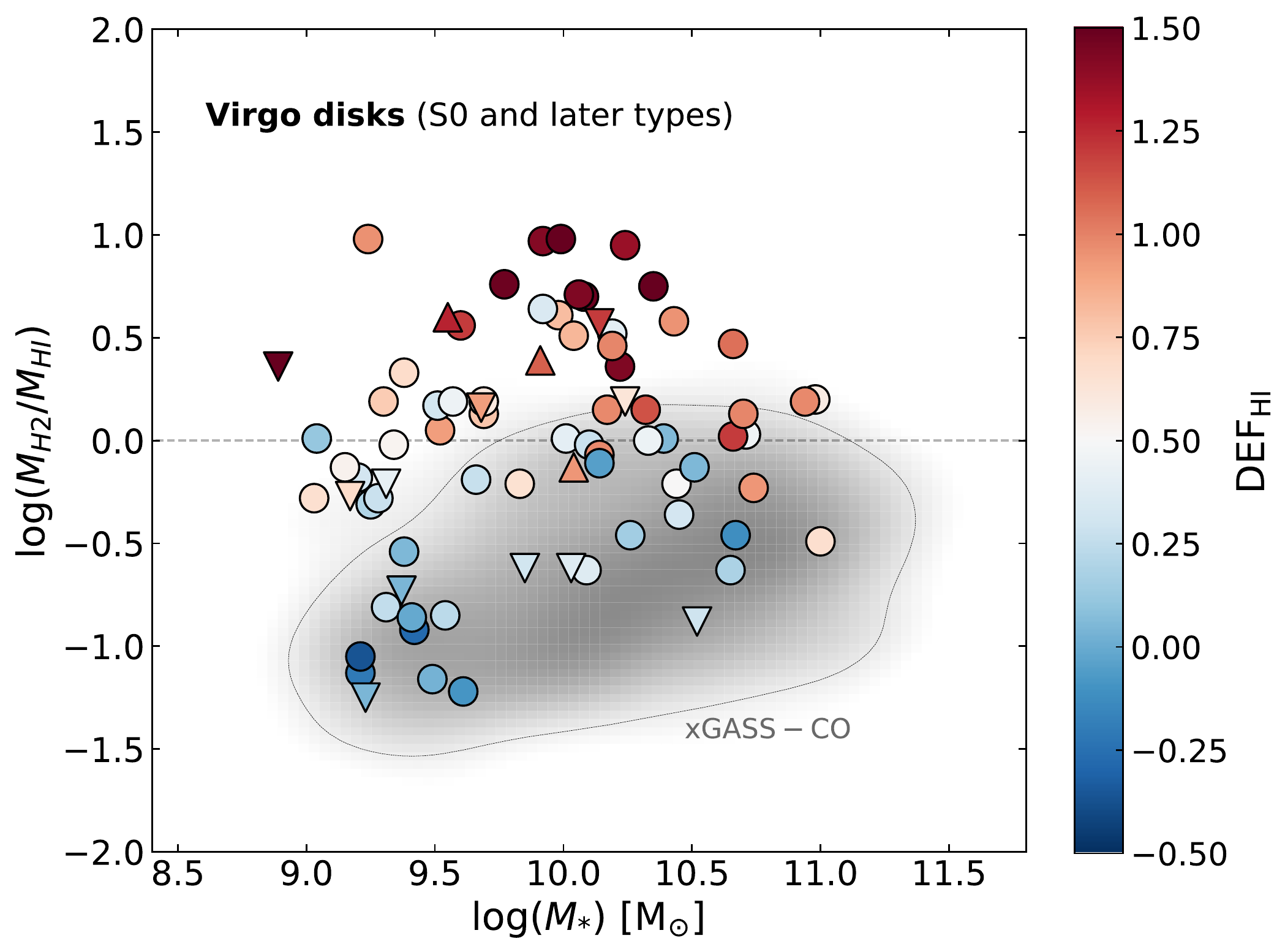}
\caption{The molecular-to-atomic hydrogen mass ratio for disk galaxies (S0 or later types) in the Virgo cluster (\citealp{hrsco}). Circles indicate detections in both \hi\ and \htwo, upward and downward triangles are upper limits in \hi\ and \htwo, respectively. Galaxies are color-coded by \hi-deficiency. The grey region in the background shows the typical parameter space covered by local galaxies as traced by the xGASS-CO sample. It is clear that \hi-deficient galaxies in Virgo have 
elevated molecular-to-atomic gas ratios,
confirming that \hi\ is much more affected by the cluster environment than \htwo.} \label{hih2}
\end{figure}

The fact that \htwo\ is significantly less affected by the cluster environment than \hi\ is evident in Fig.~\ref{hih2}, where we show \htwo-to-\hi\ mass ratios as a function of stellar mass for disk galaxies (S0 or later types) in the Virgo cluster, taken from \cite{hrsco}.
These are compared with results from xGASS-CO \citep[grey region;][]{catinella18}, a representative sample with CO($J=1\rightarrow 0$) measurements from xCOLD GASS \citep[the extended CO Legacy Database for GASS survey;][]{saintonge17}.
While \hi-normal galaxies (blue symbols) are consistent with galaxies from xGASS-CO,
\hi-deficient systems in Virgo (red symbols) are clearly \htwo-dominated (see also \citealp{mok16}; \citealp{loni21}). So, is \htwo\ affected at all, or are cluster galaxies simply not \htwo\ deficient?

Thanks to improvements of millimeter facilities, recent years have seen growing observational evidence in support of \htwo\ deficiency being 
a natural step in the path of satellite galaxies toward quenching.
One of the first and best direct examples of \htwo\ stripping is probably NGC4522 in the Virgo cluster (\citealp{vollmer4522}, but see also \citealp{vollmer05}), which shows extra-planar CO emission up to a few kpc from the disk, consistent with displacement due to ram pressure. \cite{fumagalli09} extended this work by providing the first statistical evidence for a population of \htwo-deficient galaxies in the Virgo cluster. By combining resolved \hi\ and \htwo\ observations for 47 galaxies across different environments, they showed that, {\it while \hi\ deficiency by itself is not a sufficient condition for molecular gas depletion, \htwo\ reduction is associated with the removal of \hi\ inside the galaxy's stellar disk}. This is consistent with the idea that the cold ISM is removed from the disk outside-in and molecular hydrogen, being more centrally concentrated than \hi, is affected only when the stripping becomes efficient within the radius of the stellar disk. Intriguingly, they did not interpret \htwo\ deficiency as evidence for direct stripping of CO, but more as evidence that -- when \hi\ is stripped -- the remaining atomic gas is at too low column densities to condense and form molecules. 

A similar trend was more recently presented by \cite{boselli14c}, who took advantage of single-dish CO observations for the volume- and K-band-limited Herschel Reference Survey (HRS, \citealp{HRS}) to quantify \htwo\ normalcy in a sample of nearby, \hi-normal late-type galaxies (based on either stellar mass or size). They then quantified the degree of \htwo\ deficiency in Virgo galaxies and showed that the most \hi-deficient galaxies are also \htwo-deficient. Importantly, galaxies are \htwo\ deficient only when they have already lost $\sim$80-90\% of their original atomic hydrogen reservoir (i.e., $DEF_{HI}>$1). Contrary to \cite{fumagalli09}, they interpreted this trend as mainly due to active stripping of CO by ram pressure. 

\begin{figure*}
\centering
\includegraphics[width=16.8cm]{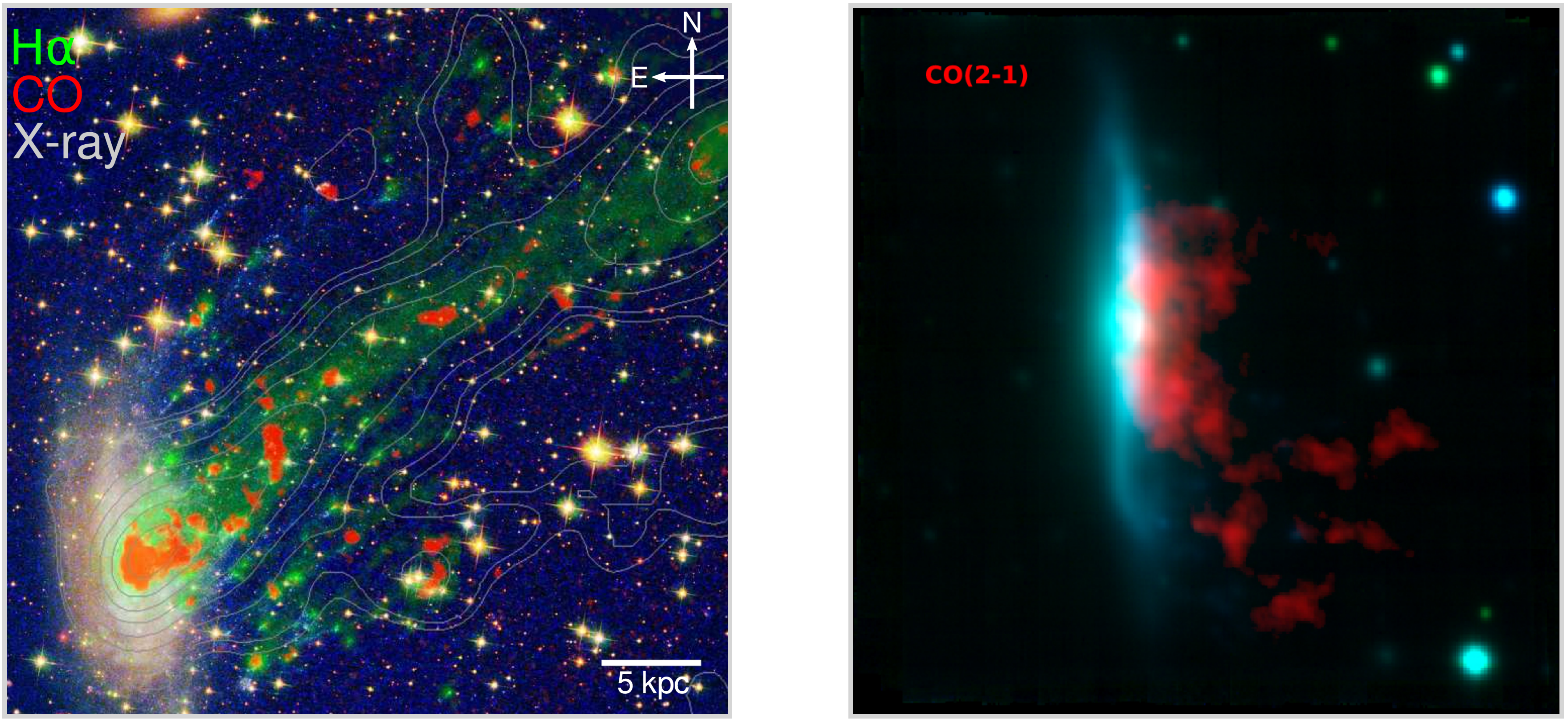}
\caption{Spectacular examples of stripping of molecular gas, as traced by the CO(2-1) transition line (in red). 
{\it Left:} ESO 137-001, a disk galaxy infalling into the Norma Cluster. CO(2-1) emission, along with H$\alpha$ emission in green and X-ray contours, are overlaid on a Hubble Space Telescope image (adapted from Figure 3 in \protect\citealp{jachym19}). {\it Right:} JW100 in Abell 2626, with CO(2-1) emission overlaid on a composite V- (blue) and I-band (green) image extracted from Multi-Unit Spectroscopic Explorer (MUSE) observations (adapted from Figure 8 in \protect\citealp{moretti20}).
Images reproduced with permission from the authors. Copyright by AAS.}\label{h2stripping}
\end{figure*}

The difference in the interpretation between these two works, despite their similar observational results, reflects the gradual paradigm shift that was beginning to take place in the community a decade ago, based on the growing evidence for individual objects showing direct signs of molecular hydrogen stripping, via the presence of extraplanar CO and/or CO tails (Fig.~\ref{h2stripping}; e.g., \citealp{corbelli12,scott13,sivanadam14,jachym14,jachym17,jachym19,lee17,lee4522_18,chung17,moretti19,moretti20}). 
Similar evidence supporting wide-spread CO stripping has also been reported for galaxies in the Fornax cluster \citep{zabel19}. While it is possible that part of the CO observed in the stripped tails was formed in situ (e.g., \citealp{jachym14}),  
the fact that these extra-planar features/tails do not show a clear separation from the CO disk suggests that at least part of the CO is being directly stripped from the disk.

Furthermore, it is now clear that cold dust is removed during the interaction with the ICM (e.g.., \citealp{cortese10c,cortese10b,cortese12,cortese16b,pappalardo12,bianconi20}) and that global oxygen abundance in the ISM 
of cluster galaxies shows clear enhancements with respect to isolated systems, with radial profiles sometimes steepening (e.g., \citealp{skillman89,shields91,petropoulou,hughes13,gupta16}).
This is consistent with expectations in the case where metal-poor cold hydrogen, mostly found in the outer parts of the star-forming disk, is stripped more preferentially than the metal-enriched ISM, which is typical of the inner parts of the stellar disk. In other words, we now have plenty of evidence that the low density, diffuse, atomic gas is not the only ISM component that is affected by the cluster environment.

{\it How can this conclusion be reconciled with the theoretical expectation that ram pressure can hardly directly strip molecular clouds?} 
A possible explanation is that not all CO emission from galaxies is bound within giant molecular clouds, but a significant fraction of it could trace a diffuse, thick, low volume density disk, as such more susceptible to environmental effects. 
While the first evidence for a CO thick disk associated to a star-forming galaxy dates back nearly three decades \citep[NGC 891;][]{garciaburillo92}, observational support for this has been growing in recent years.
On one side, the detailed analysis of M51 performed as part of the Plateau de Bure+30m Arcsecond Whirlpool Survey \citep[PAWS;][]{schinnerer13} demonstrated that practically 50\% of the CO emission originates from structures on spatial scales larger than $\sim$1 kpc, consistent with the presence of a diffuse disk of molecular hydrogen with a typical scale height of $\sim$200pc \citep{pety13}. On the other side, detailed studies of the velocity broadening of the CO line in star-forming disks have shown that the line widths are significantly wider than that expected from a cold, thin molecular disk (i.e., $\sim$10-12 km s$^{-1}$ instead of $\sim$5 km s$^{-1}$, \citealp{calduprimo13,calduprimo15}), and more in line with those observed for the \hi\ component, again suggesting that a significant fraction of the CO emission could come from warmer and more diffuse gas (e.g., $n\sim$ 100-500 cm$^{-3}$ and $T\sim$50-100 K; see also \citealp{liszt10}) than generally thought. This seems consistent with Milky Way observations, where diffuse CO emission associated with gas at surface densities more typical of \hi\ gas (e.g., < 10 M$_{\odot}$ pc$^{-2}$; \citealp{romanduval16}) appears to account for $\sim$25\% of the total mass of CO-emitting gas, but can even dominate the mass budget at large galactocentric distances (e.g., $>$8 kpc). If so, it would be easy to explain cases of extraplanar CO emission, CO tails and/or truncated CO disks observed in cluster galaxies. Indeed, this scenario seems to be consistent with recent high-resolution ALMA observations of NGC 4402 in Virgo \citep{cramer20}.

However, 
it is important to acknowledge that there are also cases where no evidence for a significant thick CO component has been found \citep[e.g. M33;][]{koch19}, and that we are still missing an extensive investigation of the frequency and properties of CO diffuse, thick disks, primarily due to the lack of CO observations able to trace emission at all physical scales. Moreover, it remains unclear if the fraction of diffuse CO emission maps directly into diffuse \htwo\ or not, which would clearly impact the overall physical scenario of direct molecular hydrogen stripping. Hopefully, the advent of facilities such as ALMA will soon fill this important gap in the field.

In summary, there is mounting evidence supporting the idea that CO-emitting gas can be directly removed from the disk, even by ram pressure, at least in 
environments similar to (or harsher than) those of the Virgo cluster. However, it seems clear that satellite galaxies can become \htwo\ deficient only if environmental processes are strong enough to be efficient within the optical radius of infalling 
galaxies. Otherwise, despite becoming \hi\ deficient, satellite galaxies could remain \htwo\ normal for a long time after infall (i.e., at least a couple of billion years after their first pericentre passage).
This fully supports a scenario of outside-in removal of the ISM, where each component is affected differently, primarily depending on its radial distribution across the disk. As discussed below, this is a key element for our current picture of galaxy quenching, and for reconciling apparent contradictions within the literature.

\subsection{The connection between cold gas stripping and star formation quenching}
The evidence for cold gas stripping in clusters presented above may naively suggest that there is a one-to-one correlation between gas deficiency and star formation quenching. As we discuss in this section, the situation is slightly more complicated. 

Due to the initial lack of CO observations, most of the early focus of the community has been on the link between \hi\ deficiency and star formation quenching. From the early work by \cite{kenn83virgo}, and since then gradually confirmed with better data and larger samples \citep{haanna,ha06,gavazzi13}, it is now clear that 
satellite galaxies with at least a factor of $\sim$2-3 less \hi\ than expected from their size and morphology (i.e., $DEF_{HI}\sim0.3-0.5$) have lower global SFRs than isolated systems. However, recent works have highlighted the presence of a significant population of \hi-deficient galaxies that still lie on the blue/main-sequence of star-forming galaxies \citep{cortese09,boselli14}. This nicely confirms what is seen for the molecular hydrogen component of the ISM: i.e., as long as the environment is only able to affect the ISM outside the optical radius, its immediate effect ($<$1Gyr) on star formation is minor or negligible. Indeed, for an outside-in gas removal scenario, those levels of \hi\ deficiency correspond to \hi\ disks affected up to one optical radius (see Fig.~\ref{viva}).

However, the clear proof of a direct connection between cold gas stripping and quenching of the star formation comes again from resolved studies. Particularly powerful have been narrow-band H$\alpha$ imaging surveys of nearby clusters of galaxies, which showed that the typical truncation of the gas disk in \hi-deficient galaxies is mirrored by a similar truncation in the SFR surface density profile. This field was pioneered by \cite{koop1,koop2} and \cite{koop06}, who surveyed a sample of 55 Virgo cluster galaxies, and showed that their star formation activity is primarily truncated in the outer parts, with the inner parts of the disk showing normal - or sometimes slightly enhanced - SFR surface densities. While, statistically, these results are consistent with active removal of the ISM, they stressed that the physical mechanisms at play might not be the same for all galaxies in their sample. 

\begin{figure}[t]
\centering
\includegraphics[width=6.5cm]{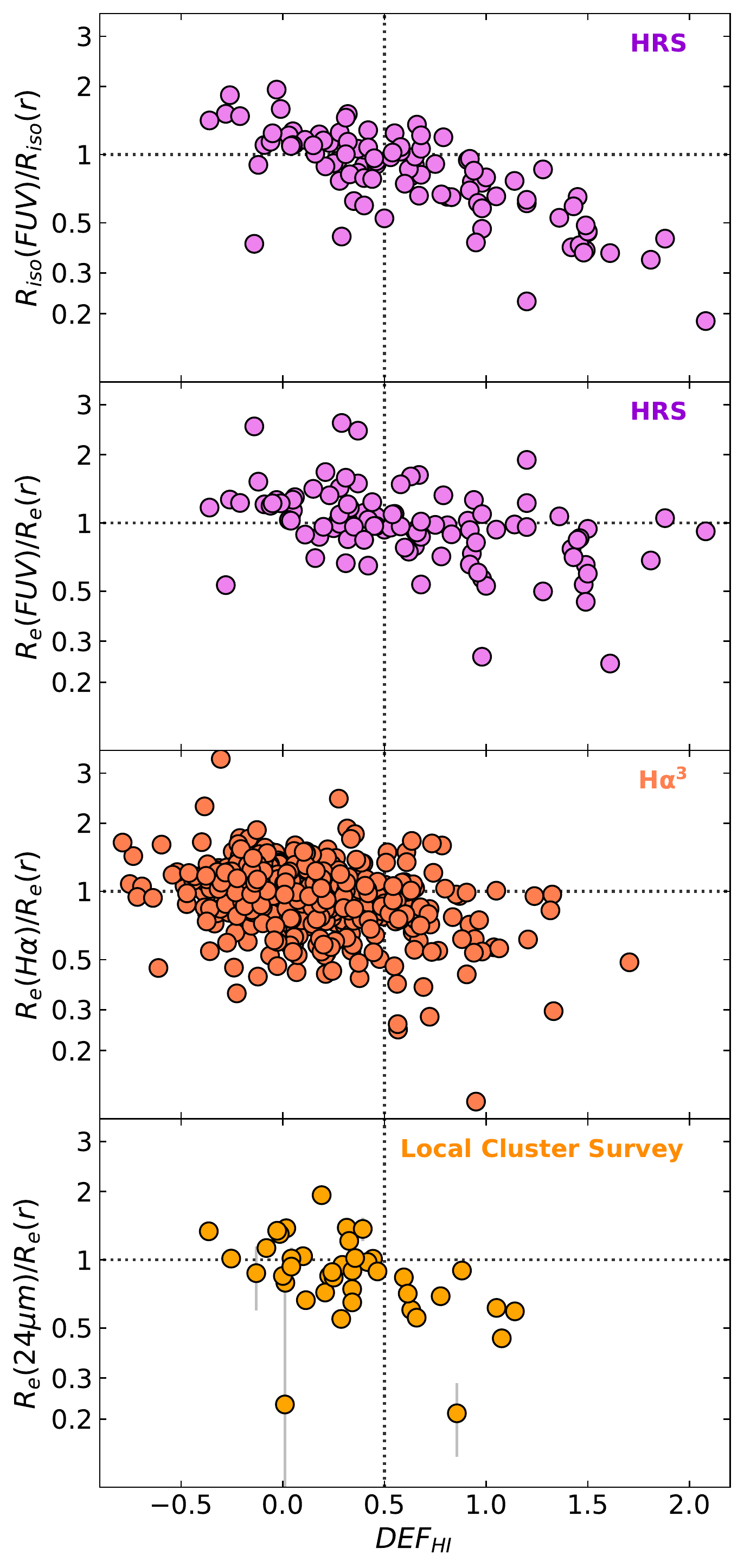}
\caption{Star-forming-to-optical ($r$-band) disk size ratio as a function of \hi\ deficiency. The panels show different SFR indicators and/or size estimates. Namely, from top to bottom: FUV isophotal and effective radii (Virgo cluster, \protect\citealp{cortese12}), H$\alpha$ effective radii (Virgo and Coma supercluster, \protect\citealp{fossati13}), and 24$\mu m$ effective radii (Local Cluster Survey, \protect\citealp{finn18}). } \label{isoradii}
\end{figure}

The existence of a strong correlation between \hi\ deficiency and extent of the SFR density profiles (normalised to the extent of the optical disk) has now been confirmed by several independent works looking at different clusters and using H$\alpha$, ultraviolet emission or 24$\mu$m emission as SFR indicators \citep{rose10,cortese12,fossati13,finn18}. However, as shown in Fig.~\ref{isoradii}, the strength of the correlation varies significantly depending on the star formation tracer used and, most importantly, on whether disk sizes are estimated using effective or isophotal radii. The latter generally provide tighter correlations, as a result of being more directly sensitive to variations in SFR surface density in the outer parts of the disk (see also \S~3.1). 

It is important to remember that the extent of the star-forming disk alone provides limited information on the actual causes of quenching, and cannot be blindly used to discriminate between active environmental stripping and simple cessation of infall of cold gas onto the disk. This is because a reduction of the extent of the star-forming disk can take place even without a change in the shape of the profile (i.e., by systematically reducing SFR across the entire disk, see e.g., \citealp{n4569}), whereas it is the change in shape and the clear presence of a ``truncation radius'' that supports the stripping scenario. This is exemplified by the apparent contradiction in the interpretation of H$\alpha$ morphology of cluster galaxies between \cite{koop1,koop2}, \cite{moss98} and \cite{moss00} on the origin of quenching of the star formation. Both groups reported that, in clusters, star-forming disks are more compact than in the field, but while the latter interpreted this as due to funneling of gas into their centres, feeding star formation primarily in the inner regions, the former argued for a stripping scenario. It is the ability to characterize the radial variation of SFR surface densities across the disk that allowed this debate to be settled, in favour of the stripping scenario. This is something to keep in mind, as the advent of large integral-field spectroscopy (IFS) surveys allows us to push these studies beyond the cluster environment (e.g., \citealp{schaefer17}; see \S~5.3) and better isolate the physical mechanisms affecting the shape of SFR surface density profiles in satellites.

\subsubsection{How quickly is star formation quenched following cold gas stripping?}
The link between gas stripping and quenching has been further reinforced by the ability to time the star formation quenching in the regions of the disk where gas has been stripped. This approach was pioneered by \cite{crowl2008}, who combined \hi, H$\alpha$ and ultraviolet imaging with resolved optical spectroscopy to determine how long ago star formation was quenched (i.e., the quenching {\it time}) just outside the \hi\ truncation radius for 10 Virgo cluster galaxies. Assuming nearly instantaneous quenching ($\lesssim$50 Myr), they found that, for all galaxies with truncated star-forming disks, the ages of the stellar populations outside the truncation radius are consistent with quenching within the last 0.5 Gyr. Remarkably, these quenching times agree with the time since when the gas was stripped, as estimated independently by \cite{vollmer04,vollmer4522} using N-body simulations that included the effect of ram-pressure stripping. This suggests that quenching is indeed very fast, once the gas has been stripped, and that this happens primarily close to the pericentre of the orbit around the cluster, which coincides with the time of peak intensity of ram pressure.  

These results have been corroborated by several independent works (e.g., \citealp{pappalardo10,boselli16,ciesla16,fossati18,owers19}), using different techniques, timescale 
definitions and multiwavelength datasets, confirming that in galaxies infalling into a cluster star formation has been quenched relatively recently ($\sim$0.5 Gyr ago) and on short timescales ($\lesssim$100-200 Myr) {\it after} the cold ISM has been removed from the disk.
These techniques generally require knowledge of the location of the \hi\ and/or SFR truncation radius (i.e., 2D resolved maps) and this is why it has so far been applied only to a relatively small number of galaxies.
\begin{figure}[t]
\centering
\includegraphics[width=8.5cm]{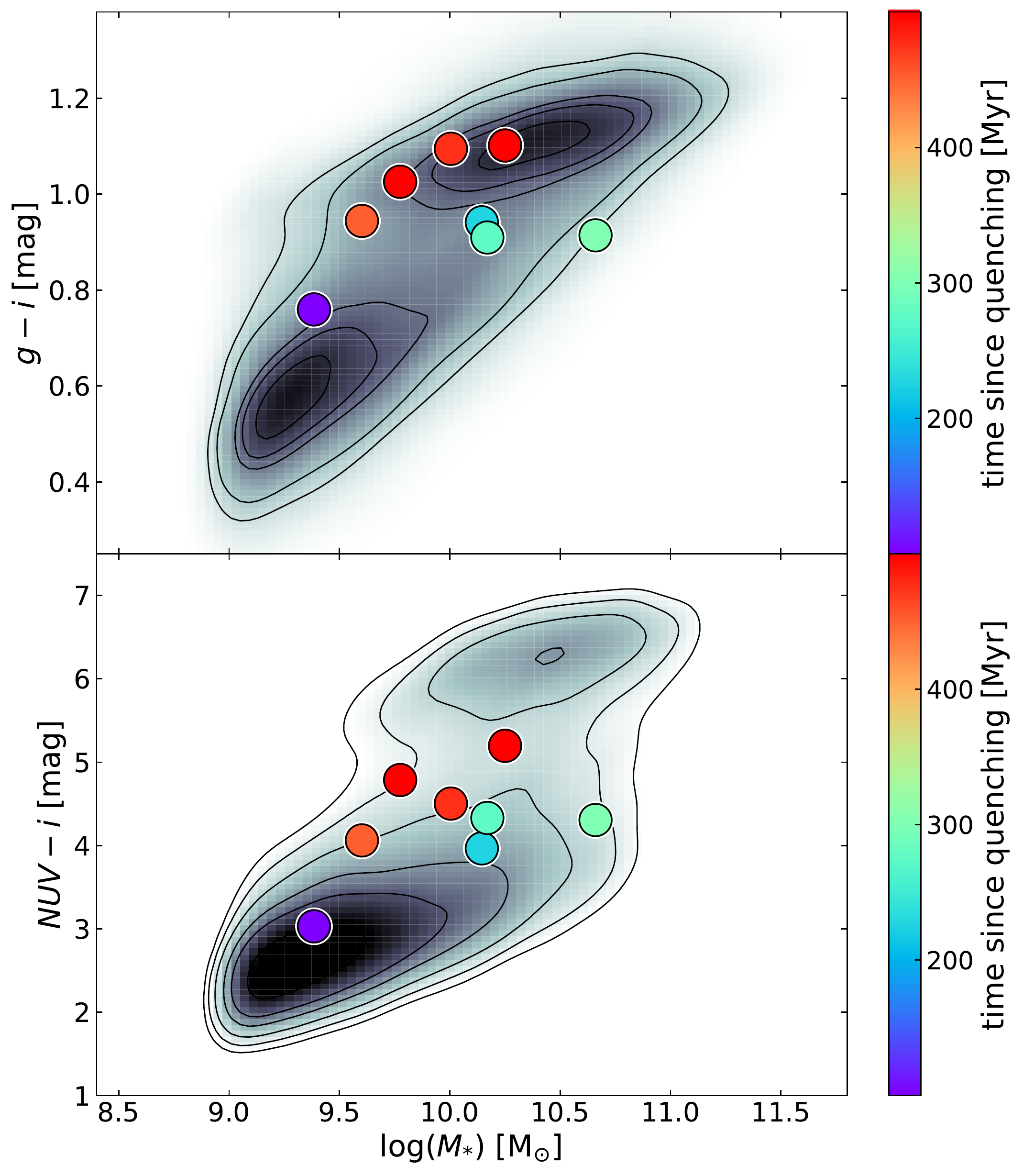}
\caption{The $g-i$ (top) and $NUV-i$ color (bottom) vs. stellar mass diagrams for the bulk of galaxies in the \protect\cite{crowl2008} sample. Points are color-coded according to the time since quenching. Stellar masses and colors are taken from \protect\cite{boselli14b}. Grey-scale and contours show the same relations for a volume-limited sample of nearby galaxies (0.02$<z<$0.05) extracted from the SDSS, shown to guide the eye on the location of the blue cloud ($g-i \lesssim 0.8$, $NUV-i \lesssim 4$) and red sequence ($g-i \gtrsim 1$, $NUV-i \gtrsim 5.5$). The region between the two is usually referred to as ``green valley''.}\label{crowl}
\end{figure}

One important point to note is that most of these works assume that, while the outer parts of the disks are consistent with rapid quenching of star formation, the inner parts do not show significant signs of quenching and their stellar populations are not dissimilar from those of unperturbed galaxies. While the sharp truncation seen in both \hi\ and SFR profiles clearly shows that star formation outside and inside the truncation radius proceeds in a different way, it is still unclear if/how environment is able to affect the star formation-cycle in the inner parts of galaxies. Is low-density gas stripped and star formation affected also in the inner parts of the disk? Or can environment compress the gas in the disk, thus increasing its star formation efficiency (e.g., \citealp{mok17})? This is something that only future resolved studies of statistical samples of cluster galaxies will be able to address. 
\begin{figure*}
\includegraphics[width=17.7cm]{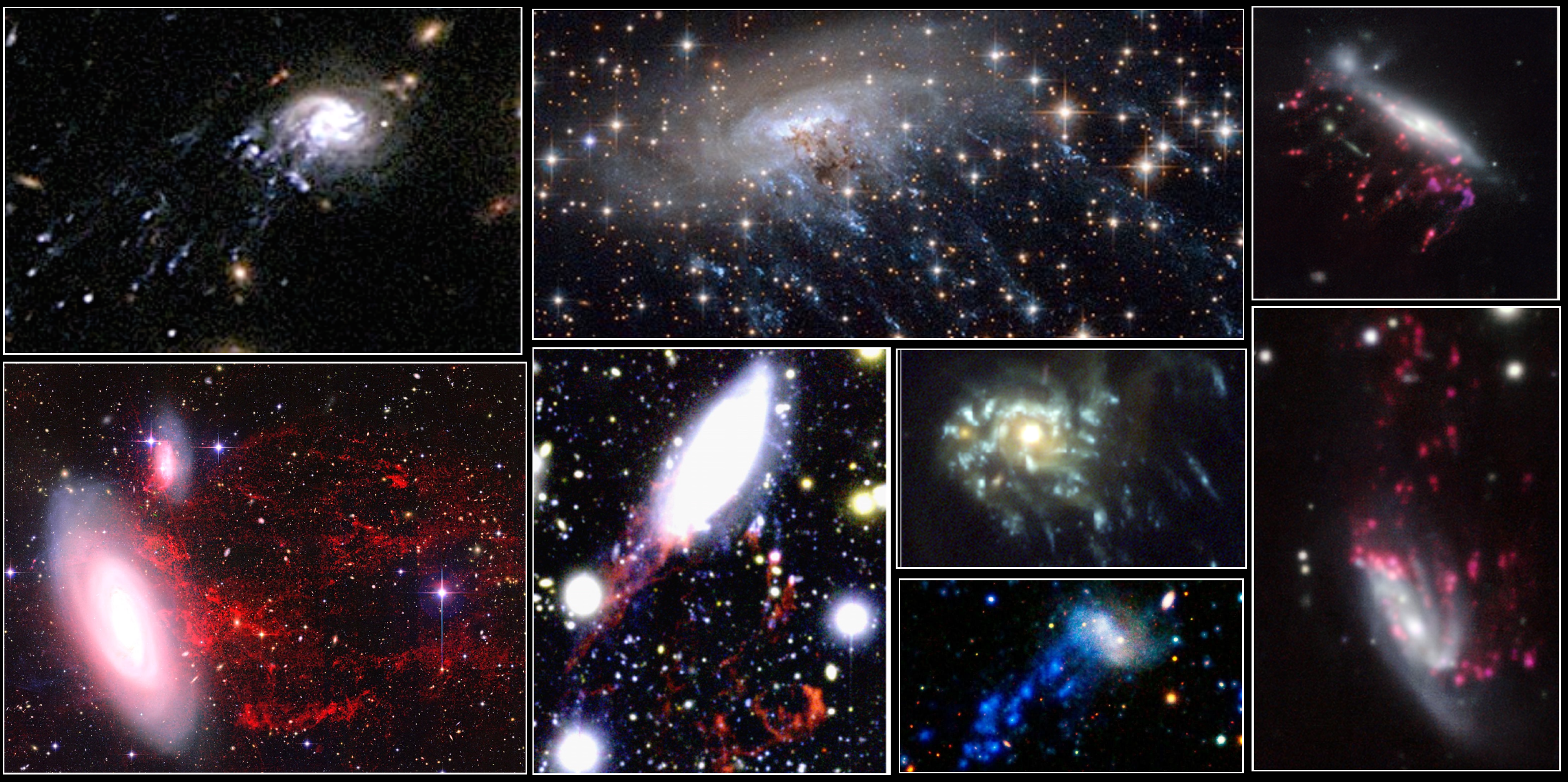}
\caption{Examples of galaxies where ionised gas and/or active star formation is observed in the stripped gas tail. Red and blue indicate ionised gas (primarily traced by the H$\alpha$ line) and star-forming regions (traced by ultraviolet or optical continuum emission), respectively. 
Images show galaxies in different clusters (starting from the top left in clock-wise order): Abell 2667 (NASA, ESA, Jean-Paul Kneib\protect\footnotemark{}, see also \protect\citealp{a2667}), Norma (NASA/ESA/STScI/M. Sun\protect\footnotemark{}, see also \citealp[]{jachym14}), Abell 957 and IIZw108 (ESO/GASP collaboration\protect\footnotemark{}, see also \protect\citealp{poggianti17}), 
Abell 2744\protect\footnotemark{} (see also \citealp{owers12}), IC 3418 in Virgo (NASA/JPL-Caltech\protect\footnotemark{} see also \protect\citealp{hester10b}), Coma (image reproduced with permission from Figure 4 in \citealp{yagi10}, copyright AAS) and NGC 4569 in Virgo (kindly provided by A. Boselli, see also \citealp{boselli16}).}\label{jelly}
\end{figure*}

Regardless, it should not be ignored that none of the galaxies analysed by \cite{crowl2008} are fully quenched, as they lie either in the lower part of the main sequence or in the green valley \citep{cortese09}. This is shown in Fig.~\ref{crowl}, where we plot the position of most of \cite{crowl2008} sample on the $g-i$ (top) and $NUV-i$ (bottom) color vs. stellar mass diagrams. Galaxies are color-coded by the time since star formation ceased in the stripped part of the disk. Even after $\sim$0.5 Gyr since quenching occurred beyond the truncation radius, these objects have total $NUV-i$ colors a couple of magnitudes bluer than red-sequence objects, while their optical colors are in most cases already consistent with passive galaxies. This is most likely because, even at pericentre, environment is not strong enough to perturb the inner parts of the disk of these relatively massive galaxies (M$_* >10^{9.5}$ M$_{\odot}$). 
Conversely, galaxies that quenched earlier appear to lie already on the optical red sequence. 
This is just a consequence of the fact that optical colours saturate, therefore galaxies with residual star formation (using ultraviolet indicators) might appear completely passive \citep{cortese12c}, and shows that optical colors should not be blindly used to time quenching.

While for high stellar mass galaxies full quenching may not occur by the first pericentre passage, it now seems established that, at low stellar masses ($<$10$^{9}$ M$_{\odot}$), most satellites close to (or past) their first pericentre passage have been fully stripped of their \hi\ reservoir and completely quenched \citep{dEale,boselli14,hester10b}. As we discuss in \S~4.5 and \S~7, this is a fundamental difference in how massive and dwarf galaxies respond to cold gas  stripping mechanisms.

\subsubsection{When cold gas stripping triggers star formation}
While in the long term cold gas stripping always leads to local quenching, on short timescales it is reasonable to wonder whether the high pressure exerted by the ICM on the ISM of satellite galaxies is able to compress the gas and trigger bursts of star formation. Indeed, several theoretical works  predict that, under particular conditions (e.g., the right combination of galaxy mass, ISM properties, inclination, infalling velocity and density of the ICM, etc.), the mechanism responsible for the stripping can also briefly enhance (either locally or globally) the star formation \citep{bekki03,bekki13c,kronberger08,kapferer09, roediger14,steyrleithner20}. While cases of galaxies showing clear examples of bow-shocks and asymmetries in their H$\alpha$ distributions (consistent with star formation triggered by ram pressure) have been known for decades (e.g., \citealp{gavazzi95,chaname00}) and recent studies have suggested that extreme environmental effects may dramatically increase the efficiency of \hi\ condensation into \htwo\ \citep{moretti20b}, the importance of these effects on the enhancement of the global SFRs of galaxies is still unclear, and most likely less than a factor of $\sim$2 \citep{iglesias04}. This seems in line with recent IFS observations showing that, even in the most extreme cases of ram-pressure stripping, the {\it global} SFR is enhanced on average by less than a factor of two (i.e., less than the typical scatter observed in the star-forming main sequence; e.g., \citealp{vulcani18,roberts20}). Of course, locally this enhancement could be larger, and the advent of large IFS surveys targeting clusters of galaxies should provide some definite quantification of environmentally-driven {\it local} bursts of star formation (e.g., \citealp{vulcani20}).

An unexpected region where it is now clear that stripping can trigger star formation is in the wake of the gas tail removed from the disk, where turbulence can compress atomic hydrogen, allowing its condensation into molecules and the onset of star formation. The presence of CO emission associated with the tails supports this scenario (e.g., \citealp{jachym14,jachym19}). Galaxies with prominent tails of star-forming knots started to be discovered in the mid 2000s, with probably the first confirmed cases presented by \cite{owen06} and \cite{a2667} in clusters at $z\sim0.2$
(although, as also noted below, evidence for 
extra-planar H{\sc ii} regions co-located with stripped gas in cluster galaxies had been around for much longer, e.g., \citealt{kenney99}).

Thanks to the improvement of optical and ultraviolet telescopes, the number of these objects (lately dubbed ``jellyfish'' galaxies) has increased dramatically, and we now know of more than one hundred cases of galaxies showing active star formation in their tails (see Fig.~\ref{jelly}; e.g., \citealp{sun07,yoshida08,smith10,hester10b,owers12,fumagalli11,ebeling14,mcpartland16,poggianti19}). The increased focus of the community on jellyfish galaxies has also been driven by the `GAs Stripping Phenomena in galaxies with MUSE' (GASP, \citealp{poggianti17}) survey, which is providing us with one of the most detailed views on the properties of ionised gas and star formation in this unique class of galaxies (e.g., \citealp{poggianti19,bellhouse19}).

\addtocounter{footnote}{-5}
\stepcounter{footnote}\footnotetext{\url{https://www.spacetelescope.org/news/heic0705/}}
\stepcounter{footnote}\footnotetext{\url{https://www.spacetelescope.org/news/heic1404/}}
\stepcounter{footnote}\footnotetext{\url{https://web.oapd.inaf.it/gasp/}}
\stepcounter{footnote}\footnotetext{\url{https://frontierfields.org/2016/11/14/the-hunt-for-jellyfish-galaxies-in-the-frontier-fields/}}
\stepcounter{footnote}\footnotetext{\url{http://www.galex.caltech.edu/newsroom/glx2010-02f.html}}

Interestingly, the increase in the number of satellite galaxies with star-forming tails, and ensuing popularity of the ``jellyfish'' term, has led part of the community to label as such every system showing a hint of a tail of stripped material, regardless of whether this material is star-forming (as in the original definition) or not (e.g., \citealp{chung09,boselli16}, see also Fig.~\ref{jelly}). If one adopts this broader definition, then the first clear cases of jellyfish galaxies date back at least three decades, with the first detections of radio continuum tails in clusters \citep{GAVJAF87,gavazzi95}, and it is pretty safe to assume that every satellite galaxy in clusters will go through some kind of jellyfish phase during its evolution\footnote{Adding to the ambiguity in the definition of a jellyfish system, it is important to remember that the likelyhood of detection of a tail is also subject to projection effects.}. 

\begin{figure}
\centering
\includegraphics[width=8cm]{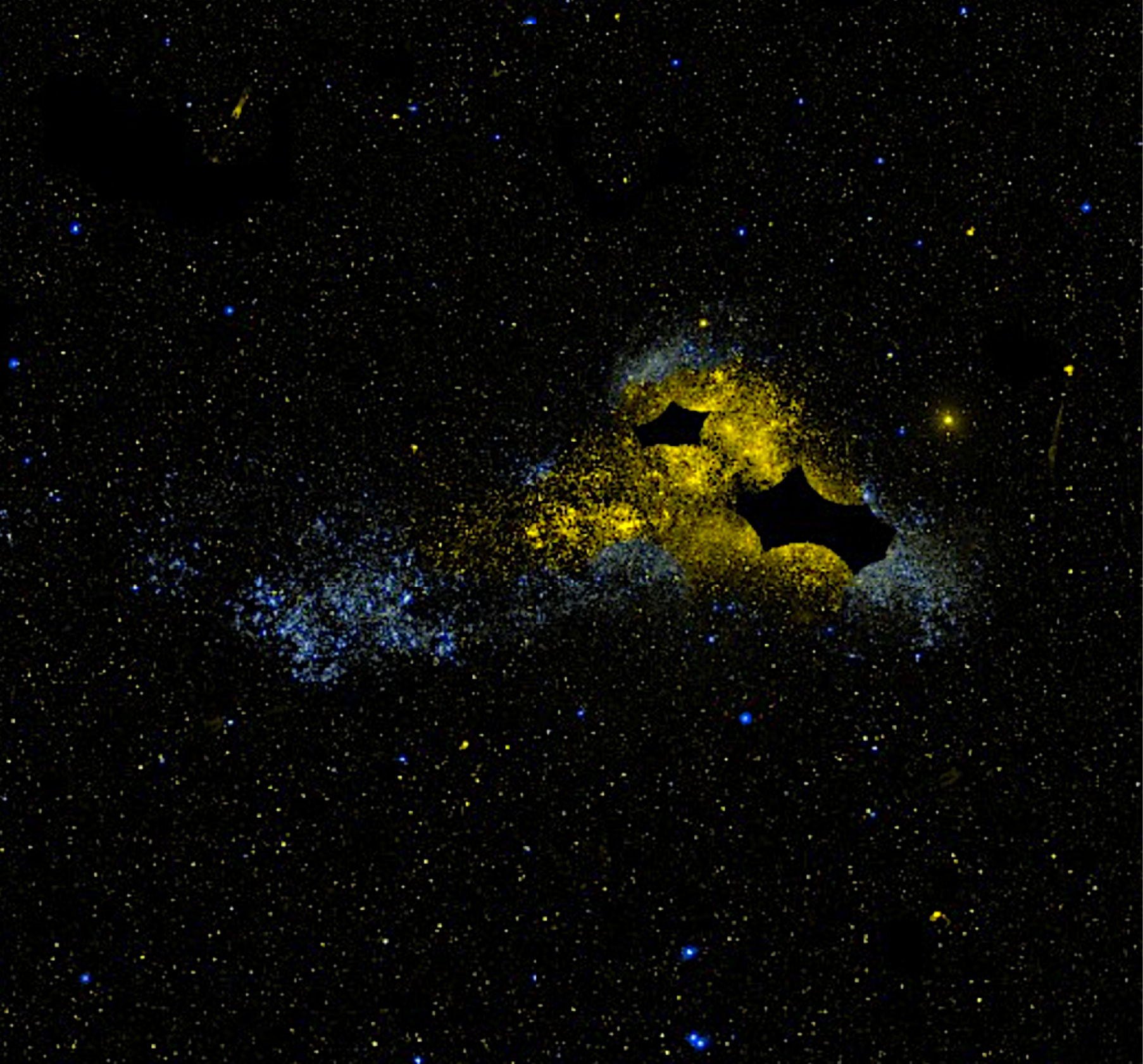}
\caption{A GALEX far- (blue) and near-ultraviolet (yellow) color composite image of the Small Magellanic Cloud. The star-forming tail associated with the galaxy closely resembles that of some jellyfish galaxies in clusters. The holes in the inner parts are due to the lack of GALEX pointings for the galaxy centre. Similarly, the yellow color of most of the main body is simply due to the lack of far-ultraviolet imaging. Credit: GALEX/NASA/JPL-Caltech\protect\footnotemark} \label{SMCjelly}
\end{figure}

However, it is still unclear what is the fraction of cases in which the stripped material `lights-up' and forms stars in the wake of the tail. Indeed, the reason why some gas tails are able to light-up, while others do not is not completely understood. Numerical simulations point to a combination of infalling velocity, ISM density and ICM density and temperature, with the confining pressure of the ICM surrounding the tail likely playing an important role \citep{Tonnesen11}. However, there are still a lot of open questions regarding the complex and intertwined roles of turbulence, magnetic fields, and thermal conduction within the wakes \citep{Tonnesen2014,vijayaraghavan17}. Multi-phase, high-resolution observations able to resolve individual giant molecular clouds, measure the properties of star-forming complexes in the tails and trace magnetic fields \citep{mueller20} should help with making further progress in this area.

Lastly, it is important to note that, while triggered star formation in the stripped gas tail is expected to be mainly associated with ram pressure, this does not mean that ram pressure is the only mechanism helping to light-up the tail. We now know
of several examples where gravitational interactions may also play a role. Probably the best, most generally overlooked example is represented by the Small Magellanic Cloud, whose UV tail is reminiscent of many jellyfish galaxies (see Fig.~\ref{SMCjelly}), despite the fact that ram pressure is known to be not the only mechanism at play in this case (e.g., \citealp{donghia2016}).
\footnotetext{\url{https://www.legacysurvey.org/viewer}}

\begin{figure*}
\includegraphics[width=17.3cm]{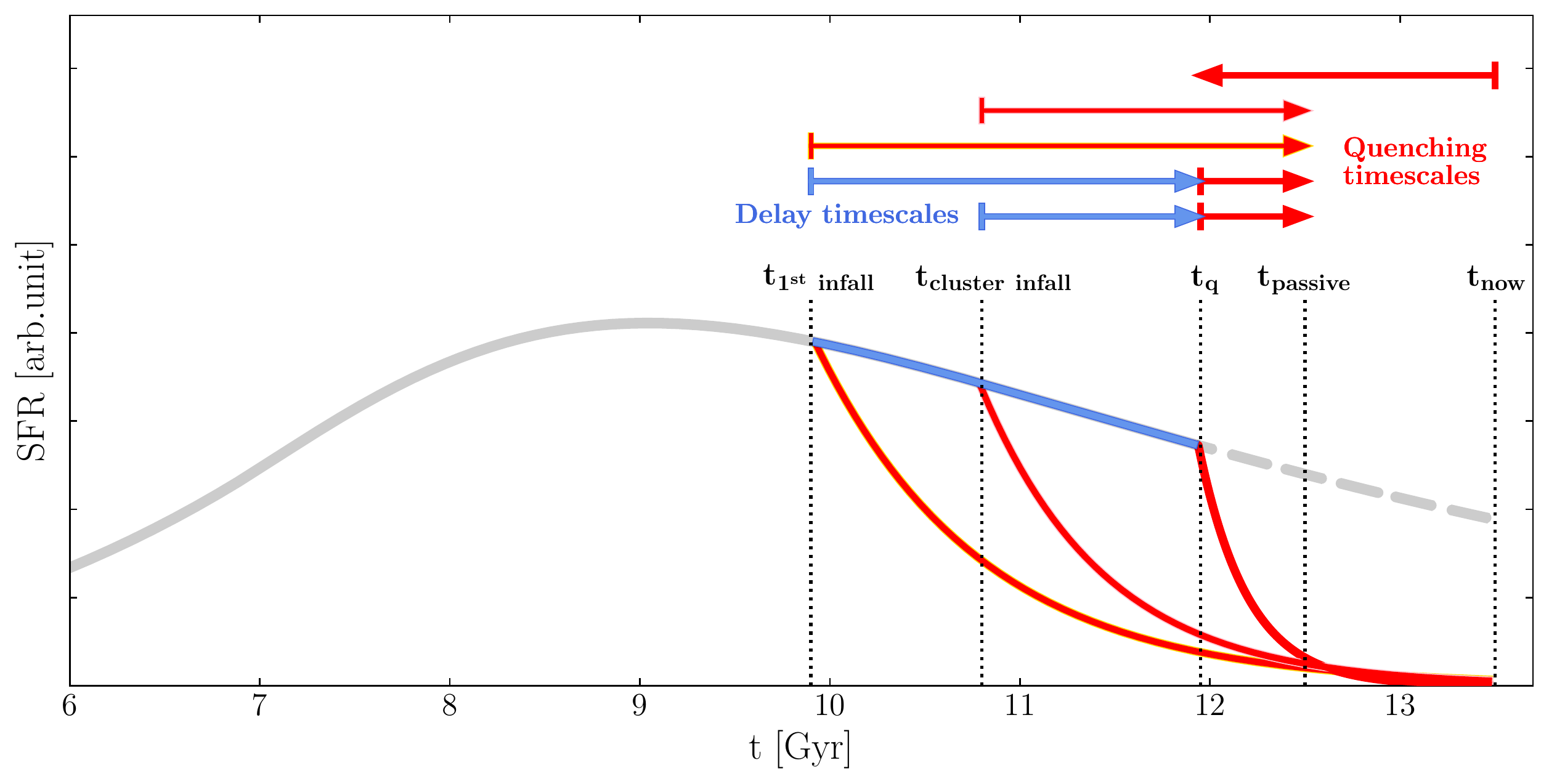}
\caption{Cartoon illustrating various definitions of satellite quenching timescales found in the literature. The solid plus dashed line shows the expected star formation history for a galaxy with stellar mass $\sim$10$^{9.5}$ M$_{\odot}$ at $z\sim$1 that stays on the main sequence until $z=$0. Five key timescales are indicated: $t_{1^{st}~infall}$ is the time when the galaxy becomes a satellite; $t_{cluster~infall}$ is the time of infall into its current host (assumed to be a cluster in this case); $t_{q}$ is the time when environment starts affecting the galaxy's SFR; $t_{passive}$ is the time when the galaxy crosses the SFR or sSFR threshold adopted to separate active and passive systems. 
The red lines show how its SFR can be affected by the cluster environment, depending on whether $t_{q}$ corresponds to the time of first infall, to the time of cluster infall or to some later time.
Red and blue arrows show the relative different definitions of delay and quenching timescales, with the direction of the arrow giving an idea of how time is generally assumed to proceed in the timescale estimate.} \label{QTcartoon}
\end{figure*}

\subsection{Star formation quenching timescales}
As discussed in the previous section, all observational evidence presented so far supports a scenario in which, when cold gas is stripped from the disk, {\it star formation in the stripped regions} is fully quenched quickly (i.e., on timescales of a few hundreds of Myr or less). This is consistent with the predictions from theoretical modeling of ram-pressure stripping, leading many to associate hydrodynamical effects to short {\it full quenching} timescales. However, as argued below, this assumption is plainly incorrect and, combined with the different definitions of quenching timescales presented in the literature, has caused considerable confusion.

The estimate of the time needed by cluster satellites to leave the main sequence of star-forming systems does not require any information on their cold gas content, but primarily a quantification of their stellar ages and/or recent star formation history. Thus, studies focused on timing the quenching of star formation in satellites have been able to benefit, in the last two decades, from the large statistics provided by wide-area photometric and spectroscopic surveys such as the SDSS, inevitably detaching themselves from the approaches adopted in the detailed studies of the few clusters with cold gas information available. In these statistical studies, the goal is not to determine quenching timescales for individual galaxies (something generally challenging with just UV/optical broad-band integrated photometry anyway), but to statistically infer them by matching the predictions of cosmological models of structure formation with the average properties of satellite galaxies, such as their passive fraction, their specific SFR distribution and/or their 3D position within the cluster (e.g., \citealp{balogh00b,wetzel13,haines15,oman16,rhee20,oman20}). The use of merger trees in a cosmological context is the key element that makes this technique orthogonal to the one applied to individual galaxies and described in previous sections. 

In order to illustrate the different definitions of quenching timescales adopted in the literature we refer to Fig.~\ref{QTcartoon}, which schematically shows the evolution of the SFR with cosmic time for a galaxy, highlighting four key epochs in its evolution: i.e., the time when it first becomes a satellite (i.e., crossing the virial radius of a host halo at $\rm t_{1^{st} infall}$), the time when it is accreted into an even bigger halo, such as a cluster of galaxies (crossing the cluster virial radius at $\rm t_{cluster~infall}$), the time when its SFR starts being quenched by the environment ($\rm t_q$) and when it becomes passive ($\rm t_{passive}$). The present day is indicated as $\rm t_{now}$.

Going back now to the quenching timescales estimated via matching to cosmological merger trees, in this case the clock starts when the galaxy becomes a satellite (either at first infall -- $\rm t_{q}=t_{1^{st} infall}$ -- or when it crosses the virial radius of the current cluster host -- $\rm t_{q}=t_{cluster~infall}$) and increases with the amount of time spent in the cluster until the object becomes passive. More recently, \cite{oman20} introduced a variant in which the clock starts at the time when a satellite reaches its first pericentre passage through the cluster (not shown in the figure).

Conversely, for the technique based on individual galaxies the timescale is generally measured from $z=0$ ($\rm t_{now}$) to the time when star formation starts being affected by the environment, and is expressed in terms of how long ago the galaxies started quenching. The additional complication here is that sometimes quenching is assumed to be instantaneous, and what is determined is how long ago star formation ceased, instead of a quenching timescale.


It is clear that, with such different assumptions, the two broad approaches can only agree under extremely rare circumstances: e.g., if a) quenching is not assumed to be instantaneous, {\it and} b) the time of infall corresponds to the time when environmental quenching starts,
{\it and} c) the object crosses the threshold used to divide star-forming from quenched galaxies at the time of the observation (i.e., $\rm t_{passive}=t_{now}$). 

Interestingly, the first applications of this technique to SDSS observations of cluster satellites (e.g., \citealp{balogh00b}) assumed that quenching starts indeed at the time of first infall, and motivated this choice by the fact that, in most semi-analytic models at that time, the primary mechanism for environmentally-driven quenching was stripping of the hot halo gas reservoir, which started as soon as a galaxy became a satellite (see also \S~8). 
Under these assumptions, despite significant differences in the technique used to estimate quenching timescales (e.g., definition of virial radius, threshold used to separate star-forming from passive galaxies, observations used to constrain the model, etc.), all studies reached the same general conclusion: it takes a few billion years (e.g., from $\sim$2 to $\sim$6 Gyr) for satellites to become passive since the time of their infall (e.g., \citealp{vonderlinden10,weinmann10,delucia12,wetzel13,hirschmann14,oman16}).   

Such long quenching timescales have always been assumed to be inconsistent with expectations from stripping mechanisms that directly affect the cold ISM in a star-forming disk. Instead, they are more in line with theoretical estimates based on starvation/strangulation, whereby quenching is caused by halting the infall of gas onto the disk\footnote{We note, however, that some models including only starvation can also lead to short ($<$1 Gyr) depletion times, as star formation can still be quenched by gas ejection via outflow (i.e., an internal process even for satellites, e.g., \citealp{zoldan17,xie20}).}.
Indeed, regardless of the uncertainties associated with the various techniques, these timescales are arguably much longer than what is observed in the clearest cases of ram-pressure stripping. Thus, either the detailed studies of individual galaxies in Virgo provide us with a biased and non-representative view of the quenching pathway followed by satellite galaxies, or the basic assumption in the statistical estimates of quenching timescales (i.e., that quenching starts at time of infall and proceeds exponentially since then) is incorrect.

Starting from the work presented by \cite{wetzel13}, the refinement in the techniques used for the statistical estimate of quenching timescales suggests that the main quenching phase does not start at the time of infall, but is significantly delayed (e.g., $\sim$2-6 Gyr; \citealp{haines15,oman16,rhee20,oman20}), as the star formation in satellites is not immediately affected by the environment after infall (indicated in light blue in Fig.~\ref{QTcartoon}). After this delay phase, quenching proceeds extremely fast, with satellites becoming passive in one billion year or less (the exact value being again heavily model-dependent). Such a delay phase does not imply that the environment is not affecting satellites, but that its effect on star formation is negligible. Indeed, \cite{maier19} find that star-forming cluster galaxies show a $\sim$0.1-0.2 dex enhancement in their oxygen abundance well before a decrease in their SFR. This supports a ``slow-then-rapid'' quenching scenario, where the ISM content and/or metal enrichment is affected before the SFR starts decreasing (e.g., infall is stopped and metals are gradually less diluted; see also \S~5.3.2). 

Regardless of whether star formation remains constant or drops mildly after infall, it seems now well established that the main quenching phase (i.e., when satellites leave the locus of the star-forming sequence) does not start at the time of infall, but closer to the first cluster pericentre passage. This is a key point that allows us to reconcile different estimates of quenching timescales and build a coherent picture of quenching in clusters of galaxies.

\begin{figure*}[t]
\includegraphics[width=17.7cm]{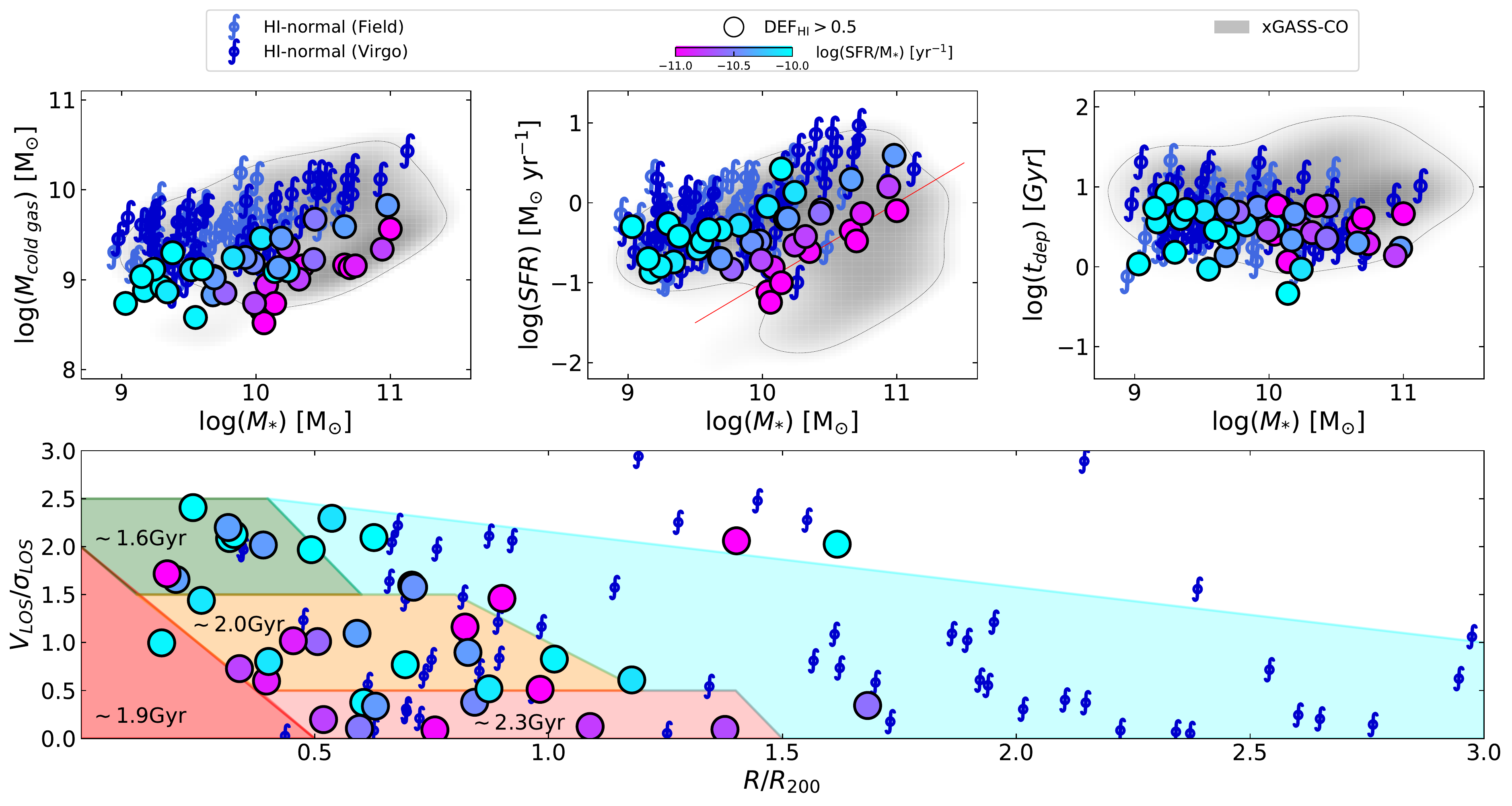}
\caption{Properties of \hi-deficient spiral galaxies in the Virgo cluster.
In all plots, large symbols show HRS star-forming, spiral galaxies observed in both \hi\ and \htwo\ and for which at least one of the two phases has been detected, separated into \hi-deficient Virgo members ($DEF_{HI}>$0.5; circles, color-coded by specific SFR) and \hi-normal systems in the Virgo volume and in the field (dark and light blue spiral symbols, respectively). {\bf Top row.} Stellar mass is plotted versus total cold gas mass (left), SFR (middle) and total gas depletion time (i.e., total gas mass-to-SFR; right); grey shaded regions show the corresponding distributions for the xGASS-CO sample (no morphological selection) as a reference.
The red line in the middle panel indicates the threshold used by \protect\cite{wetzel13} to separate star-forming from passive galaxies. {\bf Bottom row.} Position of Virgo galaxies on the projected phase-space diagram, which is split into five regions as proposed by \cite{rhee17}; these roughly correspond to increasingly longer infall time (since crossing the virial radius) populations, from first infallers (cyan), to recent and intermediate ones (green, orange and pink), to ancient infallers/virialised (red). For the latter four regions, we provide typical infall times for the `Recent Infallers' population as defined in \protect\cite{rhee17}. Symbols are as in the top row.
} \label{phasespace}
\end{figure*}

In the rest of this section, we show how tensions between long and short quenching timescales 
for Virgo cluster galaxies that are clearly affected by ram-pressure stripping can be resolved. From the HRS sample, we select late-type (Sa or later types) galaxies with active star formation (as traced by the H$\alpha$ emission line; \citealt{hrsha}), with observations of both global \hi\ and molecular hydrogen content (as traced by the CO(1-0) transition line; \citealt{hrsco}), and for which at least one of the two phases has been detected. We then focus on the \hi-deficient cluster members ($DEF_{HI}>$0.5; \citealt{cortese11}).
By selection, this sample includes the bulk of Virgo galaxies for which quenching timescales have been directly estimated, all suggesting either instantaneous or fast (a few hundred Myr or less -- e.g., \citealp{crowl2008,pappalardo10,boselli16,ciesla16,fossati18}) quenching timescales. 

In the top row of Fig. ~\ref{phasespace}, we compare the location of these galaxies (filled circles) with that of non \hi-deficient HRS late-types (blue spiral symbols) on the total cold gas mass vs. stellar mass (left panel) and SFR vs. stellar mass (middle) diagrams. The total gas mass is the sum of atomic and molecular gas masses, including a 30\% correction for helium. As in Fig.~\ref{hih2}, we also show the distribution of the xGASS-CO representative sample as a reference for nearby galaxies (grey shaded area).
Unsurprisingly, \hi-deficient galaxies are also more passive (i.e., offset towards lower SFR at fixed stellar mass) than \hi-normal systems but, by selection, none of them have stopped forming stars completely. Only few have already crossed the threshold separating passive from active galaxies, here indicated by the red line as defined in \cite{wetzel13}.

The fact that some of these galaxies have short quenching timescales (e.g., $\lesssim$0.2-0.5 Gyr) is not at odds with them not being fully quenched. As already noted, estimates obtained for individual galaxies are either local timescales for complete quenching (i.e., where star formation has already stopped, e.g., \citealp{crowl2008,fossati18}) or times since star formation started to decrease, regardless of whether the galaxy is already fully quenched (e.g., \citealp{ciesla16,boselli16}). Thus, it is clear that quick -- even instantaneous -- local (i.e., in the outer disk) quenching via stripping does not automatically imply rapid global quenching of the star formation. This is a key misconception that has led to confusion in the literature.
The reason why these galaxies are still star-forming is simple: while they have already lost most of their cold gas reservoir in the outer parts of the disk, cold gas is still present in the inner parts, where star formation is still ongoing.   
Indeed, as shown in the top right panel of Fig.~\ref{phasespace}, the cold gas 
global depletion times for these galaxies are not too dissimilar from those observed in main sequence star-forming galaxies, and at least a couple of billion years, without even considering gas recycling (see also \citealp{cortese09} and \citealp{boselli14b}). Thus, without any additional major stripping event, billions of years are needed for these galaxies to become fully passive, despite the fact that they were selected to be \hi\ deficient. 

The next step is to estimate how long ago these galaxies became satellites of the Virgo cluster. To do so, we take advantage of the approach developed by \cite{rhee17} and look at the position of our \hi-deficient sample on the Virgo phase-space diagram\footnote{The phase-space diagram combines information on the projected cluster-centric distance and line-of-sight velocity of a satellite relative to the typical dispersion velocity of the cluster. This is a powerful tool used for decades to identify cluster members and objects at different infall stages (e.g., \citealp{kent82}).}, practically extending the analysis recently published by \cite{yoon17}. By using cosmological hydrodynamic N-body simulations of clusters of galaxies, \cite{rhee17} provide an easy statistical way to assign infall timescales to different galaxy populations according to their position on the projected phase-space diagram. This is divided into five different regions, based on the typical fractions and relative infall times of different satellite populations (namely, first infallers, recent infallers, intermediate infallers, and ancient infallers). First infallers have not yet crossed the virial radius for the most part, whereas the other classes are already cluster satellites with gradually increasing infall times. 

In the bottom panel of Fig.~\ref{phasespace}, we show the position of the \hi-deficient Virgo satellites in phase-space, with the coloured boxes highlighting the regions defined by \cite{rhee17}. By selection, almost all our galaxies are at least recent infallers (i.e., they are already Virgo cluster members) and occupy the three regions (green, orange and pink) typical of galaxies that are still completing their first orbit into the cluster (i.e., either just before or after first pericentre passage), in agreement with the detailed dynamical modelling of Virgo galaxies by \cite{vollmer09b}. There is a dearth of galaxies in the ``fully-virialised'' zone at the centre of the cluster (red region). This is simply by selection, as we do not include passive (and thus most likely virialised) galaxies in these plots. It is interesting to note that, when moving from the green to the pink region, the specific SFRs of \hi-deficient galaxies tend to decrease from levels typical of the star-forming main sequence (cyan circles) to lower values (violet and purple circles), supporting the idea that less active galaxies have spent longer times in the cluster. According to \cite{rhee17}, galaxies in these regions have an 80\% likelihood to have crossed the cluster virial radius at the very least $\sim$1-1.5 Gyr ago (i.e., the average value for `recent infallers' varies between $\sim$1.6-2.3 Gyr, see bottom panel of Fig.~\ref{phasespace}), consistently with the typical expectation that it takes $\sim$1 Gyr for a satellite to get from the virial radius to first pericentre (e.g., \citealp{lotz19}). Intriguingly, the picture emerging from this exercise is qualitatively consistent with recent work by \cite{oman20}, who combined numerical simulations, SDSS optical data and ALFALFA \hi\ observations to revisit the connection between \hi\ stripping and full star formation quenching.

In summary, we have shown that the bulk of the Virgo \hi-deficient late-type galaxies with residual gas reservoir and star formation in their disk are consistent with stripping of cold (atomic) hydrogen by the cluster environment, they have been cluster satellites for at least 1 Gyr and have enough cold gas to keep forming stars (at the current rate) for at least another couple of billion years. Thus, if we had started the clock at the time of infall into the cluster, we would infer quenching timescales that are significantly longer than what one might expect from a simple ram-pressure stripping scenario. In other words, while still a useful quantity, {\it the global quenching timescale provides us with limited information on the physical process(es) responsible for halting star formation in satellite galaxies.}


\subsection{Drawing a coherent picture of the quenching of cluster satellites at $z\sim$0}

Having reconciled the apparent tension between different estimates of quenching timescales, showing that observations of fast ``local'' stripping are not in contradiction with claims of full quenching timescales of a few Gyr since infall, we can now combine all the evidence summarised above into a coherent picture of the typical pathway to quenching followed by cluster satellite galaxies. 
For simplicity, here we focus on the simple case of a galaxy becoming satellite for the first time when entering the cluster. We discuss how pre-processing might alter this picture in \S~6, after reviewing the role of cold gas stripping in the quenching of group satellites in \S~5.

{\bf{First infall.}}
After entering the cluster halo, it takes some time for the star-forming disk to feel the effect of the environment, and star formation does not drop immediately after crossing the virial radius. It is still unclear if gas accretion onto the disk remains active for some time after infall or if it stops immediately. 

{\bf{Approaching pericentre.}}
The first major quenching phase starts only when active stripping of the cold ISM becomes efficient at least in the outer parts of the disk, which generally coincides with a few hundred Myr before pericentre passage. Stripping happens predominantly outside-in, perturbing first the outer parts of the disk that are less bound to the galaxy. While the primary mechanism at play seems to be hydrodynamical, in some cases gravitational interactions and/or internal mechanisms, such as feedback from supernovae or AGN, could increase the stripping efficiency (see e.g., \S~4.1). 
How far stripping gets inside the optical disk during the first pericentre passage likely depends on the properties of both satellite (e.g., its mass, orbit, ISM properties, etc) and cluster (e.g., ICM density, dynamical state, etc), though we are still missing a detailed characterisation of the physical parameters that determine the gas truncation radius. Moreover, parts of the ISM might be affected by the environment even inside the truncation radius, but testing this requires observations of the inner parts of the cold gas disk of satellite galaxies at kpc or sub-kpc spatial resolution.

{\bf{Mass-dependent quenching.}}
At least for a Virgo-like cluster it appears that, to first order, low-mass galaxies ($M_{*}<$10$^{9}$ M$_{\odot}$) can be fully stripped at first pericentre (e.g., \citealp{dEale}), but massive systems ($M_{*}>$10$^{10}$ M$_{\odot}$) are not, as a significant fraction of their gas reservoir survives in the inner parts of the disk and keeps feeding star formation. This implies that, for massive galaxies, full quenching takes place significantly after the first-pericentre passage and that ram-pressure stripping alone is likely not enough to explain the properties of passive satellites. This also appears to be consistent with the presence of small molecular gas reservoirs in the inner parts of early-type galaxies in clusters (e.g., \citealp{young11,davis13}). Admittedly, what happens to a galaxy's remaining gas after pericentric passage is one of the most important missing pieces to the quenching puzzle. While environmental effects may still be playing a role, it is possible that the remaining gas is to a large extent exhausted internally via star formation or expelled/heated by feedback. 

We note that a scenario in which low-mass satellites are quenched faster than higher mass ones may appear in contradiction with some claims of longer quenching timescales for low-mass systems (e.g., \citealp{wetzel13}, see also \S~5.4). However, this is again a matter of quenching timescale definitions. First, if the clock starts at time of first infall, it is natural that low-mass galaxies will take longer to be quenched as, on average, they spend longer in smaller groups before infalling into a cluster \citep{delucia12}. Second, the use of a simple threshold in specific SFR to separate passive and active galaxies may introduce systematic effects in the estimate of quenching timescales (e.g., \citealp{pasquali19,rhee20}). Indeed, even on the star-forming main sequence, high-mass galaxies have lower specific SFRs than low-mass ones. As such, once star formation stops, their specific SFR will reach any fixed threshold sooner than lower mass systems, even if the rate of quenching is the same. Indeed, when individual star formation histories are examined, both observations and simulations seem to agree that low-mass galaxies are quenched more rapidly (e.g., \citealp{pasquali19,lotz19,smith19}).       

{\bf{Is stripping of cold ISM from the disk sufficient to halt star formation?}}
The availability of cold gas in the inner part of massive disks that continues to feed star formation implies that, at high masses, direct stripping of the cold ISM from the disk of satellite galaxies is a {\it necessary but not always sufficient} condition to halt star formation. At face value, this means that neither the stripping of the cold ISM (e.g., ram pressure), nor the cessation of infall (e.g., starvation) alone are sufficient to explain the quenching of massive galaxies in clusters, but that both are required: environmental effects strip part of the cold gas {\it and} are responsible for stopping the infall of gas, but gas consumption via star formation likely affects the inner parts of the disk, where environment is less effective. In addition, it is possible that feedback may further contribute to emptying the inner cold gas reservoir of galaxies (see also \S~8). This highlights, once more, how inadequate it would be to invoke one single mechanism as the sole responsible for star formation quenching. 

\begin{figure*}
\includegraphics[width=17.7cm]{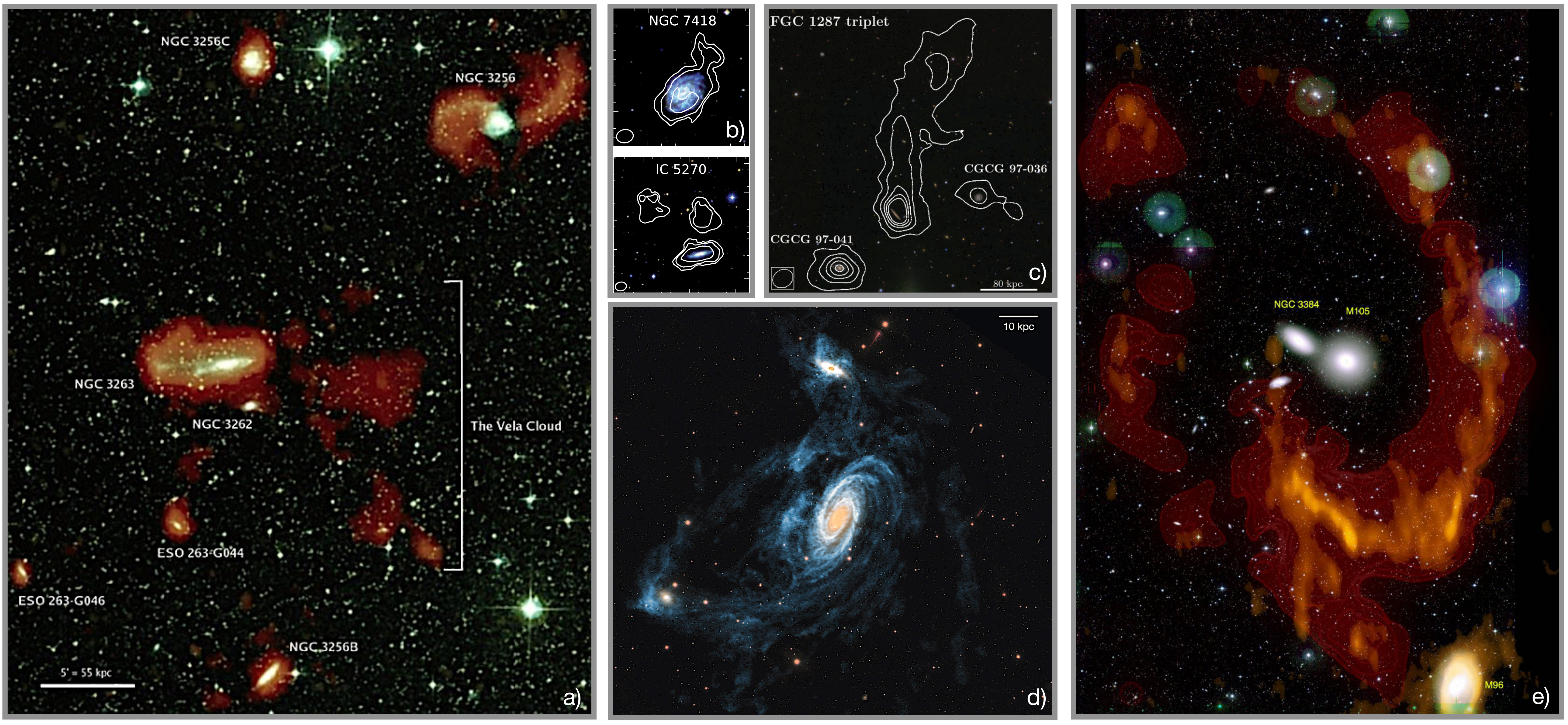}
\caption{Examples of \hi\ stripping in galaxy groups. In all panels, \hi\ emission (in orange, blue or white contours) is superposed on optical colour images. The images show the Vela group (a; from Fig. 1 in \citealp{english10}), the galaxy group IC1459 (b; from Fig. 6 in \citealp{serra15}), the FGC1287 triplet (c; from Fig. 1 in \citealp{scott12}), the M81/M82 group (d; from Fig. 6 in \citealp{deblok18}) and the Leo ring (e; from Fig. 1 in \citealp{michel-dansac2010}). All images are reproduced with permission. Copyright by AAS or the authors.} \label{groupstripping}
\end{figure*}

\section{Cold gas removal in groups -- M$_{halo}<$10$^{14}$ M$_{\odot}$}

The observational evidence for a direct connection between stripping of the cold ISM and quenching for satellites in groups with $M_{halo}<$10$^{14}$ M$_{\odot}$ is significantly less rich than in the case of clusters of galaxies. As discussed in \S~4.1, this is primarily due to the limitations of current \hi\ surveys, which are not sensitive enough to probe the \hi-deficient regime for large numbers of galaxies (including those in groups). The situation is worse for \htwo\ studies, which have not well characterized even the star-forming regime across environments, as the largest sample currently available comprises $\sim$500 galaxies, nearly half of which are non-detections (e.g., \citealp[][see \S~4.2]{saintonge17}).

While some of the conclusions drawn 
for clusters may be valid also for massive groups (e.g., $M_{halo}\sim$10$^{13.5}$ M$_{\odot}$), it is natural to expect some potentially major changes due to the different conditions experienced by satellites in groups. On one side, the temperature and density of the intra-group medium ($\rho$), as well as the orbital velocity ($v$) of the satellites are lower than in clusters, hence the efficiency of ram pressure (proportional to $\rho v^{2}$, see also \S~5.2) may significantly decrease. On the other side, thanks to the lower relative velocities of group members, gravitational interactions become more efficient than in clusters at fixed satellite mass. 
Despite these differences, as we discuss in this section and summarise in \S~7, the overall picture emerging in \S~4.5 seems to be applicable also to galaxies in groups.

\subsection{\hi\ stripping in groups}
Intriguingly, the first evidence for stripping of cold atomic hydrogen in groups emerged before that for clusters of galaxies. Since the first interferometric observations of the Magellanic Clouds \citep{kerr54,hindman63}, and of pairs and group galaxies in our nearest neighborhoods \citep{roberts68,rots75}, it was clear that environment can directly remove \hi\ and create striking trails and streams. As showcased in Fig.~\ref{groupstripping}, we currently have a plethora of examples of \hi\ stripping in groups, with most of the evidence pointing towards gravitational interactions as primarily responsible for creating these features (e.g., \citealp{yun94,hibbard01,williams02,koribalski03,kantharia05,english10,michel-dansac2010,serra13,serra15,leisman16,hess17,deblok18,leewaddell19}). This is due to the fact that, in most cases, stellar tidal features and/or peculiar optical morphologies are also clearly visible in these systems, suggesting a prominent role of tidal forces in affecting both their gas and stellar distributions.

However, one fundamental difference between the evidence for direct stripping in group vs. cluster environment is that the vast majority of the galaxies in groups for which \hi\ streams and tails have been detected are still actively star-forming and not \hi-deficient (regardless of the exact definition of  deficiency). This is primarily a selection effect, because interferometric observations of groups very rarely probe the \hi-poor regime, and thus preferentially pick gas-rich (or gas-normal) systems, whose outer gas disks are easily perturbed by the gravitational pull from other group members. Moreover, given the mild level of stripping, in some cases it is unclear how much of the displaced gas is no longer bound to the galaxy and will not ``rain down'' onto the disk (or accrete onto another satellite), as we know of at least a few examples where this seems to be the case (e.g., \citealp{hibbard95,struck03,koribalski04}). Thus, these might not be examples of a direct link between stripping and quenching in groups. Access to the critical gas-poor regime is still primarily limited to single-dish telescopes, at the cost of losing spatial information. Even so, in the past few decades, a consistent picture is starting to emerge from \hi\ studies.

Since the late 20th century there have been claims of the presence of \hi-deficient galaxies in loose or compact groups (e.g., \citealp{gerin94,hutchmeier97,verdes01}), but the installation of the multibeam receiver on the 64m Parkes telescope and the HIPASS \hi-blind survey 
made large statistical studies more feasible, allowing us to quantify the \hi\ content of galaxies in different types of groups (e.g., \cite{hickson82} compact groups, loose groups, X-ray selected groups, etc.; \citealp{stevens04,omar05,sengupta06,kilborn09}). The vast majority of these works found evidence for \hi-deficient galaxies in groups (but see e.g., \citealp{stevens04} for an opposite view), suggesting that, even in halos smaller than those of clusters, environmental processes are able to remove a significant fraction of the cold ISM from star-forming disks. 
Although different definitions of \hi\ deficiency (see \S~3) make quantitative comparisons difficult, all major statistical studies \citep{verdes01,sengupta06,kilborn09} agree that \hi-deficient galaxies are  preferentially found in X-ray bright groups and that, while \hi-poor galaxies are more common in the inner parts of groups, there is no strong correlation between gas content and distance from the centre of the group. This is not entirely surprising, given that such a correlation is weak even in clusters, as well as strongly affected by projections effects.

Interestingly, while both \cite{verdes01} and \cite{kilborn09} support gravitational interactions as the primary driver for \hi\ deficiency in groups, \cite{sengupta07} show that the \hi\ morphology of gas-deficient galaxies is similar to that observed in clusters and propose a scenario in which hydrodynamical mechanisms such as ram pressure may play an important role in removing cold gas from the disk. Similarly, \cite{rasmussen06,rasmussen08,rasmussen12} claim that ram pressure is needed to explain the \hi\ properties of group galaxies, although it may not always be the dominant stripping mechanism.
\begin{figure*}
\includegraphics[width=17.7cm]{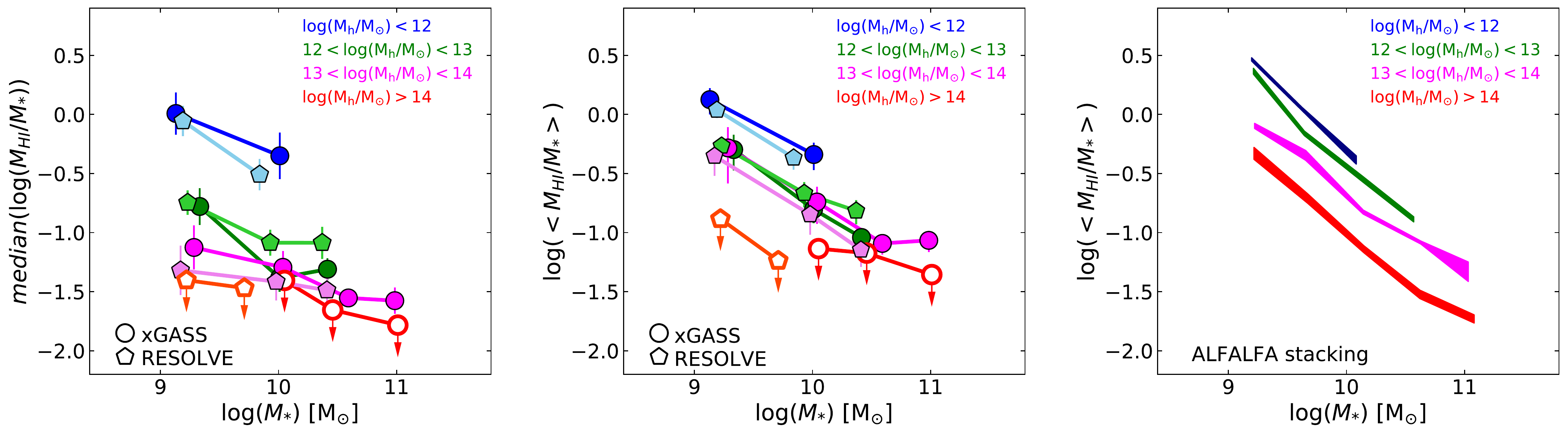}
\caption{The \hi-to-stellar mass ratio vs. stellar mass relation for satellite galaxies as a function of mass of their host group halo. Results from targeted single-dish \hi\ surveys are shown in the left and middle panel, with circles and pentagons indicating xGASS \citep{catinella18} and RESOLVE \citep{stark16}, respectively. Colors correspond to different halo masses (as noted on the top-right corner of each panel), with arrows indicating bins dominated by upper limits. Only bins with at least 10 galaxies are shown. While the left panel shows medians values, the middle panel shows linear averages in each bin. The right panel shows the same plot obtained by \cite{brown17} via stacking of ALFALFA galaxies. By construction, this technique only provides linear averages, and cannot be blindly compared with median scaling relations.} \label{groupHI}
\end{figure*}

At the same time, a similar picture emerged from studies of the Local Group, which showed a dramatic transition between gas-rich and gas-poor satellites across the virial radius of the Milky Way \citep{grebel03,grcevich09,spekkens14,westmeier15,putman21}, with very little or no \hi\ found in satellites inside the virial radius. This was shown to be consistent with the expectation for ram-pressure stripping by the Milky Way halo, although it is possible that gravitational interactions, supernova winds and heating by the cosmic ultraviolet background could contribute by increasing the stripping efficiency (e.g., \citealp{mayer07,kazantzidis17}).
 
 More recently, the advent of deep targeted \hi\ surveys such as xGASS and the REsolved Spectroscopy Of a Local VolumE (RESOLVE) atomic gas survey \citep{stark16}, along with the second-generation \hi-blind ALFALFA survey, have made it possible to extend studies of atomic hydrogen in groups to significantly larger samples and take advantage of the overlap with wide-area optical surveys such as SDSS. 
 Perhaps the first evidence supporting the idea that \hi\ deficiency may be widespread in groups across a wide range of halo masses came from GASS \citep{catinella13}, a representative sample of $\sim$800 galaxies with stellar masses greater than 10$^{10}$ M$_{\odot}$, limited an \hi-to-stellar mass fraction of $\sim$1.5\%, which was later extended to smaller stellar masses (xGASS). \cite{catinella13} showed that satellites in groups have significantly lower gas content than central galaxies of the same stellar mass and/or stellar mass surface density. Similar results were found by \cite{stark16} using RESOLVE, a targeted \hi\ survey designed to be baryonic (i.e., stellar plus atomic hydrogen) mass limited, as opposed to stellar mass selected as GASS/xGASS. Intriguingly, as shown in Fig.~\ref{groupHI} (left), the agreement between RESOLVE and xGASS is also quantitative, as both surveys provide a consistent view of the decrease of gas fraction at fixed halo mass from low- to high-mass groups. Unfortunately, most of the satellite population in large groups is still not detected even in these deep surveys, making it difficult to gain insights into the physical drivers of the observed trends.

Similarly, works based on galaxies detected by ALFALFA \citep{hess13,yoon15,odekon16} found that the fraction of \hi-detected systems gradually decreases towards the centre of groups, with the effect being stronger in more massive halos. However, as also pointed out by \citet{yoon15}, environmental studies based on \hi\ detections alone have limited informative power apart from simply confirming the presence of a morphology-density relation, thus providing little evidence for the presence of stripping and any links with star formation quenching.  

The real breakthrough in this field came from spectral stacking, a technique used to constrain the statistical properties of a population of galaxies that lack individual detections in a survey.
When applied to an \hi-blind survey, spectral stacking is a powerful tool to reach sensitivities well below the nominal one, provided that independent redshift measurements are available for large numbers of galaxies in the survey's sky footprint. Briefly, this requires to extract individual \hi\ spectra at the known positions of the galaxies (regardless of whether these are formally detected or not), align in redshift and co-add (or {\it stack}) -- the spectrum thus obtained gives an estimate of the total and average \hi\ content of the sample that was co-added. By binning galaxies according to a given property (e.g., stellar mass, colour, morphology, etc.) and stacking the corresponding \hi\ spectra in each bin, it is possible to study gas scaling relations (e.g., \citealp{fabello11,fabello12,gereb14, brown15,brown17,healy19,guo20,hu20}).

Naturally, the gain in sensitivity afforded by spectral stacking comes at a cost, in terms of two important drawbacks. First, we only know the average \hi\ content of the population that was co-added, and nothing about its dispersion (although the scatter of a relation can be explored by binning according to multiple properties simultaneously, if the sample is large enough). Second, stacking is an intrinsically {\it linear} operation, thus only provides linear averages. This is in contrast to scaling relations based on targeted observations, which are typically presented in terms of medians or logarithmic averages (reflecting the fact that almost all galaxy properties follow a log-normal distribution).
This is illustrated in Fig.~\ref{groupHI}, where we show both medians and linear averages for xGASS and RESOLVE: the two averaging techniques do not provide the same answer, hence one cannot blindly compare results from targeted observations (left panel) with stacking (right panel), or even values for single galaxies with averages derived from stacking.
This means that the power of spectral stacking for scaling relations lies in the determination of {\it relative offsets between various samples or subsets}, and that care must be taken when comparing scaling relations (and their detailed shape) from stacking with ones based on targeted observations and/or predictions from theoretical models.

Despite its limitations, stacking has provided us with the strongest observational evidence supporting cold gas stripping in groups. \cite{fabello12} first applied this technique to a sample of $\sim$5000 galaxies with stellar mass $M_*>10^{10}$ M$_{\odot}$ and redshift $0.025<z<0.05$, extracted from the overlap between ALFALFA and SDSS surveys. They found that, for galaxies with $M_*<10^{10.5}$ M$_{\odot}$, the \hi-to-stellar mass ratio decreases with increasing local environmental density faster than the specific SFR, suggesting that galaxies first lose their cold gas, and then have their star formation reduced. They interpreted this result as evidence for outside-in gas stripping/star formation quenching, favouring a scenario in which hydrodynamical mechanisms like those observed in clusters of galaxies play a role even in groups with halo masses down to $\sim$10$^{13}$ M$_{\odot}$.

This work was further extended by \cite{brown17}, who stacked the ALFALFA \hi\ spectra of $\sim$10,600 satellite galaxies extracted from the SDSS-based group catalog of \cite{yang09} to quantify the variation of \hi\ content as a function of halo mass and galaxy projected density. They confirmed that at fixed stellar mass, or stellar mass surface density, or specific SFR, the \hi-to-stellar mass ratio of satellites monotonically decreases when moving from low to high halo masses and/or projected galaxy densities (Fig~\ref{groupHI}, right panel). 
Incidentally, the fact that similar trends are present at fixed specific SFR or stellar surface density (a proxy for morphology) implies that the results shown in Fig.\ref{groupHI} are not simply a consequence of the well-known morphology/star formation rate-density relation (i.e., the fact that, at fixed stellar mass, groups have a higher fraction of gas-poorer, early-type galaxies than the field).

Very interestingly, the large number statistics allowed \cite{brown17} to go one step further and look for variations in \hi\ content as a function of environment and more than one galaxy property at the same time.
First, they showed that \hi\ content decreases from small to larger halos even when controlling for {\it both mass and SFR}, which confirms that their findings are genuinely tracing environmentally-driven gas removal. They interpreted this trend as evidence that \hi\ is removed without any strong effects on star formation over timescales of $<$1 Gyr. They also showed that the decrease in \hi\ content depends more strongly on halo mass than local density, suggesting that the physical mechanism at work is more closely related to the size of the halo, rather than the galaxy distribution within it -- again favouring active outside-in stripping, most likely by ram pressure, initiating the quenching phase of satellite galaxies.

It is important to note that, while qualitatively the results of the ALFALFA stacking agree with linear averages from xGASS and RESOLVE, there are still systematic differences, as clearly seen by comparing the middle and right panels of Fig.~\ref{groupHI}. While the origin of such differences is still unclear, this highlights how more work in this area is much needed.

In summary, we have now plenty of evidence that cold \hi\ stripping is ubiquitous also in the group environment, although the degree of \hi\ deficiency is on average significantly lower than in clusters. However, the physical mechanism responsible for the removal of the gas remains unclear. Historically, gravitational interactions were the preferred option, as the dramatic effect of galaxy-galaxy encounters on the \hi\ distribution is very easy to spot in groups. In recent years, though, the popularity of hydrodynamical mechanisms as a cause for \hi\ stripping in groups has been increasing. But should we really expect ram pressure to be efficient in the group environment?

\subsection{Is ram pressure efficient in the group environment?}
Unlike in galaxy clusters, the observational evidence in support of ram pressure playing a significant role in groups is, arguably, circumstantial.
The higher frequency of optically-disturbed systems (highlighting the importance of gravitational interactions), combined with the lack of large samples of highly \hi-deficient galaxies, prompted the question of whether the interaction with the intra-group medium is strong enough to strip any gas from the satellites, in particular in groups without an IGM component detected in X-rays. 

As mentioned above, perhaps the best case of ram pressure-driven gas deficiency in groups is our Local Group, where the lack of \hi\ detections in any of the dwarf satellites within the virial radius of the Milky Way is fully consistent with the predictions from a simple ram-pressure stripping model (e.g., \citealp{gatto13,spekkens14,caproni17}). This indicates that, for dwarf galaxies, ram pressure can be efficient even for IGM densities of $\sim$1-2 $\times$ 10$^{-4}$ cm$^{-3}$, as expected for the corona of the Milky Way. This also seems to be consistent with the presence of \hi\ tails seen in isolated Local Group dwarf galaxies (e.g., \citealp{mcconnachie07}). Outside the Local Group, similar or even lower IGM densities have been shown to be sufficient to explain the \hi\ asymmetries and/or kinematic disturbances observed in group galaxies (e.g., \citealp{bureau02,bernard10,westmeier11,elagali19}), but admittedly it is not clear if ram pressure is powerful enough to reproduce the observed trends between \hi\ gas fractions and group halo mass. 

The lack of an X-ray emitting IGM in many groups with clear signs of gas stripping has been commonly used to argue against ram pressure stripping (e.g., \citealp{scott12}), on the basis that they are thought to lack a dense medium that could provide enough ram pressure to remove cold ISM from star-forming disks.
However, this is definitely not the case. Indeed, individual groups with halo mass smaller than $\sim$10$^{13}$ M$_{\odot}$ are still partially out of reach of current X-ray missions (e.g., \citealp{mulchaey00,finoguenov09,miniati16}), and can also host dense IGM at lower temperatures (e.g., 10$^{5-6}$ K) than larger groups and clusters.    
Interestingly, indirect proof of the presence of a significant IGM component in X-ray faint groups comes from the work of \cite{freeland11}, who looked for 
bent-double radio sources in groups and estimated the density of the IGM, under the assumption that the radio jets are bent by ram pressure (see also the X-ray stacking results of \citealp{anderson13,anderson15}). 
They probed projected group-distances between $\sim$20 kpc and 2 Mpc and targeted groups with velocity dispersions between 250 and 570 km~s$^{-1}$, finding clear evidence for a significant IGM even at large group-centric distances, with volume densities between 3$\times$10$^{-3}$ and 2$\times$10$^{-4}$ cm$^{-3}$ -- i.e., a factor of $\sim$1-10 lower than the typical density observed in the centre of clusters such as Virgo and Coma (e.g. \citealp{review}). While 
maybe not enough to fully strip a Milky Way-size galaxy in a group, these densities should be sufficient to remove the outer parts of the \hi\ disk, as well as a significant fraction of the gas reservoir of dwarf galaxies.  
To show this, we rewrite the standard \cite{GUNG72} formalism for ram pressure ($\rho v^{2} \geq 2\pi G \Sigma_{g} \Sigma_{s}$, where $\rho$ is the volume density of the IGM, $v$ is the orbital velocity of the galaxy, G is the gravitational constant and $\Sigma_{g}$, $\Sigma_{s}$ are the surface densities of gas and stars) by rescaling it to the conditions typical of the outer disk of a galaxy in a group:
\begin{equation}
\small
\frac{\rho~[\rm~cm^{-3}]}{5\times10^{-4}} \big(\frac{v[\rm km~ s^{-1}]}{250}\big)^{2}\geq 0.88 \frac{\Sigma_{g}[\rm~ M_{\odot}~pc^{-2}]}{5} \frac{\Sigma_{s}[\rm~ M_{\odot}~ pc^{-2}]}{5}
\end{equation}
%
It is clear that, for surface densities typically observed at a galaxy's optical radius (e.g., \citealp{leroy08,sanchezalmeida20}), ram pressure can be strong enough, even for IGM densities a factor of a few smaller and orbital velocities 5 times slower than 
for galaxies passing through the centre of massive clusters (see also the modelling presented by \citealp{hester06} and \citealp{koppen18}).

In summary, \hi\ stripping by hydrodynamical processes such as ram pressure can be a viable scenario to explain the \hi\ properties of group galaxies, and should not be dismissed simply based on a lack of an X-ray emitting IGM. However, even more than in clusters of galaxies, multiple physical mechanisms are likely playing a significant role in removing cold gas from the disk, and it is plausible that the primary mechanism responsible varies depending on the properties of the individual group and its satellites. 
Nevertheless, as discussed so far, it is clear that direct stripping of the cold ISM from the disk is widespread even in groups. So, the next step is to determine whether stripping is physically linked, or even driving, the quenching of star formation in group satellite galaxies.

\subsection{Does cold gas stripping cause star formation quenching in groups?}
The limited amount of \hi\ observations available for groups, combined with the lower efficiency of stripping, makes it more challenging to obtain direct observational proof of a tight connection between cold gas stripping and quenching. As will soon become clear in this section, this issue is made even worse by the lack of adequate molecular hydrogen information {\it and} resolved studies of star-formation activity for large statistical samples of group galaxies. Thus, in addition to considering \htwo\ and SFR properties of satellites in groups (as we did for galaxy clusters in \S~4), we are forced to resort to more indirect tracers, such as gas and stellar metallicities, in order to tease out the impact of environmental effects on the star-formation cycle of galaxies in groups, and establish a connection between \hi\ gas stripping and quenching.

\begin{figure*}
\centering
\includegraphics[width=8.6cm]{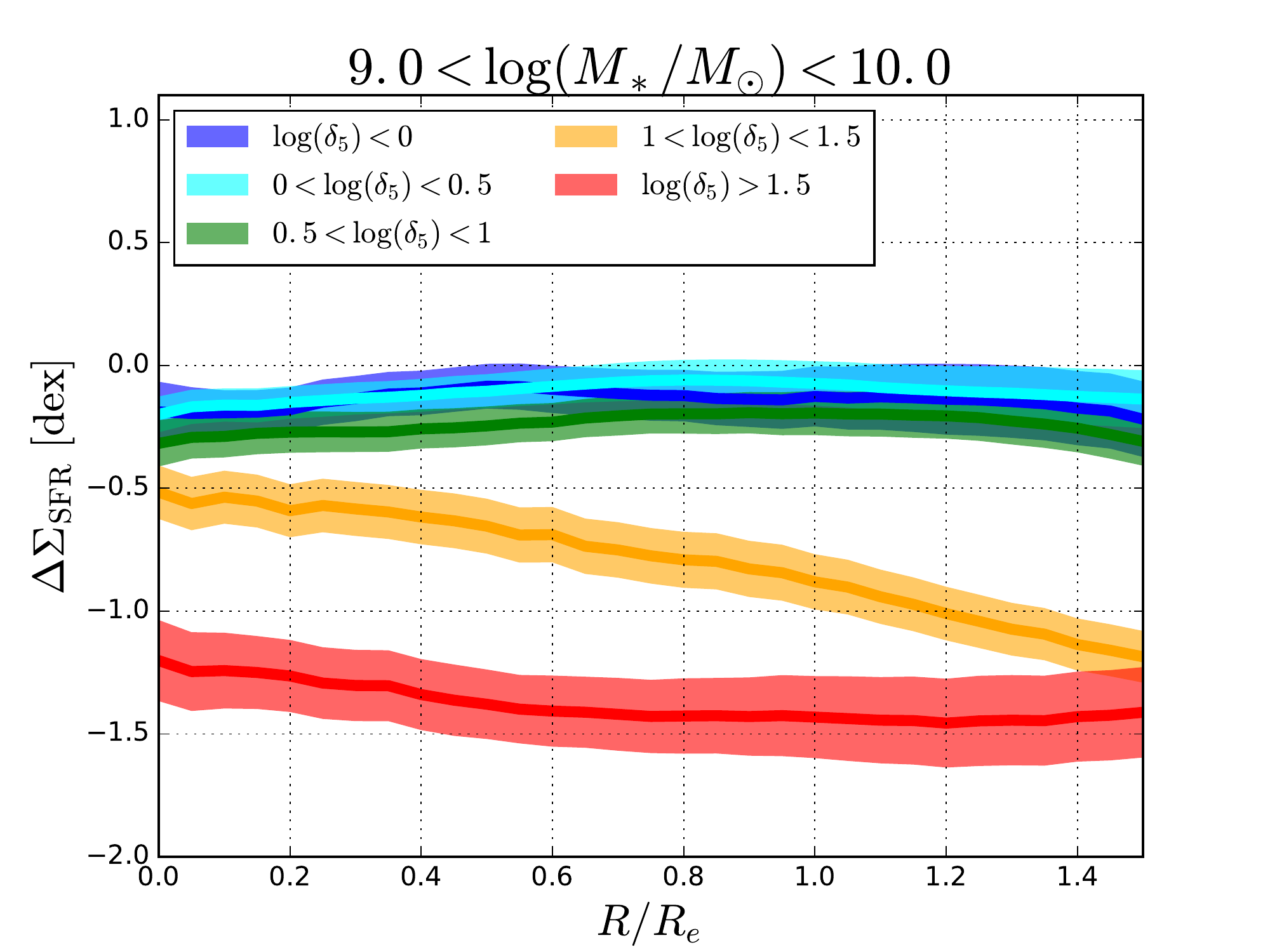}
\includegraphics[width=8.6cm]{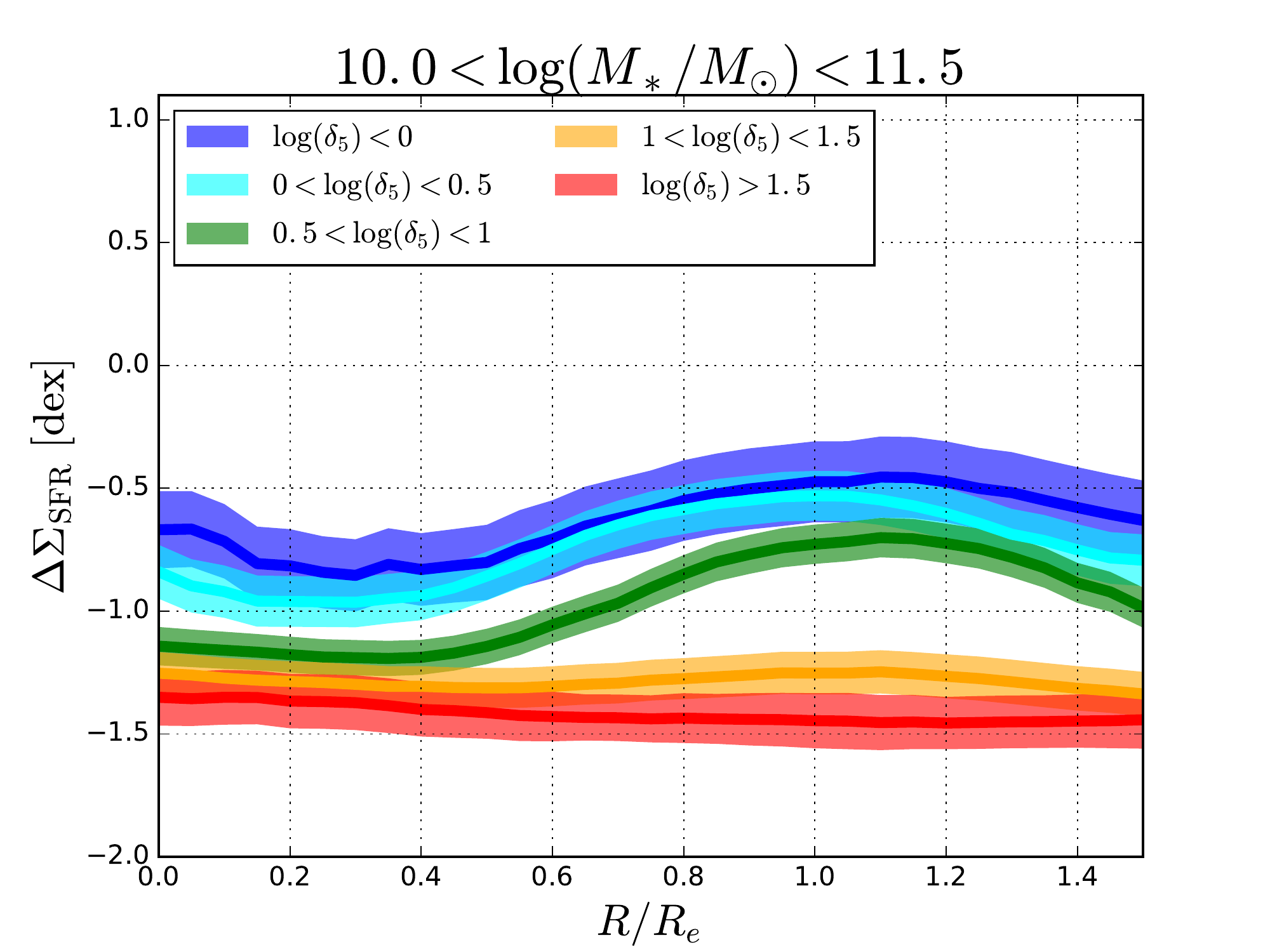}
\caption{SFR surface density radial profiles for satellite galaxies in different environments. SFR surface density is expressed in terms of deviation from the resolved star-forming main sequence ($\Delta\Sigma_{SFR}$), with $\Delta\Sigma_{SFR}=0$ indicating regions that are forming stars as expected for their stellar mass surface density.  Median $\Delta\Sigma_{SFR}$ radial profiles for low-mass (9$<\log(M_{*}/M_{\odot})<$10, left panel) and high-mass (10$<\log(M_{*}/M_{\odot})<$11.5, right panel) galaxies are split into ranges of local galaxy over-density evaluated at the 5th nearest neighbour ($\delta$5). The width of each coloured region indicates the 1$\sigma$ uncertainty on the population average. While low-mass satellites show gradual ``outside-in'' quenching, high-mass satellites are more consistent with either flat radial profiles or even ``inside-out'' quenching. Adapted from \cite{bluck20} and kindly provided by A. Bluck.}\label{sfprof_manga}
\end{figure*}

\subsubsection{Molecular hydrogen in groups}
Molecular hydrogen investigations for representative samples of satellite galaxies in groups are almost nonexistent. So far, CO observations have primarily focused on pre-selected samples of Hickson/compact groups. In these environments, molecular hydrogen appears significantly less affected (if at all) than atomic hydrogen, cases of \htwo\ deficiency are rare and star formation efficiency appears to be the same as observed in isolated systems (e.g. \citealp{lisenfeld14,martinezbadenes12,lisenfeld17}). In some cases, galaxies whose star formation has already been significantly reduced show significant \htwo\ reservoirs but low star-formation efficiencies, perhaps suggesting that \htwo\ suppression (either by stripping or starvation) may not always be needed to quench star formation \citep{alatalo15}, but higher turbulence and/or gas heating may reduce the efficiency with which gas is turned into stars.. However, it is still unclear if these findings are applicable to the average satellite galaxy in a group. 
Thus, it is still impossible to directly connect \hi\ to \htwo\ in groups in a statistical sense, and to use observations of cold gas in groups as direct evidence for star-formation quenching.

\subsubsection{SFR surface density profiles}
The situation is slightly better when it comes to the resolved star formation properties of satellites. While we lack representative samples for which both resolved \hi\ and SFR maps are available, the advent of large optical IFS surveys is gradually providing us with important insights on how group environment affects the SFR surface density profiles of satellites. Intriguingly, while pretty much all studies so far (e.g., \citealp{schaefer17,schaefer19,medling18,lian19,spindler18,coenda19,bluck20} but see also \citealp{eigenthaler15} for a narrow-band-based approach) suggest that the SFR density distribution in group satellites is significantly less affected by environment than observed in nearby clusters, the agreement between these works stops here. 
\cite{schaefer17} take advantage of the Sydney-AAO Multi-Object Integral-Field Spectrograph (SAMI) Galaxy Survey \citep{bryant15} to show that the H$\alpha$ SFR profiles of galaxies (with no distinction between centrals and satellites) become steeper with increasing environmental density at fixed stellar mass, consistently with what is expected if star formation is quenched outside-in. This trend is particularly strong for massive (M$_\star >$10$^{10}$ M$_{\odot}$) systems. In other words, they suggest that quenching is due to a milder version of the same process acting in clusters. This work is further expanded by \cite{schaefer19} to show that massive satellite galaxies in groups (M$_{h}>10^{12.5}$ M$_{\odot}$) have star formation more concentrated in their inner regions than what observed in central/isolated galaxies, whereas lower mass satellites do not show any evidence of SFR concentration. 

Conversely, \cite{spindler18} perform a similar analysis using data from the Mapping Nearby Galaxies at Apache Point Observatory \citep[MaNGA;][]{bundy15} survey and find no variation in the SFR (or specific SFR) profiles of centrals and satellite galaxies, nor a variation in the SFR density profiles with environmental density. This seems confirmed by a more recent analysis of MaNGA presented by \cite{lian19}, where a correlation between environmental density and shape of SFR profiles may be present, but not at a statistically significant level. Intriguingly, the only stellar mass bin where the correlation may be significant covers the range 9$<\log(M_{*}/M_{\odot})<$9.7, in contradiction with the SAMI-based results. Similar findings have been reported by \cite{coenda19} using the Calar Alto Legacy Integral Field Area \citep[CALIFA;][]{califa} survey, who find that the specific SFR profiles of group satellites are not different from those of field galaxies, except for galaxies in the stellar mass range 9$<\log(M_{*}/M_{\odot})<$10, where they find evidence for outside-in quenching. 

Apart from potential differences in data sets and environmental metrics used, there are some key limitations associated with the use of H$\alpha$ SFR profiles extracted from IFS data, which are most likely behind the differences between the investigations discussed above. By selection, most of these works are primarily focused on the high SFR surface density regime (i.e., $>10^{-3}$ M$_{\odot}$ yr$^{-1}$ kpc$^{-2}$), thus mainly tracing galaxies that are still on the main sequence. This is due not only to the sensitivity of the observations, but also to the fact that, below the star-forming main sequence, the dominant source of ionisation in the ISM is no longer star formation, but either more evolved stars, shocks and/or accreting super-massive black holes. This makes it very challenging to derive SFR density profiles for a representative sample of satellite galaxies below the main sequence \citep{belfiore17}. Thus, it is not too surprising that actively star-forming satellites do not show strong signs of environmental perturbation  
(note that a similar conclusion would be reached also for cluster galaxies), and that those studies pushing into the passive population do find more tantalising evidence for outside-in quenching in group satellites. 

Excitingly, \cite{bluck20} have recently been able to push the analysis of resolved star formation in satellite galaxies to lower SFRs with a two-stage approach: i.e., using H$\alpha$ emission where available, and the strength of the 4000 \AA\ break otherwise. Their findings seem to confirm 
that low-mass ($M_{*}<$10$^{10}$ M$_{\odot}$) satellites are quenched outside-in, whereas higher mass systems may be more consistent with inside-out quenching (see Fig.~\ref{sfprof_manga}). Moreover, they argue that environment may not even play a significant role in the quenching of high-mass satellites, potentially driven by `internal mechanisms'. This technique is very promising, and could become a powerful tool for establishing how star formation is quenched in group satellites. 

In summary, the star formation in the inner parts (i.e., within $\sim$1-1.5 effective radii) of the optical disk of group satellite galaxies does not seem to be significantly affected by environmental effects. This is in contrast to the outer parts of the disk, where the quenching of the star formation seems to be consistent with the evidence for stripping of the cold ISM up to the optical radius. Thus, it remains unclear whether there is a direct connection between stripping of \hi\ in the outer parts of the disk and full quenching of the star formation. In order to gain more insights into this matter, we next discuss what gas-phase and stellar metallicities can tell us about the recent and past enrichment history of the ISM.

\subsubsection{Gas-phase metallicities} 
\begin{figure}
\includegraphics[width=8cm]{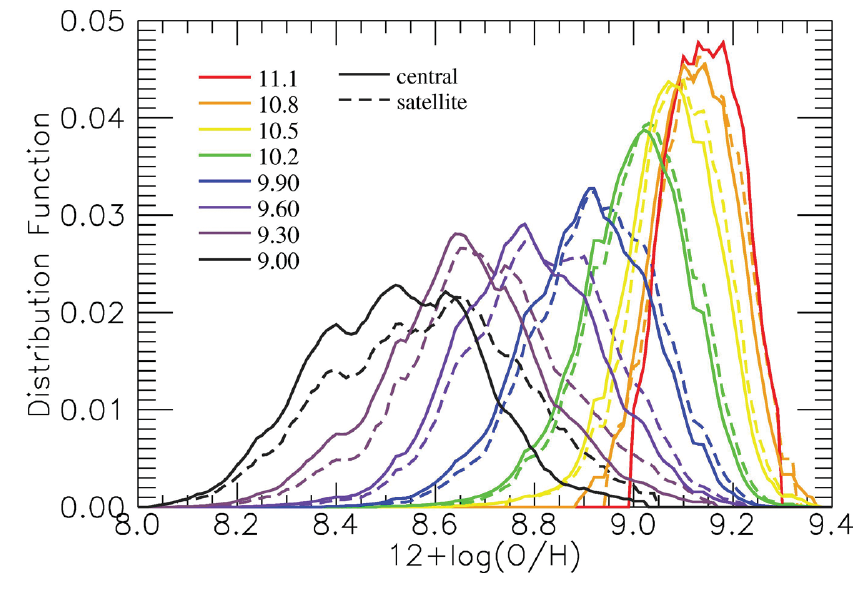}
\caption{Gas-phase oxygen abundance distribution for star-forming centrals (solid lines) and star-forming satellites (dashed lines) for different stellar mass bins (colored lines), based on SDSS. At the same stellar mass, the average metallicity of satellites is higher than that of centrals, especially at low stellar masses. Image reproduced with permission from Figure 2 in \cite{peng14}. Copyright by the authors.}\label{satmet}
\end{figure}

It is now well established that, even when still forming stars at rates typical of what is observed in the star-forming main sequence, satellite galaxies in groups show gas-phase metallicities significantly higher ($\sim$0.1-0.2 dex) than central galaxies of the same stellar mass. This increase in metallicity of the ISM has been reported by studies based on both fiber spectroscopy (e.g., \citealp{pasquali12,peng14,wu17}) and IFS data (e.g., \citealp{lian19,schaefer19b}), and appears to become more and more prominent with decreasing galaxy stellar mass (see Fig.~\ref{satmet}). While these observational findings are overall in agreement with results from clusters of galaxies and generally interpreted as evidence for the cessation of infall of pristine gas and stripping of cold ISM from the outer disk (see \S~4.2 and \S~4.4), for satellites in groups multiple scenarios have been proposed. Most importantly, it is still debated whether such an increase in metallicity is directly linked to stripping. \cite{pasquali12} put forward a scenario in which, at the same time, a) the halting of gas infall prevents the galaxy from accreting pristine gas; b) ram pressure removes the gas in the outer disk and prevents radial inflow of low-metallicity gas into the disk, and c) metal-enriched outflows are prevented due to the external pressure of the IGM. A similar conclusion was reached by \cite{wu17}, who linked the increase in metallicity to the lack of cold gas in satellites in high-density environments.

Conversely, \cite{peng14} argued that the increase in gas-phase metallicity has little to do with stripping or halting of gas inflow, but that it is primarily a consequence of the fact that satellites in groups accrete preferentially metal-enriched gas. They used the gas regulator model presented by \cite{lilly13} to show that a metallicity enhancement at fixed mass and SFR can simply be obtained by assuming that, in denser environments, the IGM is more metal enriched and, therefore, its infall into satellites would boost the metal content in their ISM.  One of the key elements of their argument is indeed the presence of a clear metallicity offset between centrals and satellites having the same SFR. A similar conclusion was recently reached by \cite{schaefer19b}, who looked at the resolved mass-gas-phase metallicity of satellite and central galaxies extracted from the MaNGA survey.  

The recent analysis presented by \cite{lian19}, and based again on MaNGA data, may provide an avenue to gradually discriminate between the different scenarios presented above. They found that, at least at low stellar masses (9$<\log(M_{*}/M_{\odot})<$9.7), the enhancement in gas-phase metallicity is associated with a gradual flattening of metallicity gradients with increasing environmental density. This is accompanied by a weak steepening of the SFR density profiles. Using a chemical evolution model they showed that the observed changes in SFR and gas-phase metallicity profiles are more consistent with outside-in quenching and halting of gas infall primarily in the outer parts of the star-forming disk, rather than accretion of pre-enriched material. However, they stressed that larger number statistics is needed to fully reject the metal-enriched gas accretion scenario. 

As argued by \cite{schaefer19b}, the \cite{lian19} sample extends to lower SFRs than the one used in other works, suggesting that they are much more sensitive to environmental effects than analyses focused on satellites that are pre-selected to lie on the star-forming main sequence. In other words, there may be multiple channels affecting the enrichment of satellite galaxies and our ability to trace them may heavily depend on sample selection criteria. 

Intriguingly, a similar debate has recently emerged also from the analysis of cosmological simulations. For example, \cite{bahe17} take advantage of the Evolution and Assembly of GaLaxies and their Environments \citep[\textsc{EAGLE};][]{eagle15} simulation to show that gas-phase metallicity enhancements are directly linked to environmental quenching, and due to the removal of metal-poor gas from the outskirts of the star-forming disk and suppression of gas inflows. Conversely, \cite{gupta18} show that  \textsc{Illustris-TNG} \citep{pillepich18} satellite galaxies in groups accrete more pre-enriched gas than galaxies in isolation, and suggest that this can easily account for the observed difference in gas-phase metallicity between satellites and centrals. However, as extensively discussed in \cite{bahe17}, even in the case of theoretical simulations, the selection criteria used to match simulated and observed data can significantly influence the results. 

In summary, while further analysis is needed to determine the origin of the gas-phase metallicity enhancement observed in satellite galaxies, all evidence suggests that stripping of \hi\ can directly affect the outer parts of the star-forming disk, whereas for most of the optical disk changes in star formation activity and ISM enrichment appear more consistent with just a reduction of the gas infall rate onto the disk. This is particularly true for galaxies with stellar masses M$_{*}\geq$10$^{10}$ M$_{\odot}$.

\subsubsection{Stellar metallicities} 
A scenario in which cold gas stripping marginally affects the inner star formation activity of group satellites appears consistent with the stellar abundance properties of satellite galaxies (e.g., \citealp{pasquali10,peng15}). In particular, \cite{peng15} show that passive satellite galaxies have higher stellar metallicities than star-forming ones at fixed stellar mass (see also \citealp{bluck20,trussler20}). They argue that this is consistent with a scenario in which star formation is quenched slowly, allowing significant further enrichment of stellar populations after infall (see also \citealp{bahe17} for a theoretical view on this). This appears also in line with the lack of any difference in the shape of stellar age and stellar metallicity gradients in galaxies as a function of environment (e.g., \citealp{goddard17,zeng17}) and the fact that already quenched satellite galaxies rarely support an outside-in quenching of their star formation within $\sim$1-1.5 effective radii (e.g., \citealp{lin19}). One potential caveat in trying to link studies of stellar metallicities with those focused on SFRs, gas-phase metallicites and cold gas is that these two approaches generally target different populations, with gas studies focusing on star-forming satellites only, while stellar ones also including galaxies that are already quenched and have likely spent most of their life as satellites (e.g., \citealp{pasquali19,smith19}). Thus, the two approaches may be sensitive to different physical processes and infall times.

\subsection{Star formation quenching timescales in the group environment}
It should be now clear to the reader that, even in groups, the picture emerging on the potential connection between stripping and quenching is not dramatically different from what we have seen for clusters. Thus, the key points on quenching timescales highlighted in \S~4.4 are applicable also to the group environment. However, here it is worth going into greater detail on issues that are directly relevant to group satellites. 

Even in groups, it is natural to expect that, for low-mass galaxies ($\lesssim$10$^{8}$ M$_{\odot}$), ram pressure is strong enough to remove all the cold gas reservoir soon after the crossing of the virial radius or at first pericentre passage. Thus, for these objects, quenching timescales (measured from the time of infall until the time of quenching) should be relatively short (i.e., $\sim$1 Gyr). 

This is indeed the consistent picture emerging from all estimates of quenching timescales for galaxies in the Local Group (e.g., \citealp{fillingham15,wetzel15,fillingham19}). Excitingly, with the advent of the Gaia mission \citep{gaia16}, it has even become possible to reconstruct the orbits of satellites around the Milky Way and determine quenching timescales with unprecedented accuracy. By doing so, \cite{fillingham19} find quenching timescales (measured from the time when the galaxy crossed the virial radius of the Milky-Way halo) $<$2 Gyr for all Milky Way satellites with stellar masses lower than $M_{*}\sim$10$^{8}$ M$_{\odot}$. Moreover, as expected, the exact timescale strongly depends on the detailed orbital parameters of each satellite, with orbits with unusually large pericentre passages corresponding to significantly longer (i.e., several Gyr) quenching timescales. Interestingly, \cite{wetzel15} report a trend of increasing quenching timescale with increasing stellar mass, consistent with the idea that the efficiency of the quenching mechanism decreases with increasing mass, as in the case of ram-pressure stripping.     

It would be natural to assume that, at fixed group mass, the trend of increasing quenching timescale with increasing mass can be extrapolated to higher stellar masses. However, the picture appears a bit more complicated. If we combine all most popular estimates of quenching timescales in satellites (see e.g., \citealp{wheeler14,fillingham15,fillingham16,wetzel15}) an unexpected trend emerges, where quenching timescales increase with stellar mass up to $\sim$6-8 Gyr for stellar masses of $\sim$10$^{9}$ M$_{\odot}$ and start decreasing for higher stellar masses. The favoured interpretation of this  trend is that M$_* \sim$10$^{9}$ M$_{\odot}$ represents the transition mass above which stripping of the cold ISM becomes inefficient and starvation becomes dominant. Thus, the decrease in quenching timescale at higher masses is supposedly just a consequence of shorter total gas depletion times in more massive galaxies (e.g., depletion time scaling with M$_{*}^{-0.3}$; \citealp{dave11}). 

While future investigations are needed to settle this issue, we recommend caution in interpreting such trends at face-value. Not only are different group masses combined but, as discussed in \S~4.4, there are many caveats in the use of quenching timescales to isolate physical processes driving quenching, and mass trends could be a simple by-product of the threshold used to separate passive and active galaxies. Moreover, longer timescales may not mean less efficient quenching, but different orbital parameters/accretion histories. Quenching likely does not start at the time of crossing the virial radius, and there is evidence suggesting that it is this time delay between infall and on-set of quenching that drives the mass dependency of quenching timescales in both groups and clusters (e.g., \citealp{wetzel13,mika19}). Indeed, as already mentioned, when quenching timescale estimates do not depend on assumptions on delay-time, mass dependencies either disappear or revert back to less massive systems being quenched faster (e.g., \citealp{smethurst17,pasquali19}). 
This conclusion appears in line with the recent findings by \cite{oman20}, who provide an extensive discussion on the biases and limitations of the various techniques used so far to infer quenching timescales and on how these could artificially introduce mass dependencies. 

From a quantitative point of view, a more detailed analysis is required to determine whether a decrease in quenching timescale of a factor of $\sim$4 from stellar masses of 10$^{9}$ to 10$^{11}$ M$_{\odot}$ (e.g., from 8 to 2 Gyr; \citealp{fillingham15}) can be explained with just a variation in depletion time. At the time of writing, any mass dependence in the total gas depletion time observed in nearby galaxies rarely surpass a factor of $\sim$2, with a large scatter (see also Fig.~\ref{phasespace}). Indeed, 
previous works have reported total gas depletion times even increasing with decreasing specific SFR \citep{boselli14b}, and molecular gas depletion times monotonically increasing with increasing mass and decreasing specific SFR \citep{saintonge17}, i.e., opposite to what is expected if cold gas depletion times drive quenching timescales. 

Regardless of the exact mass dependence of quenching timescales for stellar masses larger than $\sim$10$^{9}$ M$_{\odot}$, it is clear that in groups, as in clusters, a few billion years (i.e., multiple pericentre passages) are needed to fully quench intermediate- and high-mass satellites. Most importantly, even in groups it seems clear that both stripping of the cold ISM (for dwarf galaxies and the outer parts of disks in more massive systems) and slower quenching, at least partially associated with the halting of gas accretion onto the star-forming disk, are playing a role in driving satellites out of the star-forming main sequence.

\section{Pre-processing and the role large-scale structure}
So far, we have looked at satellites in groups and clusters under the hidden assumption that they spend all their life as satellites in the host halo in which they are found at z$\sim$0. 
This is definitely useful to set the scene, and provide an overview of the observational evidence supporting the link between cold gas removal and quenching. This is also motivated by the fact that we are still lacking the ability to trace environmental effects on cold gas beyond $z\sim$0.1. However, it is important to note that this is an oversimplification. 

In a hierarchical framework, structures grow through accretion of lower mass halos. Thus, satellites observed today in massive groups and clusters may have spent a significant part of their past evolutionary history as satellites of smaller groups, and may have been affected by nurture in environments different from those they currently inhabit, a scenario generally referred to as {\it galaxy pre-processing} (e.g., \citealp{zabludoff98,FUJI04}). Similarly, galaxy groups are not isolated entities but are embedded into the large-scale structure of the Universe and, during their growth, move along cosmic filaments. Growing support is emerging from cosmological simulations for a potential physical connection between cold gas accretion, disk spin orientation and location of galaxies within filaments (e.g., \citealp{codis12,dubois14,kraljic20}). 

Thus, it is important to review the current observational evidence for the role of pre-processing and large-scale structure on cold gas stripping, and discuss how this fits into the picture emerging from the previous sections.

\subsection{Pre-processing}
Given the focus of this review on the quenching of star-forming satellites, here we are interested in the potential role of pre-processing in starting (but not completing) the quenching phase of satellites before their infall into the current cluster (or large group; for simplicity, we refer only to clusters in the rest of this section). Indeed, while important for understanding the variation of the passive satellite fraction with environment and mass (e.g., \citealp{delucia12,wetzel13,haines15}), pre-processed satellites infalling into the cluster already fully quenched are irrelevant for what was discussed in \S~4, and their quenching history (as well as the role of pre-processing) is fully encapsulated in \S~5, where we have seen how groups can quench star formation in their satellites.

\begin{figure}
\includegraphics[width=8cm]{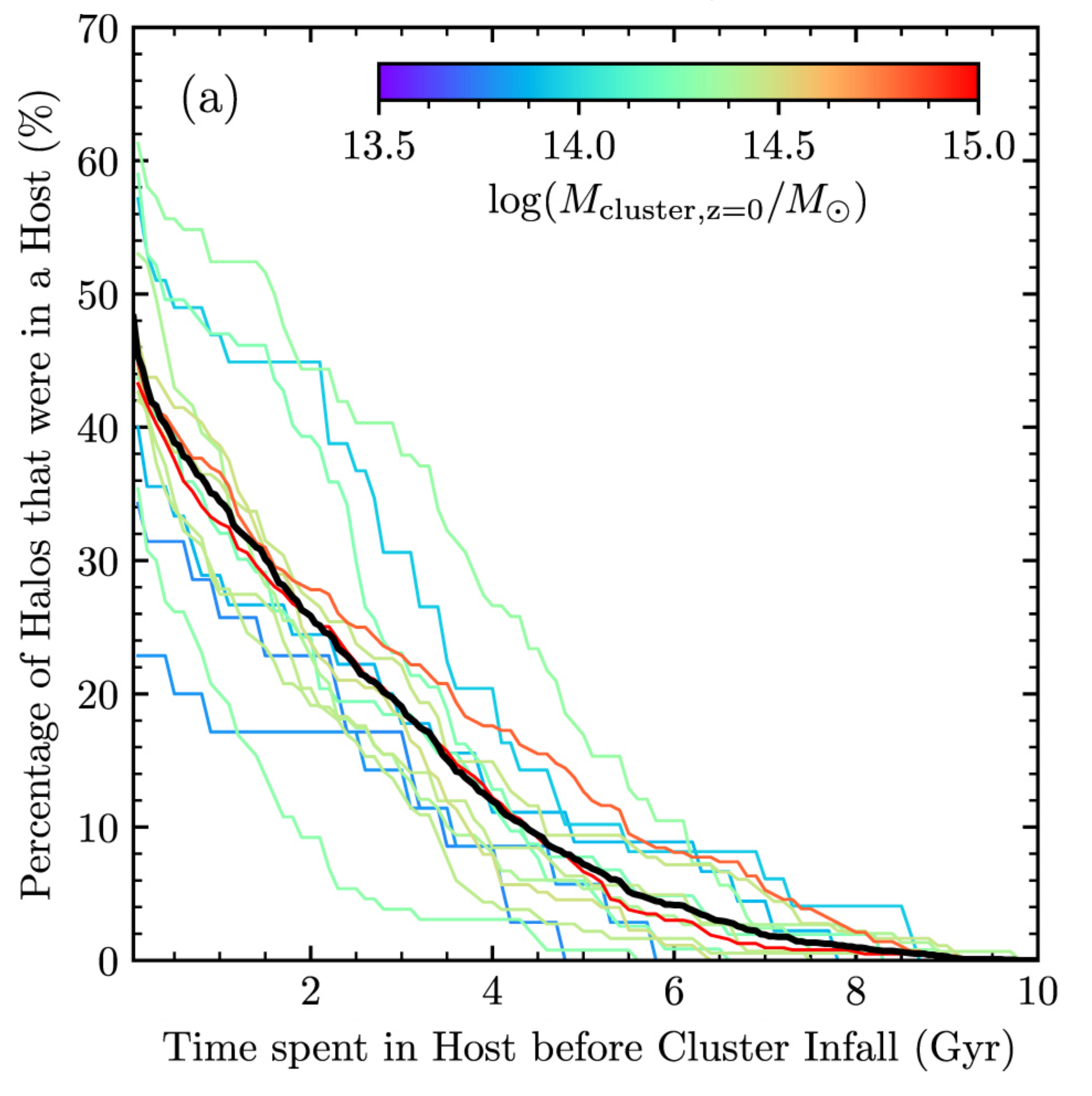}
\caption{Cumulative percentage of cluster satellites at $z=$0 that have spent more than a certain amount of time in a host before infall, as predicted by the Yonsei Zoom-in Cluster hydrodynamic simulation. The thick dark line shows the percentage of all members in the 15 clusters studied, while thinner lines (color-coded by cluster mass) show individual clusters. Image reproduced with permission from Figure 2 in \cite{han18}. Copyright by AAS.}\label{preproc}
\end{figure}

From a theoretical point of view, there is general agreement that roughly $\sim$30-40\% of the satellite population found in cluster-size halos (e.g., 10$^{14}$ M$_{\odot}$) fell into the cluster as a satellite of a halo of mass $\sim$10$^{13}$ M$_{\odot}$ (e.g., \citealp{mcgee09,delucia12,han18}). However, this fraction strongly depends on stellar mass, with more massive galaxies having an increased likelihood of infalling into clusters as centrals than lower mass systems \citep{delucia12,wetzel13}. Most importantly, the time spent as a satellite before infalling into the cluster can vary dramatically (e.g., from just $\sim$0.1 to 8 Gyr or more), but is generally $\sim$2-3 Gyr or less (e.g., see Fig.~\ref{preproc}). This is of the same order of magnitude, or even shorter, than the typical delay time before the onset of the major quenching phase in satellites ($\sim$2-6 Gyr). In this time frame, pre-processing could definitely impact on the ability of satellites to accrete new gas, and could also contribute to the removal of cold gas from the very outer disks. But it seems unlikely that, for the bulk of star-forming galaxies infalling into today's clusters (which are the focus of this review), pre-processing plays a key role in their path towards quenching. This is simply because they are still in the star-forming main sequence when they reach the cluster environment and, by definition, their star formation activity has not yet been significantly affected by the environment experienced before infall.

Indeed, from an observational point of view, apart from peculiar cases of groups infalling into clusters (e.g., \citealp{cortese06,dzudar19}), the few observational pieces of evidence suggesting an effect of pre-processing on the gas reservoir of infalling star-forming satellites indicate just very small variations in \hi\ content ($\sim$0.1-0.2 dex, e.g., \citealp{yoon15,odekon16}). These are generally within the typical variation in the \hi\ content of galaxies on the star-forming main sequence \citep{janowiecki20} and,  indeed, the bulk of the population of infalling satellites in clusters has SFR and \hi\ content within $\sim$0.2 dex of those observed in isolated systems (e.g., \citealp{cortese09,boselli14b}). 

Admittedly, the limitation of current \hi\ and \htwo\ surveys makes it impossible to further investigate the 
role of pre-processing in any detail and quantify its effect on cold gas content and SFR {\it simultaneously} as a function of group and satellite mass. However, it is clear that, when star-forming satellites infall into cluster-size halos ($\sim$10$^{14}$ M$_{\odot}$ or higher), the key paths driving the quenching of the star formation are those summarised in \S~4.5 (i.e., outside-in stripping and quenching), with group pre-processing potentially only making them more efficient.  

\subsection{Large-scale structure filaments}
Even more circumstantial is the observational evidence for a connection between cold gas content, star formation quenching and position of galaxies within filaments. Only a handful of studies have investigated the correlation between \hi\ mass and distance from the spine of large-scale structure filaments, and samples are still relatively small and generally biased towards gas-rich systems. While there seems to be common agreement that gas richness may vary with distance from the filaments \citep{kleiner17,odekon18,bluebird20}, additional work is needed to properly quantify these differences and their dependence on galaxy properties. For example, while \cite{odekon18} find an increase of \hi\ content with increasing distance from the spine of the large-scale structure, \cite{kleiner17} report the opposite trend for galaxies with stellar masses greater than 10$^{11}$ M$_{\odot}$. Perhaps more importantly for the focus of this review, both theoretical and observational studies seem to imply that the overall effect of filaments on gas content and star formation (at fixed group environment) should be relatively small ($<0.3$ dex) and well within the scatter in properties observed for the star-forming galaxy population (indeed, the large-scale environment could well be one of the contributing causes of this scatter).

However, much like in the case of pre-processing, understanding the role played by filaments is most likely critical in regulating the inflow rate of gas into star-forming disks. We still know very little about how cold gas gets into the disk and whether, for fixed satellite and group properties, large-scale structure plays an important role in regulating when and how accretion occurs and stops. Looking ahead, it will be first necessary to carefully control for the effect of the group environment, in order to isolate filament-driven effects.

\section{A unified view on the role of cold gas stripping in satellite quenching}
The observational evidence presented in this review suggests that stripping of the cold ISM is practically {\it always} associated with quenching. So far, we have never observed a clear example of quenching in a satellite galaxy where there was no removal of at least the very outer parts of the \hi\ disk. Thus, we can at least exclude the ``stability'' scenario described in \S~2 as a key pathway for satellite galaxies.
At the same time, active stripping rarely appears sufficient to fully quench galaxies, 
unless {\it all} the cold gas is stripped during the first orbit through the centre of the parent halo. For a typical Milky Way-like galaxy, the inner parts of the cold disk remain bound to the galaxy and can continue to feed star formation for at least a few billion years longer.

{\bf Cold gas stripping: key parameters.}
The mass of the satellite and of the parent halo are most likely the two primary parameters determining whether or not full gas stripping is efficient, as testified by the fact that all Local Group dwarf satellites within the virial radius of the Milky Way are fully quenched, whereas massive  disks (i.e., M$_{*}>$10$^{10}$ M$_{\odot}$) in Virgo are able to retain a significant amount of gas in their disk after first pericentre passage. However, the value of the stellar mass threshold below which gas is fully stripped, and its dependence on halo mass, is still to be determined. In addition, it is natural to expect that the orbital properties of satellites as well as the accretion history of the groups may play a secondary, yet significant, role. 

{\bf Outside-in gas removal.}
Stripping of cold gas is preferentially outside-in, with the outer parts of the star-forming disk quenched first. This does not necessarily imply that the inner parts of the disk are immune from environmental effects, as diffuse/extraplanar gas may still be affected within the truncation radius. But, more simply, it means that the overall stripping efficiency varies significantly with galactocentric distance. The details also depend on the physical mechanism(s) responsible for the stripping and, while hydrodynamical processes are clearly the dominant players when the ISM is affected within the optical disk, multiple physical processes (e.g., including gravitational interactions) are most likely at work when gas is removed only from the outer edges of the \hi\ disk, and their relative role and efficiency still need to be quantified. This is something that may even be nearly impossible as they often occur simultaneously.

{\bf Starvation and stripping.}
One key implication of the widespread presence of cold ISM stripping is that the physical mechanism(s) responsible for removing the cold gas from the disk also prevent the accretion of new gas onto the disk. Thus, both quenching channels (i.e., active cold gas removal and cessation of gas infall) are happening at the same time, and may even be driven by the same physical processes. What changes is primarily the importance of the two channels for full quenching, depending on how far inside the optical disk is stripping efficient.

In this regard it is clear that, at least for star-forming satellites with $M_{*}\sim$10$^{10}$ M$_{\odot}$ in a typical galaxy group (e.g., M$_{h}\sim$10$^{13}$ M$_{\odot}$), their cold gas content is almost never fully stripped after one pericentre passage. Thus, star formation will be fully quenched only gradually (i.e., over timescales of at least a few billion years after crossing the virial radius of their host halo), most likely due to the consumption of gas via star formation and lack of additional infall. Pre-processing in smaller groups might make stripping and quenching more efficient but, at least at $z\sim$0, its role appears secondary. 

It is important to stress that the fate of the inner parts of the star-forming disk (where stripping cannot be dominant) is probably one of the most important unknowns in the picture emerging from this work. While the general assumption in the literature is, indeed, that star formation will be quenched gradually and the galaxy will starve, this still needs to be proven as it is possible that other processes (e.g., outflows) may additionally contribute to regulating the last stages of the quenching phase. 

In summary, as already hinted by several studies (e.g., \citealp{cortese09,fillingham16,roberts19}), satellite galaxies in groups at $z\sim$0 seem to follow a well defined path towards full quenching, where both direct stripping and halting of infall play a key and {\it simultaneous} role and might even be driven by the {\it same} physical process(es). The main variable is how deep into the star-forming disk does stripping affect the cold ISM, which determines its relative importance with respect to gas exhaustion via star formation in driving full quenching, and in the end sets the timescale needed for satellites to transition from the active to the passive population.

\section{The theoretical perspective}
Before concluding, it is natural to wonder if the observationally motivated scenario presented here is in line with the predictions from theoretical models of galaxy formation and evolution. While it is beyond the scope of this work to review all the theoretical work on this topic carried out in the last decades, we believe that is important to provide the reader with an overview of how current simulations fit in the observational picture that we have put forward. Hence, we specifically focus only on three key points: a) whether cold gas stripping is ubiquitous in satellites galaxies; b) if theoretical models can help identify the driving physical mechanism stripping the gas, and c) if satellites are fully stripped, and thus quenched, after the first pericentre passage. 

\subsection{Do we need cold gas stripping?}
In the last two decades, one of the most popular tools
for studying the gas content of galaxy populations in a cosmological context has been Semi-Analytical-Models (SAMs). Most modern SAMs take the merger trees of dark-matter only cosmological N-body simulations {\it to follow the growth of structure through gravitational instability}, and then include a semi-analytical treatment of the various physical processes affecting the evolution of baryons (e.g., gas cooling, star formation, feedback, etc) to derive the properties of galaxies hosted within their dark matter halos. Because of the limited computational cost, they are extremely efficient at exploring large parameter space and investigating wide simulated volumes (e.g., \citealp{baugh2006,somerville15}). 

The vast majority of early SAMs implemented gas loss in satellite galaxies by removing only the hot gas content of galaxies instantly after they become satellites \citep{Baugh96, Springel01, Lagos08}. However, it soon became clear that the combination of instantaneous stripping with very efficient supernova feedback produced galaxies that tended to use up (or eject)  
their remaining cold disk gas too rapidly, resulting in fractions of red galaxies in groups and clusters that were too high compared to observations \citep{weinmann06, baldry06, Kimm09}. This also explains why early efforts to include ram-pressure stripping of the cold ISM in SAMs (e.g., \citealp{okamoto03,lanzoni05}) reported no significant improvements in the ability of models to match the star formation properties of satellites (i.e., galaxies were quenched before ram-pressure stripping of cold gas could play a role). 

As a result, \cite{mccarthy08} provided new recipes derived from hydrodynamical simulations describing gradual hot gas stripping that, combined with an improved treatment of how the gas is re-heated by feedback, improved the match with observations (e.g., \citealp{font08,lagos14,delucia19}). However, these changes were still insufficient to match the observed passive fractions (e.g., see \citealp{guo11,hirschmann14,henriques15}). Intriguingly, more recent work focused on the relative effect of hot gas stripping on the atomic-to-molecular hydrogen ratio has also highlighted that hot stripping alone would produce satellite galaxies with lower H$_2$-to-\hi\ ratios than star-forming centrals of the same stellar mass (e.g., \citealp{xie18,xie20}), contrary to what is observed in clusters. 

These issues, combined with the increasing number of observational constraints on \hi\ and \htwo\ properties of satellite galaxies, have prompted several groups to include recipes for cold gas stripping in their SAMs (e.g., \citealp{tecce10, stevens17, cora18, xie20}).  To do so, the general approach is to calculate the cold gas stripping radius analytically by balancing the galaxy's restoring forces against the drag expected from ram pressure. Commonly, the hot halo gas shields the cold disk until it has been stripped, although the exact implementation varies somewhat between studies, as do the timescales for the gas to be removed, generally related to the time resolution of the simulation. 

Although these studies concur that hot gas stripping is still the primary factor dictating the strength of quenching in most galaxies\footnote{However, we note the caveat that, due to hot halo shielding of the cold gas, it is difficult to directly see the effect of each mechanism independently. For instance, if hot halo stripping is switched off, it will disable cold stripping as well.}, some works are starting to suggest that the inclusion of recipes for cold stripping is necessary for improving the match to observed quenching fractions (at least in clusters, e.g., \citealp{xie20}), supporting the idea that cold gas stripping is important for the evolution of satellite galaxies. More generally accepted is the fact that gas stripping appears indispensable to better reproduce the \hi\ content of satellites, the variation of \hi\ deficiency with cluster-centric distance, and the increase of H$_2$/\hi\ fractions in clusters (e.g., \citealp{cora18,stevens18,xie20}). Nevertheless, the match between observations and SAMs when it comes to multi-phase gas properties of satellite galaxies is not yet perfect (e.g., 
it is still challenging to match the \hi\ and \htwo\ content of satellites simultaneously, across all ranges of halo masses; \citealp{xie20}), implying that more work is needed to understand the interplay between stripping and quenching in these models.

In parallel to SAMs, in the last decade, large-scale hydrodynamical cosmological simulations have become a commonly used tool to study the evolving gas content of galaxy populations, with a clear explosion in the number of simulations readily available for a comparison with observations (e.g. \textsc{EAGLE},  \citealp{eagle15}; \textsc{Illustris}, \citealp{vogelsberger14}; \textsc{Illustris-TNG}, \citealp{pillepich18}; \textsc{Horizon-AGN}, \citealp{dubois14}; \textsc{Magneticum},  \citealp{hirschmann14b}). As opposed to SAMs, cosmological hydrodynamical simulations follow gravitational and hydrodynamical processes in a self-consistent manner, so that cold gas stripping is inherently built-in to these simulations. 
Nevertheless, their typical resolution is still insufficient to trace the formation of the cold gas phase of the ISM, hence they have to rely on subgrid prescriptions to separate the atomic and molecular cold gas phases.

Contrary to SAMs, in cosmological hydrodynamical simulations we cannot simply switch on and off cold stripping in order to answer the question ``is cold stripping required to better match observations?'' What we can determine is whether cold gas stripping commonly happens in the path of a satellite towards quenching. Not surprisingly, the answer to this question appears to be a resounding `yes' for both cosmological (e.g., \citealp{bahe13,bahe15,cen2014,marasco16,jung18,arthur19,stevens19,stevens20,yun19,mika19}) and controlled hydrodynamical simulations (e.g., \citealp{QUIM00,ABAM99,roediger05,roediger07,bekki09,tonnesen09, bekki14}). However, while hydrodynamical simulations include stripping by default, admittedly they have not always been able to fully reproduce the properties of satellites, yet. 

Tremendous progress has been made in matching the global gas scaling relations of galaxies (i.e., the properties of centrals; e.g., see \citealp{Dave2020}), but in many instances satellite galaxies appear over-stripped and/or over-quenched (e.g., \citealp{brown17,dave17,stevens19,xie20,donnari20}). 
The mismatch between simulations and observations varies as a function of both host and satellite mass, and the exact dependency changes if we consider quenching fractions instead of gas fractions, as well as varying from model to model.
The reasons for such a mismatch are still unclear, and can be multiple. Most certainly, too efficient feedback that would eject a large fraction of the ISM out of the disk (e.g., \citealp{stevens19}) and/or insufficient resolution (i.e., 1kpc or lower), which makes it impossible to conduct a proper partition between the different cold gas phases in order to trace the denser (and more difficult to strip) gas within the multi-phase ISM of their disks (e.g., \citealp{hu16}), are among some of the most likely issues.

Thus, as in the case of SAMs, we are finding general support for the idea that cold gas stripping is part of a satellite's pathway towards quenching, though more work is clearly needed to improve the ability of cosmological hydrodynamical simulations to match the observed properties of satellites.

\subsection{Do simulations provide insight on mechanisms behind cold gas stripping?}
Keeping in mind the limitations described above, it is nevertheless tempting to see if current simulations can provide insights on the nature of the mechanism(s) responsible for cold gas stripping.

In the case of SAMs, even if hydrodynamical mechanisms cannot be directly implemented, processes like stripping due to ram pressure appear to be preferred to tidal stripping by a host's potential well. Indeed, SAMs typically calculate a tidal stripping radius and a ram-pressure stripping radius for their satellite halos as they orbit through their hosts. These radii mark the boundary beyond which each stripping mechanism is effective, 
and most works (including those based on hydrodynamical simulations) agree that the ram-pressure radius is typically smaller than the tidal stripping radius \citep{mccarthy08, font08, bahe15, cora18}. This implies that, by the time a galaxy reaches the region within its host halo at which tidal stripping becomes efficient, ram pressure has typically already removed most of the gas.

Naively, we might expect that hydrodynamical cosmological simulations should provide the best opportunity to clearly isolate the physical processes responsible for cold gas stripping, given that detailed information about all of a galaxy's mass components, and their time-evolution is, in principle, available. However, the ability to self-consistently trace multiple and interconnected physical processes makes it even more challenging to pin-point what observers would call the `dominant mechanism' at play. For example, the implementation of internal mechanisms such as star formation consumption and various forms of feedback may contribute to the stripping efficiency of different environmental mechanisms, meanwhile external factors such as the background UV field, tidal interactions or hydrodynamical processes may all influence each other's abilities to remove cold gas in complex and non-linear ways.

\begin{figure}
\includegraphics[width=8.5cm]{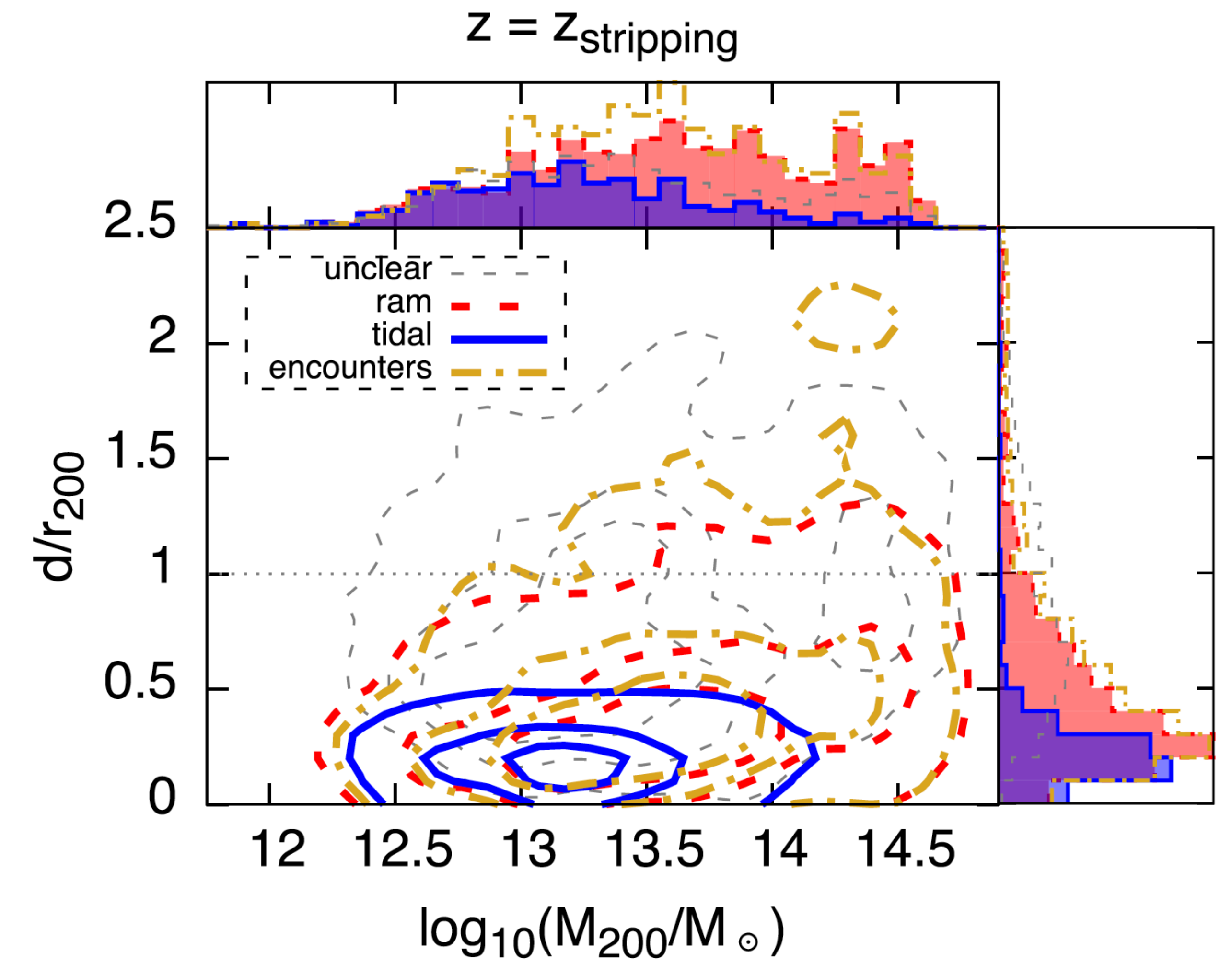}
\caption{The halo mass ($M_{200}$) vs. group-centric distance ($d/r_{200})$ distribution of \hi-poor satellites in the \textsc{EAGLE} simulation, evaluated at the time of \hi\ loss (i.e., $\rm z=z_{stripping})$. Lines show the parameter space occupied by satellites affected by different environmental processes: i.e., tidal stripping by the host halo (blue solid), ram-pressure stripping by the IGM (red thick dashed), satellite-satellite interactions (yellow dot-dashed) or none of the above (thin dashed), with contours enclosing 30\%, 60\% and 90\% of the satellites. Only systems with stellar mass (at $z=0$) larger than $10^9$ \Msun\ are considered. Image reproduced with permission from Figure 13 in \cite{marasco16}. Copyright by the authors.} \label{marasco}
\end{figure}

The complexity of this problem was clearly showcased by \cite{marasco16}, who tried to identify the responsible for cold gas stripping in groups and cluster galaxies (see also \citealp{xie20} for a similar approach from the SAMs point of view). While they emphasise that, at $z\sim$0, ram pressure is the most common physical process stripping \hi\ from satellites, they also highlighted the importance of satellite-satellite encounters in stripping \hi\ from the disk (especially at large cluster-centric distances and higher redshift), and showed that there is rarely a single cold gas stripping mechanism acting individually in a cosmological context. Instead, the most frequent source of \hi\ gas mass loss is a combination of simultaneous ram pressure and satellite-satellite encounters (see Fig.~\ref{marasco}), with the balance between the two processes changing as a function of redshift (i.e., satellite-satellite interactions more important at earlier epochs; see also \citealp{hwang18}). Interestingly, this suggests that ram pressure is efficient also in groups of halo masses of $\sim$10$^{13}$ M$_{\odot}$, consistently with what found in other cosmological (e.g., \citealp{bahe13,bahe15}) and non-cosmological simulations (e.g., \citealp{bekki09,bekki14}).  
Overall, this is in line with the key message of this review: i.e., environmental mechanisms that influence cold gas content are typically most efficient at the highest densities, for example near the pericentre passage. Therefore they tend to peak in strength at the same time, and rarely occur in isolation. 

A potential (and complementary) way forward to break this complexity is provided by controlled hydrodynamical simulations. Their controlled nature means that the effect of just one individual mechanism can be isolated. For example, in ram-pressure stripping-only simulations (e.g., \citealp{roediger05,roediger07,tonnesen09,steinhauser16}) it has been possible to clearly identify long, one-sided gas tails as observed in cluster satellites, providing some of the strongest theoretical support for the key role played by ram pressure in removing cold gas from star-forming satellite galaxies. By systematically adjusting input parameters, it has also been possible to quantify how the effects of ram pressure depend on disk inclination and orbits \citep{vollmer01,roediger07}, presence of a bulge \citep{Steinhauser12}, or cluster dynamical state \citep{Vijayaraghavan13,Roediger12}. This approach has been crucial for the development of accurate analytical recipes of gas stripping mechanisms, which can then be implemented in SAMs \citep{font08}, or used to identify when a mechanism plays a dominant role in hydrodynamical cosmological simulations \citep{marasco16}.

By limiting the simulation volume and duration, controlled simulations can also reach higher resolution or consider complex physics that would be otherwise computationally prohibitive. As such, they have been able to model and highlight the potential importance of viscous, turbulent stripping \citep{roediger05,roediger07}, a multiphase treatment of the ISM \citep{Tonnesen11}, the interplay between thermal evaporation and magnetic fields \citep{Tonnesen2014,vijayaraghavan17}, and the importance of mixing \citep{tonnesen21} for the overall cold gas stripping process. This clearly confirms that, even when only stripping by hydrodynamical processes is considered, a complex set of poorly understood processes interact, including turbulence and magnetic fields, meaning that such modelling is highly challenging. Of course, as with every approach, even controlled simulations are limited by how well they are able to properly include and interconnect all the key physical processes at play.

In summary, theoretical models are overall providing a view consistent with that emerging from observations, where multiple physical processes generally act simultaneously on the cold gas reservoir in the ISM. While hydrodynamical stripping mechanisms appear to be the most efficient in Virgo-like and more massive clusters (at least at $z\sim$0), in smaller structures the picture is much more complex and additional work is required to characterise the net effect of the various physical processes affecting the ISM in satellite galaxies.

\subsection{Are satellites fully stripped after first pericentre passage?}
While the general scenario emerging from both SAMs and hydrodynamical simulations is that cold gas stripping is wide-spread in satellites and is needed to reproduce the observed properties of group and cluster galaxies, this does not necessarily imply that satellites are fully stripped (and then quenched) after first pericentre passage. 
Indeed, as for observations, models cast doubts that stripping completely halts star formation after one pericentre passage, even in Virgo-like clusters. 

High-resolution controlled simulations suggest that $\sim$Milky-Way mass spirals would typically lose a substantial amount of their cold gas in a cluster, but would become completely stripped only in extreme cases, with the amount of stripping being a sensitive function of galaxy orbit (e.g., \citealp{roediger05,roediger07}). Similarly, \cite{steinhauser16} find that disks galaxies with V$_{\rm circ}$ $\sim$110-170 km s$^{-1}$ are only fully stripped for very plunging orbits (R$_{\rm peri} \sim$100 kpc), while star formation is only weakly affected for the rest of the orbits. 
Moving to cosmological simulations, \cite{bruggen08} applied the results found by \cite{roediger07} to the Millenium simulation and showed that only a minority of disk galaxies with V$_{\rm circ}$>200 km s$^{-1}$ are expected to be fully stripped in clusters ($\sim$11-32\% dependent on cluster mass), but that more than half ($\sim$54-64\%) would be heavily stripped. This is also consistent with the results by \cite{cen2014}, 
who finds that in hydrodynamical simulations most of the cold gas within the inner parts of satellite galaxies (at least above M$_{*}\sim3\times10^{9}$ M$_{\odot}$) is unaffected by environmental effects.

Nevertheless, some cosmological simulations do seem to provide a different picture. For example, \cite{jung18} find that almost half of their galaxies lose their cold gas before reaching the cluster pericentre, although many are already gas poor prior to cluster infall due to pre-processing. Using the \textsc{Magneticum} simulation, \cite{lotz19} report the removal of all the gas in more than 90\% of cluster satellites after a single pericentre passage, with only some of their most massive galaxies (i.e., stellar masses a few times 10$^{11}$ M$_{\odot}$) able to remain star forming for more than 1 Gyr after infall. It is not yet clear what is the origin of these differences in cold gas stripping efficiency. Simulation resolution may play a role, limiting the ability to resolve both the denser phases of the ISM and the hydrodynamical stripping processes themselves. Moreover, differences in the treatment of sub-grid physics such as feedback are also likely important, affecting both the density of ICM that galaxies pass through and the ease with which the cold disk gas can be removed.

Thus, while there is not a complete agreement between different theoretical models, we can conclude that the fact that full stripping is not widespread even in clusters does appear as a possible (if not likely) scenario. Thus, it is natural to ask how - in models - the remaining cold gas stops feeding star formation. Curiously, to our knowledge, there are few simulation papers specifically focused on this issue in the context of satellite galaxies in clusters, but we can use results obtained on quenching in a broader context to provide some insights on this issue.

It has been shown that enhanced star formation, either as a result of tidal shocking near pericentre, or due to compression of one side of the gas disk by ram pressure \citep{bekki14, steinhauser16, lotz19, troncosoiribarren20,lee2020} can potentially aid cold gas consumption. However, as discussed in \S~4.3.2, observationally such bursts appear to be relatively minor and thus unlikely to consume a significant fraction of the gas reservoir left in the disk. Feedback could also play a key role. SAMs already crucially rely on reheating and ejection of gas via supernova feedback with an efficiency that depends on redshift, in order to reproduce stellar mass functions at high redshift \citep{hirschmann14, cora18}. Stellar feedback may also heat cold disk gas into more easily stripped and/or less star-forming phases, or push the gas from inside the truncation radius to radii where it can be stripped, although so far this has only been modelled in the context of Local Group dwarf galaxies (\citealp{kazantzidis17}, but also see \citealp{emerick16}), whose shallow potential wells may enhance the amount of material lost in this manner.  In more massive galaxies, AGN triggered in ram-pressure stripped galaxies may also provide a source of feedback to the remaining cold disk gas \citep{ricarte20}. 
Finally, it should not be excluded that a more continuous loss of cold gas from the disk may also occur after pericentre passage via viscous and turbulent stripping at the ICM/disk boundary, as modelled for cluster satellites \citep{roediger07}.

In conclusion, while numerical simulations generally agree that cold gas stripping is a normal path for satellite galaxies towards quenching, there are still many open questions also from the theory side. Specifically, the relative importance of various physical processes in driving gas stripping, as well as when and how the cold gas reservoirs of satellites stop forming stars are still highly debated questions, which are also model dependent. 
In the case of hydrodynamical simulations, significant improvement will hopefully come from a more realistic multi-phase treatment of the ISM, including the role of magnetic fields, which should  provide interesting clues on these open questions. At the same time, in the case of SAMs, the development of new analytical prescriptions, such as a more multi-phase treatment of the ISM \citep{tonnesen09}, differing treatment of sub- and super-sonic disks \citep{roediger05}, and inclusion of continuous gas mass loss mechanisms \citep{roediger07} could potentially reduce some of the tensions currently present between observations and simulations. 

What is definitely already clear is that understanding the individual significance of the physical processes driving stripping in simulations remains at least as challenging as it is for observations. This is an area in which controlled simulations may yet provide the missing key for improving cosmological runs and our understanding of what can be learnt from them. On one side, they can help developing better prescriptions for SAMs and investigate issues of numerical convergence. On the other side, their higher spatial and mass resolutions, combined with their versatility, make controlled simulations able to include (and switch between) various complex baryonic physics that might be numerically unfeasible (or have unexpected consequences) in cosmological simulations, but could be critical for understanding environmental effects.

\begin{figure*}[t]
\includegraphics[width=17.7cm]{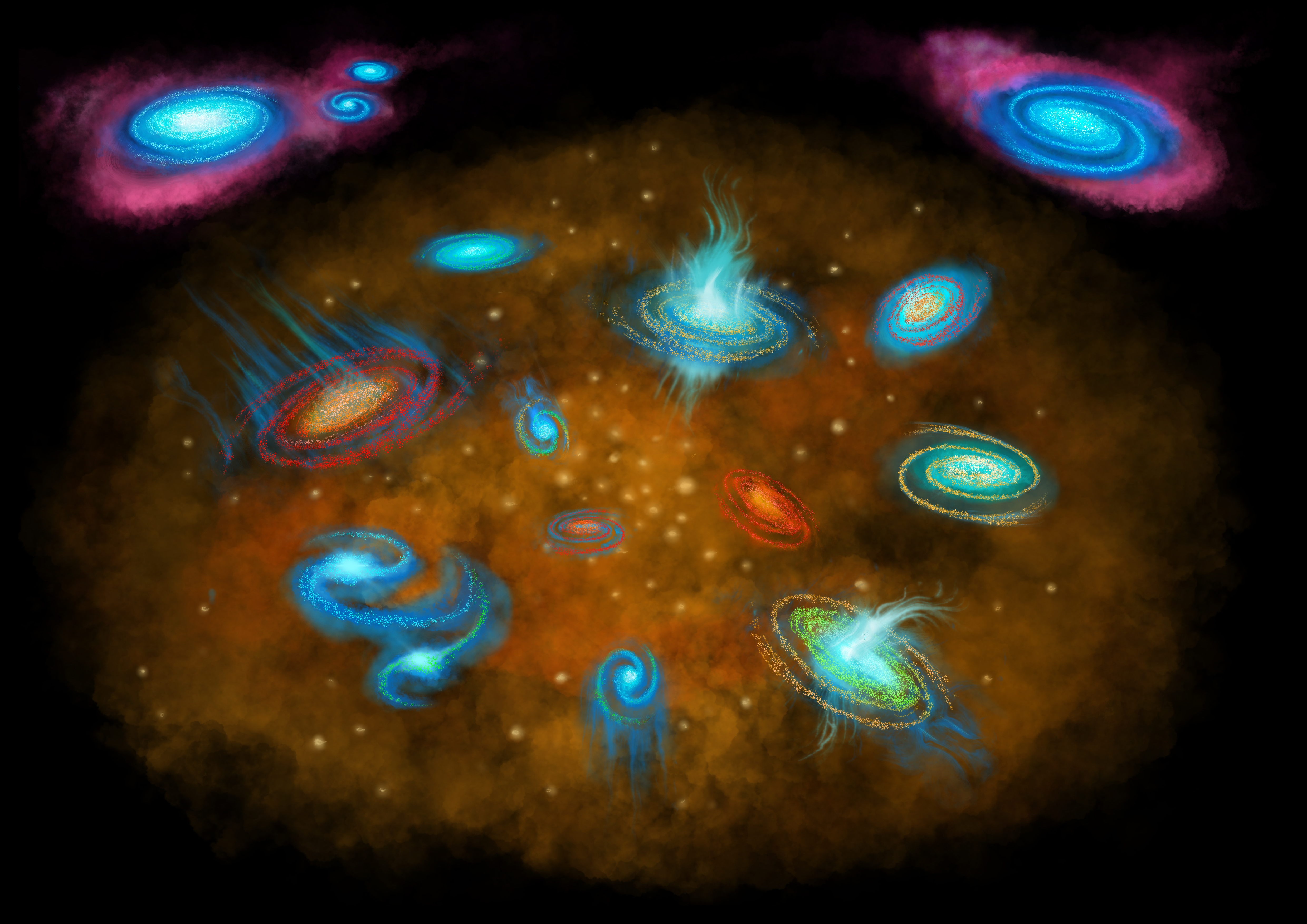}
\caption{Illustration summarising the picture emerging from this review. When galaxies become satellites, their access to CGM gas (pink) is removed, and cold gas stripping appears widespread. However, there is no single physical mechanism driving the loss of gas and, consequently, quenching. Instead, multiple mechanisms can be at play, sometimes even simultaneously. }
\label{cluster_cartoon}
\end{figure*}

\section{Closing thoughts and outlook}

In this work we have discussed {\it how}, {\it when} and {\it for how long} the gas star-formation cycle in satellite galaxies at $z\sim$0 is affected by the environment. Specifically, we have reviewed the observational evidence for the stripping of cold gas from the disk of star-forming satellite galaxies, and argued that this is a necessary step towards quenching their star-formation activity. However, both observations (e.g., \S~4 and \S~5) and numerical simulations (\S~8) suggest a complex picture, where different physical processes may act simultaneously and identifying a single culprit is neither feasible nor correct (see Fig.~\ref{cluster_cartoon}). While hydrodynamical mechanisms - such are ram pressure stripping - are likely dominant in the central regions of galaxy clusters, gravitational interactions (i.e., the interaction with the central host and/or satellite-satellite encounters) must play an important role in the lower-density environments of groups, and what regulates the balance between various stripping mechanisms remains unclear. 
While, locally, quenching is fast (hundreds of Myr or less) {\it after} the cold gas has been stripped, it is clear that stripping alone does not always lead to full quenching, as not all the gas is stripped in one orbit. This means that: a) timescales for full quenching can be long (e.g., a few Gyrs, in particular if the clock starts at time of infall) and b) other physical processes such as gas consumption by star formation or feedback can play an important role in fully quenching satellites after the first pericenter passage.

Despite substantial progress having taken place in the past decades on this topic, much remains to be explored before we can obtain a complete picture of the effects of nurture on satellite galaxies. This is in large part due to the difficulty in mapping such an extended and complex parameter space, which requires a detailed knowledge of the various components of the ISM and star formation properties for large, representative samples of galaxies spanning all the different environments. As discussed in several points throughout this review, differences in sample selections, quantification of cold gas normalcy/poorness and definitions of quenching (and related timescales) have led to apparently conflicting results and misconceptions. But even taking all this into account, existing samples lack the {\it statistics, spatial resolution and sensitivity} required to quantify the impact of environmental processes on galaxies, critically on their cold gas component, and to link this to quenching of their star formation. 

The need for statistical samples stems from the fact that environment not only spans orders of magnitude of density, but is also a second-order effect, thus it is necessary to control for stellar mass or another property scaling with galaxy size. Spatial resolution is key to understand the physics on a local basis, allowing for instance to investigate in detail what happens to the metallicity and star formation activity in a region where the gas has just been stripped, and whether an adjacent region has been affected at all. Sensitivity is critical to detect small amounts of cold gas in galaxies below the star-forming sequence, on their way to becoming passive. State-of-the-art, representative cold gas samples not only are still limited to global measurements, but are not sensitive enough to probe the gas-poor regime across all stellar masses and environmental densities, simultaneously.

Fortunately, the facilities on the horizon in the next decade will allow giant leaps forward by addressing at least some of the above limitations. This was indeed the main motivation behind this work, as it seemed timely to review the current state of the field and highlight future challenges before the next-generation radio and optical facilities begin accumulating new data. Below, we briefly focus on a few areas where progress seems particularly promising.\\

{\bf The fate of the inner parts of star-forming disks.}
Perhaps the most important unknown in the picture emerging from this work is what happens to the cold gas residing in the inner parts of galaxies (within the truncation radius), where the effects of stripping might be negligible. As previously discussed (see \S~5.4), stripping (regardless of the exact physical process responsible for it) appears not efficient enough to completely remove the cold gas disk in galaxies with M$_*> 10^{8}$~M$_{\odot}$ in groups like our Local Group, and M$_*> 10^{10}$~M$_{\odot}$ in clusters such as Virgo, at first pericentre passage (see \S~4, \S~5 and \S~7). As a result, part of the cold gas remains available after the first passage through the group/cluster, where it can sustain residual star formation for at least a few billion years. It is generally assumed that star formation will gradually cease and the galaxy will starve, as its ability to acquire new gas from the surrounding CGM and IGM has been negated by the environmental process that stripped the gas. 

However, given the length of the timescales involved and the fact that most satellites remain embedded in their group/cluster after the first pericentre passage, it seems unlikely that their inner parts will remain completely shielded from any environmental effects. In this scenario, it becomes particularly important to understand what is the interplay between atomic and molecular hydrogen, whether \htwo\ is directly stripped, or the formation of \htwo\ is simply inhibited by the partial removal of atomic hydrogen from the outer parts of the disk.  
Unfortunately, a {\it simultaneous} characterisation of both \htwo\ and SFR surface density distributions (e.g., radial profiles) for large statistical samples of satellites across environments is still missing. As highlighted in this review, large IFS surveys like MaNGA and SAMI are arguably advancing this field, but they have not been able to properly investigate the outer parts of the star-forming disks, where most of the observational signatures of environmental effects (particularly in groups) may be found. This is an area where wide-field IFS, such as MUSE \citep{bacon14}, have helped and will likely continue to do so. Moreover, next-generation large statistical IFS surveys such as HECTOR \citep{bryant18} may be key in pushing to the outer parts of disks and to start investigating second-order effects, as satellites remain a small fraction of the population targeted by any of these surveys. 

Nevertheless, there is still a crucial lack of any resolved surveys of \htwo\ content in satellites, even in clusters. While the situation is slowly improving for clusters like Fornax and Virgo, where ALMA programs such as the ALMA Fornax Cluster Survey (AlFoCS, \citealp{zabel19}) and the Virgo Environment Traced in CO (VERTICO\footnote{ \url{https://sites.google.com/view/verticosurvey/home}}) survey should provide important insights into the interplay between cluster environment and \htwo, nothing is on the horizon when it comes to molecular hydrogen properties of satellite galaxies outside clusters. This is a major issue that will most likely hamper our ability to make progress in this field. Our hope is that 
state-of-the-art millimetre facilities will be able to provide us with surveys of satellite galaxies with a number statistics and spatial resolution on a par with what IFS optical surveys have produced in the last decade.\\

{\bf A more physically-motivated characterisation of environment for cold gas studies.}
When it comes to the cold gas properties of satellites, much of the observational work so far, and hence this review, has focused on the distinction between groups and clusters of galaxies. Indeed, contrary to optical surveys that have provided stellar masses, sizes and SFR estimates for millions of galaxies across a wide redshift range, atomic and molecular gas studies of satellites are still limited to the very local Universe and have imaged galaxies only in well-known nearby groups or a handful of clusters. 

However, this is clearly an oversimplification. There is no obvious threshold between large groups and clusters, clusters often contain sub-structures with densities that are typical of groups, and most importantly environment is more complex than what a simple dichotomy might suggest: groups and clusters are embedded in a cosmic web of large-scale filaments and structures where dark and baryonic matter flow, bringing along energy and angular momentum. This requires the adoption of environmental descriptors that are continuous and more physically motivated (such as halo mass, local density, distance from the spine of a filament and so on), thus are more suitable to capture the complexity of the real Universe. Such indicators are also more directly comparable to numerical simulations and studies at other wavelengths which, thanks to the sheer amount of available data, are typically concerned with statistical studies of environmental effects. At the same time, as discussed at the beginning of this review, we need to move beyond the \hi-normal vs. \hi-deficient dichotomy and treat the \hi\ (and \htwo) content of satellites as a continuum.

While the observational optical community has made these transitions nearly two decades ago thanks to the advent of large-area surveys such as the SDSS, this {\it paradigm shift} has yet to happen for \hi\ and \htwo\ studies of satellite galaxies in the local Universe. Undoubtedly, surveys such as ALFALFA and xGASS have already started such a change, but we will most likely have to wait for the completion of large-area surveys with the SKA precursor facilities (e.g., the Australian SKA Pathfinder, ASKAP, \citealp{askap}; MeerKAT in South Africa, \citealp{meerkat})
to see \hi\ studies routinely taking advantage of the number statistics and quality of ancillary data that has become so normal for optical and near-infrared investigations. Notably for local Universe studies, the WALLABY survey alone is expected to image \hi\ emission from $\sim$5000 galaxies and detect $\sim$600,000 systems with $z<$0.26 \citep{wallaby}. This will be particularly critical to investigate the 10$^{13}$-10$^{14}$ M$_{\odot}$ halo mass regime and more accurately quantify how much (and under which conditions) \hi\ is stripped from the star-forming disk of satellite galaxies.\\

{\bf Connecting the ISM and CGM.}
Somewhat related to the large-structure environment, it is generally accepted that star formation in galaxies is fed by inflowing gas from the surrounding CGM \citep{tumlinson17}. However, how gas is accreted onto the galaxies and reaches their ISM is still an open question. We argued that, if gas accretion is indeed widespread in star-forming galaxies, it must be stopped in order for satellite galaxies to become passive. While this argument sounds plausible, it is worth clarifying that this is just an assumption, as there is no direct observational evidence showing what physical process is responsible for halting gas accretion, how and when it operates, how long does it take, what regulates the rate of inflowing gas and so on. As discussed in the Introduction to this review, this is a crucial issue because the exchange of gas between ISM and CGM via inflows, outflows and recycling is what regulates how galaxies form and evolve.

The SKA pathfinders will potentially provide the means to trace the flow of baryons into galaxies, by combining exquisite sensitivity with wide-field capability to image atomic hydrogen down to the low-column densities (e.g., $10^{17}$~cm$^{-2}$), supposedly needed to trace cold gas flows in and out of star-forming disks. However, while with ASKAP and MeerKAT this will be possible already for the Milky Way (e.g. with the Galactic ASKAP - GASKAP - survey, \citealp{gaskap}) and nearby galaxies (e.g., with the MeerKAT \hi\ Observations of Nearby Galactic Objects; Observing Southern Emitters - MHONGOOSE - survey, \citealp{mhongoose}), only the first phase of the SKA may eventually allow us to target large samples of satellite galaxies at a similar depth. 

In the meantime, we believe that significant progress in understanding the interplay between accretion, CGM and ISM in satellite galaxies could come in the next five to ten years from the combination of absorption- and emission-line studies across the ultraviolet, optical and radio regimes. This is a rapidly growing field and, despite the limited amount of data currently available, it is already clear that the combination of wide-area integral-field spectrographs such as MUSE and KCWI \citep{morrissey18} with facilities such as ALMA and ASKAP will provide us with important insights on the interplay between different gas phases flowing in and out of galaxies (e.g., \citealp{borthakur19,frye19,peroux19}). Indeed, not surprisingly the picture emerging is already a complex one where, for example, accretion may not always be halted immediately after infall but may depend on the detailed properties of the group and infalling orbit.\\

\vskip 15pt
In conclusion, the topic of this review is a piece of a larger puzzle to understand how gas flows in and out of galaxies, regulates their evolution, and what are the physical processes that affect this balance and transform them into passive systems. Probably the most important limitation that hampers our attempts to obtain a complete picture of galaxy evolution is our inability to trace multiple gas phases at the required depth and resolution, for the same samples of galaxies and across the whole parameter space, even in the local Universe. Another important challenge for the future will be to connect detailed studies of gas and star formation that are currently possible only in nearby systems, to kpc-scale and global studies, linking galaxies to their surroundings and tracing all this back to their progenitors at higher redshift. At the same time, making sense of the avalanche of data expected from the next-generation surveys with SKA, 30m class telescopes and space telescopes will be extremely challenging if not daunting, but undoubtedly this is a stimulating challenge for the future, with lots of exciting, and potentially unexpected, discoveries on the horizon.\\

\begin{acknowledgements}
We thank the Editorial Board of PASA, and in particular Stas Shabala, for inviting us to contribute this work to the Dawes Review series.

We thank two anonymous referees for useful comments that helped improve the clarity of this review.
We thank Peter Ryan for designing the two illustrations presented in Figures 1 and 20 of this review. We thank Claudia Lagos for providing the data from the \textsc{Shark} model used to create Figure 2; 
Aeree Chung for help with the VIVA data used for Figure 5; Matteo Fossati and Rose Finn for providing the data used in Figure 8, Alessandro Boselli for providing part of the data used in Figure 9 and an updated image of NGC~4569 used in Figure 10; Jinsu Rhee for help with Figure 13; Toby Brown and David Stark for sharing the ALFALFA and RESOLVE data used for Figure 15 and Asa Bluck for providing Figure 16.  
We thank Alessandro Boselli, Benedetta Vulcani, Gabriella de Lucia and Stephanie Tonnesen for providing comments on a draft of this review, Paula Calder\'on Castillo for a careful proofreading, and Claudia Lagos for useful discussions.

Parts of this research were conducted by the Australian Research Council Centre of Excellence for All Sky Astrophysics in 3 Dimensions (ASTRO 3D), through project number CE170100013. LC is the recipient of an Australian Research Council Future Fellowship (FT180100066) funded by the Australian Government. We also acknowledge support by the Australian Government through the Australia-Korea Foundation of the Department of Foreign Affairs and Trade (`Galaxy evolution across environments: linking Australian observatories and Korean simulations' -AKF00601).

This research made use of Astropy,\footnote{http://www.astropy.org} a community-developed core Python package for Astronomy \citep{astropy:2013, astropy:2018}; and of NASA's Astrophysics Data System Bibliographic Services.

\end{acknowledgements}

\bibliographystyle{pasa-mnras}

\end{document}